\documentclass[journal=chreay,manuscript=review]{achemso}
\setkeys{acs}{articletitle=true}
\setkeys{acs}{maxauthors=10}
\setkeys{acs}{etalmode=truncate}
\usepackage{amsmath}
\usepackage{amsfonts}
\usepackage{amssymb}
\usepackage{bbold}
\usepackage{graphicx}

\usepackage{color}
\usepackage{multirow}
\usepackage[table,xcdraw]{xcolor}

\usepackage{booktabs}
\usepackage{rotating}
\usepackage{todonotes}

\usepackage{xstring} %for todo notes below, will be removed before the submiission

\newcommand{\smalltodofunc}[2]{\todo[caption={#1}, inline, color=#2]
    {\begin{spacing}{0.5}#1\end{spacing}}}

\newcommand{\smalltodo}[2]{%
    \IfEqCase{#1}{%
        {SR}{\smalltodofunc{\textbf{SR} #2}{cyan}}%
        {HO}{\smalltodofunc{\textbf{HO} #2}{olive}}%
        {KR}{\smalltodofunc{\textbf{KR} #2}{red}}%
        {NH}{\smalltodofunc{\textbf{NH} #2}{yellow}}%
    }[\PackageError{smalltodo}{Undefined option to tree: #1}{}]%
}%

\newcommand*{\plimsoll}{{\ensuremath{-\kern-4pt{\ominus}\kern-4pt-}}}

\definecolor{Gray}{gray}{0.8}

\DeclareMathAlphabet{\mathpzc}{OT1}{pzc}{m}{it}

\title{Implicit Solvation Methods\\ for Catalysis at Electrified Interfaces}
\author{Stefan Ringe}
\email{sringe@dgist.ac.kr}
\affiliation{Department of Energy Science and Engineering, Daegu Institute of Science and Technology (DGIST), Daegu 42988, Republic of Korea}
\altaffiliation{The authors contributed equally to this work}
\author{Nicolas G. H\"ormann}
\affiliation{Fritz-Haber-Institut der Max-Planck-Gesellschaft, Faradayweg 4-6, D-14195 Berlin, Germany}
\alsoaffiliation{Chair for Theoretical Chemistry and Catalysis Research Center, Technische Universit{\"a}t M{\"u}nchen, Lichtenbergstr. 4, D-85747 Garching, Germany}
\altaffiliation{The authors contributed equally to this work}
\author{Harald Oberhofer}
\affiliation{Chair for Theoretical Chemistry and Catalysis Research Center, Technische Universit{\"a}t M{\"u}nchen, Lichtenbergstr. 4, D-85747 Garching, Germany}
\author{Karsten Reuter}
\email{reuter@fhi-berlin.mpg.de}
\affiliation{Fritz-Haber-Institut der Max-Planck-Gesellschaft, Faradayweg 4-6, D-14195 Berlin, Germany}

\date{\today}

\begin{document}

\begin{abstract}
Implicit solvation is an effective, highly coarse-grained approach in atomic-scale simulations to account for a surrounding liquid electrolyte on the level of a continuous polarizable medium. Originating in molecular chemistry with finite solutes, implicit solvation techniques are now increasingly used in the context of first-principles modeling of electrochemistry and electrocatalysis at extended (often metallic) electrodes. The prevalent ansatz to model the latter electrodes and the reactive surface chemistry at them through slabs in periodic boundary condition supercells brings its specific challenges. Foremost this concerns the diffculty to describe the entire double layer forming at the electrified solid-liquid interface (SLI) within supercell sizes tractable by commonly employed density-functional theory (DFT). We review liquid solvation methodology from this specific application angle, highlighting in particular its use in the widespread {\em ab initio} thermodynamics approach to surface catalysis. Notably, implicit solvation can be employed to mimic a polarization of the electrode's electronic density under the applied potential and the concomitant capacitive charging of the entire double layer beyond the limitations of the employed DFT supercell. Most critical for continuing advances of this effective methodology for the SLI context is the lack of pertinent (experimental or high-level theoretical) reference data needed for parametrization.
\end{abstract}

\maketitle
\tableofcontents
\newpage

\section{Introduction}
\label{sec:intro}

Electrocatalysis, i.e.~potential-driven chemistry at electrified interfaces, is one of the pillars of a future sustainable energy landscape, providing a green storage of renewable energy and its conversion to valuable chemicals.\cite{seh2017combining,gur2018review,nong2020key} The concomitant increased global interest in electrochemical processes at extended surfaces and interfaces has triggered unprecedented academic and industrial research efforts to optimize catalyst materials and electrochemical cell designs for maximal efficiency, sustainability, and durability. In this development, predictive-quality computational simulations have played a key role, augmenting experimental results with atomic-scale mechanistic insights and increasingly supporting catalyst discovery and optimization.\cite{chen2019understanding,he2019density,urban2016computational,norskov2011density,hammes2021integration,khatib2021nanoscale}

\begin{figure}[ht!]
    \centering
    \includegraphics[width=0.9\textwidth]{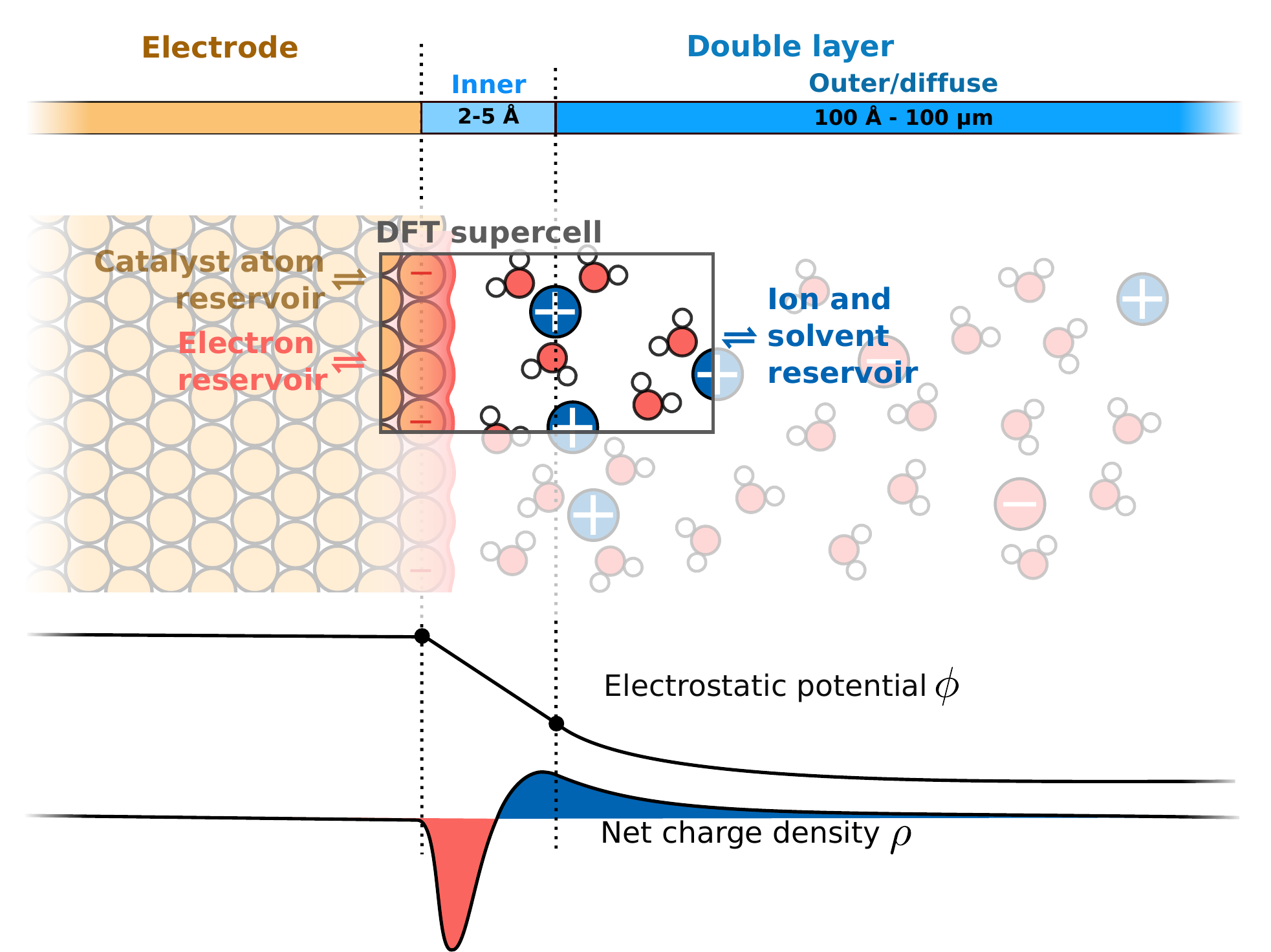}
    \caption{\textbf{\textit{Ab initio} thermodynamics approach to electrified solid-liquid interfaces as occurring in electrocatalysis}. The electrode is here negatively charged and this surface charge is compensated by the built up of counter charge in the electrolyte. The formed electric double layer (DL) can be pictured as a localized capacitor at the interface of electrode and a rather rigid layer of ions (inner DL or Helmholtz layer) and a long-range contribution (outer or diffuse DL). As in particular the spatial extent of the diffuse DL challenges efficient first-principles calculations, the {\em ab initio} thermodynamics approach considers a grand-canonical ensemble in which a finite supercell computed e.g.~with DFT is in equilibrium with appropriate reservoirs for the catalyst atoms, solvent species and electrons. Since the supercell does then generally not comprise the entire DL, it misses part of the compensating charge and does not necessarily have to be overall charge neutral.}
    \label{fig:overview}
\end{figure}

Given the fact that electrochemical reactions depend on the movement of charges, the respective computer simulations are by necessity based on a quantum mechanical description of the involved materials. Yet, while first-principles quantum chemistry provides a conceptually exact toolkit to simulate chemical reactions, current (super-)computers can even with most efficient semi-local density-functional theory (DFT) only simulate a limited amount of atoms and at time scales where chemical reactions cannot be statistically resolved.\cite{bruix2019first} 
Fortunately, energy conversion processes can often be considered as a path through thermodynamically equilibrated, meta-stable states, separated by kinetic barriers which are often in a direct, linear relation with free energy differences between those states.\cite{greeley2016theoretical} Furthermore, chemical reactions frequently occur at defined locations, the so-called active sites, and have a quite localized impact on their surrounding.\cite{reuter2017perspective} As a consequence, and as shown in Fig.~\ref{fig:overview}, to a good approximation one can in many cases carve out from the full constant-particle thermodynamic system a smaller grand-canonical sub-system which is in equilibrium with bulk reservoirs of species.\cite{reuter2016ab} 

In this {\em ab initio} thermodynamics approach to surface catalysis, this sub-system in form of a model of the active site and any adsorbed reaction intermediates can then conveniently be computed as a slab within a periodic boundary condition supercell, and a grand-canonical thermodynamic framework is used to connect the obtained first-principles energetics with the reservoirs through defined chemical potentials for the catalyst atoms and the reactants. In thermal heterogeneous catalysis,\cite{reuter2001composition,reuter2004oxide,rogal2004thermodynamic,stampfl2005surface,reuter2005ab,rogal2006ab,reuter2016ab} where this approach was pioneered and is widely used, the surrounding reactant environment and its corresponding reservoirs are generally well approximated by neutral ideal gases. Concomitantly, also the finite supercell is charge neutral and there is no necessity to explicitly include in the first-principles supercell calculation the gas-phase species that would in principle fill the finite volume between the periodically repeating slabs. Instead, the actual DFT calculations are simply performed for a slab in perfect vacuum. Unfortunately, the situation is significantly more complex in surface electrocatalysis,\cite{trindell2019well} where the solid catalyst exchanges electrons with the reactants and is in contact with a liquid electrolyte forming a solid-liquid interface (SLI). As further detailed below, this enforces the consideration of charged reservoirs (electrons, protons, or ionic species in the electrolyte) with which the then no longer necessarily overall charge-neutral supercell is in electrochemical equilibrium, cf.~Fig.~\ref{fig:overview}. Furthermore, this exchange of charge species with the respective reservoirs and potentially ongoing surface reactions are driven by applied electrostatic potentials, which directly interact with the solvent structure near the surface. Apart from the specifically adsorbed reaction intermediates there is thus now in principle also the need to describe the liquid electrolyte species within the finite volume between the periodically repeating slabs in the supercell.

It is from the objective of reducing this complexity and recovering the efficiency of {\em ab initio} thermodynamics as known from thermal surface catalysis, where much of the renewed interest in implicit solvation schemes in this field comes from.\cite{schwarz2020electrochemical,sundararaman2017evaluating,sundararaman2017grand,gauthier2019challenges,nattino2019continuum,hormann2019grand} Corresponding methodologies form in general a long-standing coarse-grained approach to describe a solvent environment on the level of a dielectric continuum. While they thus have their own history (in particular for molecular systems), their application to extended SLIs and the context of {\em ab initio} thermodynamics has its specific challenges and merits. It is from this particular angle that we here review such methodologies and discuss their recent application to the surface electrocatalysis context, especially at metal electrodes and for liquid, mostly aqueous electrolytes. We refer to excellent and comprehensive reviews for full theoretical and technical details and the more traditional uses of implicit solvation methods for molecular systems,\cite{tomasi2004thirty,tomasi2005quantum,cramer1999implicit} and content ourselves here with a focused exposition of the general concepts. Instead, we elaborate more on the specific demands, benefits and persisting issues when applying such methods to electrified interfaces. 

To set the stage for such a discussion, Fig.~\ref{fig:overview} also summarizes some key properties and specificities of the electrified SLI. Central to this is the separation of (ionic and electronic) charges that results from the interaction of the metallic electrode with the surrounding electrolyte under an applied potential. A potential-dependent amount of net charge $\rho$ is thus localized on the electrode surface and counter charges in the form of dissolved ions are redistributed to a certain depth into the electrolyte to compensate for this net charge. Additionally, rotational, translational and even vibrational degrees of freedom in particular of polar electrolyte molecules (like water in aqueous electrolytes) will be affected within this formed, so-called electric double layer (DL).\cite{eftekhari-bafrooei2011effect,nihonyanagi2013structure} As a consequence of the concomitant screening, the electrostatic potential $\phi$ drops over the width of the DL. At least in aqueous electrolytes, this drop generally occurs over two regions: the inner or Helmholtz\cite{helmholtz1853ueber} layer (iDL), where $\phi$ drops linearly, and the outer or diffuse layer, where it drops non-linearly. The capacitance $C$ arising from the charging of the DL is correspondingly also commonly separated into an inner and an outer contribution.\cite{parsons1997metal,lockett2010differential} While this was originally made without a direct reference to the actual atomic-scale nature of the DL, the different dielectric property of the iDL is now related to a crowding of counter ions directly at the charged electrode. This leads to the formation of a compact layer with almost rigid water molecules and thus a small dielectric permittivity.\cite{parsons1997metal,fumagalli2018anomalously,kornyshev2007electrochemical,nihonyanagi2013structure} In contrast, depending on the applied potential and electrolyte concentration, the more diffuse re-distribution of ions in the outer DL can extend over hundreds of {\AA}ngstroms into the electrolyte, cf.~Fig.~\ref{fig:overview}.

From this simplified capacitor picture, it becomes clear that the true amount of net surface charge on the electrode at a given applied potential is a sensitive function of the entire DL. Adsorption energies and therefore reaction pathways in turn often depend sensitively on this surface charge and the potential drop in the DL, e.g.~via electrostatic interactions of dipolar adsorbates with the electric field.\cite{chen2016electric,sandberg2016co,resasco2017promoter,ringe2019understanding,ringe2020double} 
Already this aspect alone thus reveals that electrochemical activity in the SLI is generally not merely a function of the electrode aka catalyst material. Instead, it is equally influenced by the electrolyte and the concomitant DL. Additional aspects of this influence concern also more classic solvation effects like steric or bonding interactions with electrolyte species in the inner DL (in aqueous electrolytes e.g.~prominently hydrogen bonds).\cite{dellesite2007interplay,ludwig2019solvent,heenen2020solvation,ludwig2020atomistic,oberhofer2020electrocatalysis} Capturing these multifaceted influences and in particular their net effect on reaction energetics is correspondingly a pivotal ingredient of predictive-quality computational simulations and theoretical analyses of catalysis at electrified interfaces. At the same time and as further discussed in Section~\ref{sec:FundCont} below, the outer DL's large extent plus the ions' very slow, typically nanosecond time scale diffusion render any atomic-scale first-principles calculations including an explicit and dynamical account of the full DL still prohibitively expensive.\cite{gauthier2017solvation} 

Implicit solvation schemes are at the opposite end and promise an unsurpassed computational efficiency in simulating the SLI.\cite{sundararaman2017evaluating,nattino2019continuum,schwarz2020electrochemical} In their original molecular form, these schemes define a solvation cavity in which the solute is embedded and surrounded by a dielectric continuum representing the solvent's dielectric response.\cite{tomasi1994molecular,tomasi2005quantum} 
On top of that, the contribution of ions to the overall electrostatic response can be modeled. In the application to SLIs, such implicit solvation schemes thus foremost allow to appropriately describe the capacitive charging of the DL beyond the confines of the finite supercell---though requiring the integration into an {\em ab initio} thermodynamics framework to appropriately account for the flow of particles between the subsystem and the reservoirs (cf.~Fig.~\ref{fig:overview}) as detailed in Section~\ref{sec:aitd}. Next to effectively describing the counter charge, implicit solvation models obviously also aim to capture plain solvation effects.\cite{tomasi2005quantum}
Yet, with the solvent represented by a continuum this is, of course, only on a highly effective, parametrized level, in particular in the present state-of-the-art that also includes the inner DL into the implicit description.\cite{sundararaman2017evaluating,nattino2019continuum,schwarz2020electrochemical} As further discussed in Section~\ref{sec:param}, this situation is aggravated even more by the scarcity of reliable experimental SLI data to fit the empirical parameters to. The prevalent approach to instead more or less uncritically resort to established parameters from (unbiased) molecular systems represents one of the aforementioned persisting issues in the field. It is these open challenges that are specific to the application of implicit solvation schemes to the context of SLIs that we also want to openly voice in this review, while simultaneously surveying the impressive insights that can be achieved with this at first sight admittedly rather crude approach. 

\section{Fundamentals of implicit solvation}
\label{sec:solvtheory}
\subsection{Coarse-graining the electrolyte}
\label{sec:FundCont}

Since the beginning of computational chemistry, the simulation of solid-liquid interfaces has been of particular interest to scientists. To facilitate such investigations, theorists have since developed various methods particularly to coarse grain the highly dynamic and thus complex liquid phase. Indeed, the oldest such methods go back to Kirkwood\cite{kirkwood1934theory} and Onsager,\cite{onsager1936electric} and were introduced even a few years before the invention of the electronic computer. The goal back then was essentially the same as for the here discussed contemporary SLI electrocatalysis context, namely to reduce the physical complexity of the liquid phase in such a way as to keep the essential physics of the problem intact. Practically, these theories are derived for the description of thermodynamic equilibrium states by averaging over the configurational phase space. Of course, this can be achieved at varying degrees of coarseness which we will briefly survey in the following.

\begin{figure}[ht!]
\includegraphics[width=0.8\textwidth]{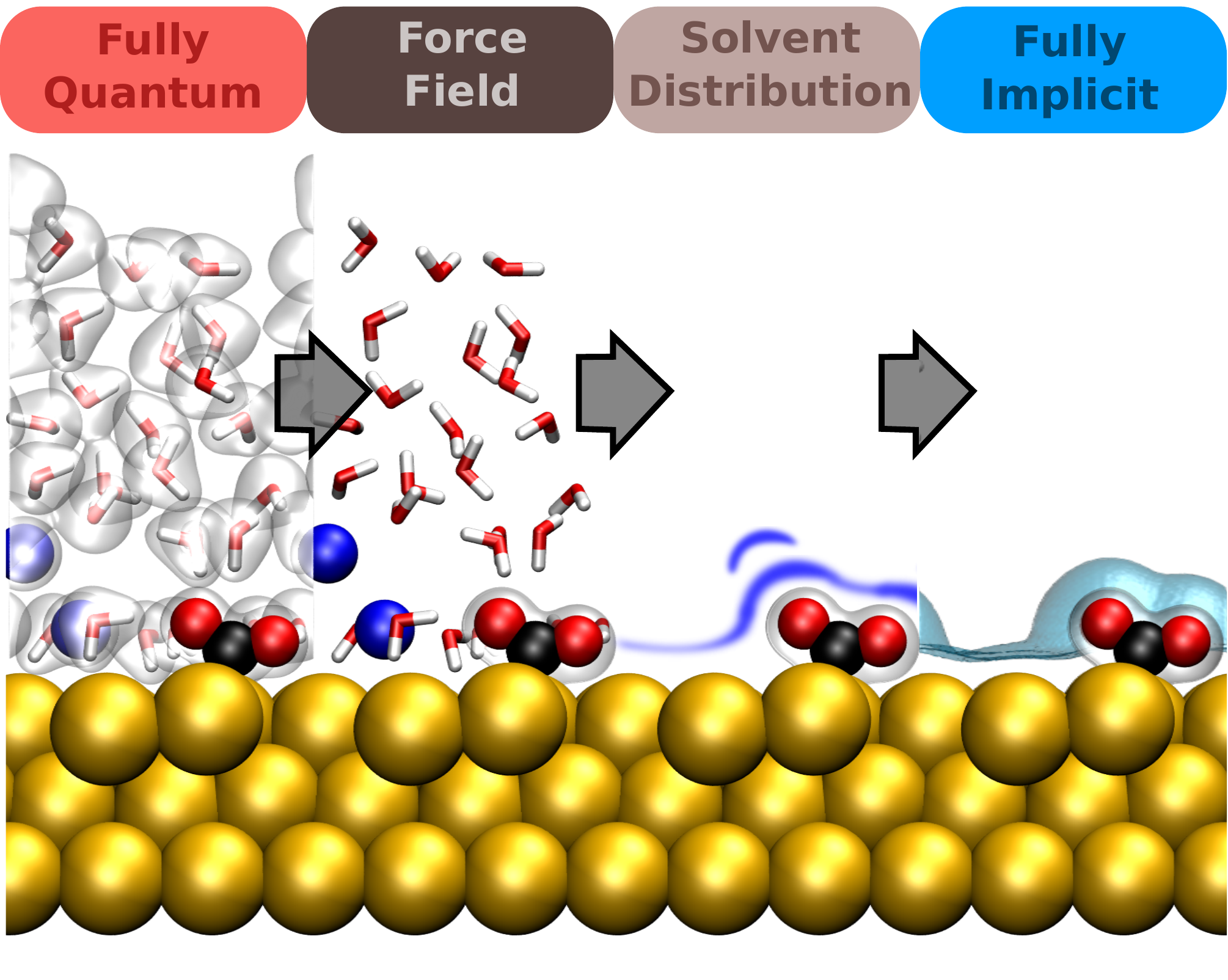} 
\caption{{\bf Hierarchy of coarse-graining approaches for the liquid phase in the context of electrocatalysis at SLIs.} The sketch depicts an aqueous electrolyte with salt ions (blue spheres) and dissolved CO\textsubscript{2} (red and black molecules) at a crystalline surface. Starting from a fully explicit quantum mechanical description (far left) one can conceptually coarse-grain away electronic DOFs to arrive at a force field or interatomic potential description (center left). From that one can gradually remove nuclear solvent DOFs to represent solvent molecules e.g.~only through their spatial distributions or correlation functions like in RISM-type models (center right). Finally, replacing even this with simply a polarizable continuum one arrives at fully implicit models (far right). Note that in the derivation and parametrization of each coarse-grained level one does not necessarily need to follow each step and can e.g.~directly parametrize an implicit model from fully explicit data.}
\label{fig:coarse_grain}
\end{figure}

The starting point of our discussion is a fully \textit{ab initio}, quantum mechanical treatment of the liquid phase, including all electronic and nuclear degrees of freedom (DOFs). Given the mobility of the molecules in the liquid phase, the evaluation of equilibrium states requires some sort of averaging or sampling over the nuclear DOFs, most often achieved in the form of \textit{ab initio} molecular dynamics (AIMD).\cite{hassanali2014aqueous,iftimie2005ab} Unfortunately, even at this fully explicit level there is still some debate which first-principles electronic structure theory is actually best suited for the task. Specifically for the description of pure water, easily the most important of solvents, there are a number of well documented failures of semi-local DFT,\cite{vandevondele2005influence,gillan2016perspective} which in terms of its computational efficiency would be the present-day method of choice to describe larger supercells and achieve longest possible simulation times.\cite{gross2014water,le2017determining}
Instead, the use of hybrid DFT with advanced dispersion corrections,\cite{gaiduk2015density,ambrosio2016structural} or the strongly constrained and appropriately normed (SCAN) meta-GGA functional\cite{chen2017ab,zheng2018structural} is often recommended, potentially even including nuclear quantum effects.\cite{ruiz2018quest} This best practice becomes challenged in the SLI context though---not only because of potentially exploding computational costs, but also because the same functional now has to describe the (metallic) solid and the liquid phase with their very different physical characteristics on the same footing. For this specific task, the use of generalized-gradient functionals, in particular the revised version of the Perdew-Burke-Ernzerhof (RPBE) functional\cite{hammer1999improved} corrected for dispersion interactions using the semi-emperical D3 approach by Grimme,\cite{grimme2010consistent,grimme2016dispersion} is presently often perceived as an acceptable compromise.\cite{tonigold2012dispersive,gross2014water,sakong2016structure,ando2013ab,sakong2018electric,le2020molecular} However, one clearly has to stress that this consensus derives more from reasonably-appearing averaged properties and functionalities computed at this level of theory, rather than from detailed experimental validation of the predicted atomic-scale structure of the electric DL.

Although AIMD simulations provide valuable insights about SLIs, they can usually only sample a few or even a single basin of the system's potential energy surface (PES) during presently computationally tractable trajectories on the picosecond time scale. Proper thermodynamic averages would instead require nanoseconds of simulations or longer, especially if the DL contains slowly equilibrating components such as ions or strongly physisorbed water.\cite{he2018statistical,gauthier2019challenges} Furthermore, the simulation cell sizes feasible even over restricted picosecond time scales can barely, if at all cover the up to $\sim 100$\,{\AA} extent of the outer DL, cf.~Fig.~\ref{fig:overview}. All these limitations can at present only be overcome by switching to more coarse-grained descriptions especially of the liquid phase as summarized in Fig.~\ref{fig:coarse_grain}. 

The first in the corresponding hierarchy of approaches focuses on eliminating the electronic DOFs. This results in a classical description of pair-wise or many-body interactions between point-like nuclei in the form of an effective force field or interatomic potential to model the high-dimensional PES.\cite{harrison2018review,cisneros2016modeling} While this is an extensive field of its own with a plethora of most advanced force fields for (bulk) water, electrolytes or materials, the crux is again in requiring them to describe the SLI within the same simulation cell. Much fewer parametrizations exist for this task, in particular for the interactions of (organic) electrolyte species with the (inorganic and heavy) elements like Pt or Cu that form the metallic electrodes. On top of that, most traditional force fields can not reliably describe bond forming or breaking events and can thus not cover the reactive surface chemistry that is central to catalysis at electrified interfaces. While there are thus only few examples where fully classical simulations were used to study the structure of SLIs,\cite{steinmann2018force,clabaut2020ten} there are currently two interesting developments to overcome these limitations. To one end, modern reactive force fields that can account for bond dissociation start being applied in SLI simulations
\cite{buckley2019electrocatalysis,senftle2016reaxff} even under applied potential.\cite{liang2018applied,onofrio2015voltage,onofrio2015atomic,xu2018pulse} To the other end, machine-learned interatomic potentials are a most promising new possibility to establish a computationally efficient surrogate to direct first-principles calculations.\cite{ko2021fourth,csanyi2021gaussian}
By construction, their reliability and range of applicability is determined by the training data fed into them. If this data contains appropriate information on the SLI and its reactive events, dynamical simulations based on such a potential would produce the same insight as direct AIMD, just orders of magnitude faster. Precisely the development of corresponding data-efficient training protocols (that would not require prohibitive amounts of first-principles training data) is presently the focus of strong research efforts worldwide. As this research is ongoing, present applications of machine-learned potentials to the SLI context are still restricted to some first case studies though.\cite{kondati2017self,natarajan2016neural,quaranta2017proton} 

An important general aspect in switching to more coarse-grained descriptions is that different levels may suitably be chosen for different spatial regions of the overall simulation cell. In the SLI context, a widespread realization of such concurrent multiscale modeling is a quantum mechanical/molecular mechanical (QM/MM) approach,\cite{bruix2019first,lim2016seamless,stecher2016first,gim2018multiscale,gim2019structure,naserifar2021artificial} in which the solid electrode and the chemical reactions thereon are kept on a quantum chemical level, while a force field or interatomic potential is employed for the liquid electrolyte. This offers significant speed-ups as much of the electrolyte sampling is done classically, while in particular the reactive surface chemistry is still described at a first-principles level. Note that the (spatial) distinction of what is described at the more coarse-grained level can be chosen flexibly, with the limitation that approaches that allow to continuously morph say a classically described atom into a quantum mechanically described one during an ongoing dynamical simulation are still in their infancy.\cite{bulo2009toward,watanabe2014size,watanabe2019quantitative}
Typically, which atoms (or molecules) are described at which level is therefore defined at the onset of a simulation, and this is kept fixed regardless of where the actual dynamical motion drives the atom or molecule to. A classical description of all electrolyte species apart from (specifically) adsorbed reaction intermediates offers thereby obviously highest computational efficiency, but is by construction unable to cover situations where the liquid phase participates actively in the reactions, e.g.~as a proton donor.\cite{stecher2016first} Furthermore, it also requires in principle specific (interface-sensitive) parametrizations to account for the overall effect of the classical solvent species on the surface reactions. Both of these limitations can instead be mitigated by including (parts of) the inner DL into the quantum mechanical part of the simulation, yet at concomitantly increased computational costs. 

Central to the value of such simulations is in any case the correct depiction of the interaction or embedding energy of solid and liquid phase, the solvation energy. In QM/MM models, the Coulomb contribution to the solvation energy is commonly described by the interaction of the QM charge distribution with fixed, fitted electrostatic charges of the classically described liquid molecules. In addition to this, non-Coulomb contributions, including Pauli repulsion, dispersion and induction forces, have to be carefully parametrized.\cite{gim2018multiscale,gim2019structure} Electronic induction of the solid phase by the liquid phase charge distribution is treated by self-consistently re-iterating the liquid distribution and electron density.\cite{lim2016seamless} Polarization of the liquid phase is instead most often only included through movement and reorientation of solvent molecules and ions. In certain situations an additional electronic polarization, i.e.~changes of the partial charges of atomic sites of the solvent molecules, has been shown to be relevant and can in principle be included using polarizable force fields.\cite{naserifar2018quantum} 
The description of the other, non-Coulomb interactions is still a topic of ongoing research. Commonly they are simply represented by pairwise interactions with parameters obtained from high-level quantum chemical calculations\cite{gim2018multiscale} or by fitting to thermodynamic or dielectric properties of the (bulk) solvent.\cite{wu2006flexible}
Nevertheless, properly parametrized force fields have actually been shown to sometimes even surpass full AIMD simulations in accuracy concerning structural and dynamic properties of the solvent.\cite{zielkiewicz2005structural}
Their still atomistic approach to representing the liquid phase also has advantages over the more coarse-grained models discussed in the following, in that they can more readily describe localized effects and directed interactions such as hydrogen bonds to surface adsorbates. 

While a QM/MM description of the SLI greatly speeds up simulations by simplifying the computational treatment of the liquid DOFs,\cite{wang2012polarizable,stecher2016first,clabaut2020solvation}
it still does not relieve the need to sufficiently sample the phase space of each solvent molecule. Combined with the need to still determine the QM polarization response to each new MM charge configuration, even such simplified models might not be computationally tractable. Recognizing the explicit sampling of the solvent dynamics as the bottleneck, a further coarse-graining step aims therefore at effectively averaging out the movement of solvent molecules and ions, and at replacing them instead with their respective spatial equilibrium distributions, cf.~Fig.~\ref{fig:coarse_grain}. A prominent representative of this ansatz is the reference interaction site model (RISM),\cite{beglov1997an} which evaluates the equilibrium radial correlation functions of each pair of species in the system through an analytical integral equation, known as the Ornstein-Zernike equation.\cite{hirata2003molecular} Within given approximations,\cite{percus1958analysis,van1959new} the equilibrium structure of the fluid around any form of solute is then fully encoded through these radial distribution functions and without further need for a costly dynamical sampling. In most variants of RISM, such as the popular 3D-RISM,\cite{kovalenko1999self} the central pair correlation functions are evaluated on a three-dimensional grid centered on the solute to yield the spatial distribution functions of each solvent site species. These distribution functions can be integrated over space and summed up to yield an excess chemical potential of solvation due to the solute-solvent interaction and solvent reorganization in the presence of the solute. In RISM theory, it is this excess chemical potential that connects the coarse-grained solvent with the explicitly treated solute. Its functional derivative with respect to the electron density yields an effective potential that can directly be included into the solute's Hamiltonian. This potential includes all the interactions used in the determination of $g_{ij}(\mathbf{r})$ such as electrostatics and, most commonly,\cite{gusarov2006self} Lennard-Jones type terms encompassing dispersion and exchange interactions. Given the implicit dependency of the solvent excess chemical potential on the electron density, the solvent response is then iterated together with the quantum-mechanical DFT-described part of the system to reach self-consistency.\cite{nishihara2017hybrid} Going beyond purely molecular solvents,
RISM-like models recently have seen very successful use in the simulations of various electrochemical processes.\cite{haruyama2018analysis,zhan2018origins,zhan2019specific,sugahara2019negative,weitzner2020toward}

Inherent to effective models, both explicit classical and RISM-based descriptions of the liquid phase depend on a series of element-specific parameters that define interatomic interactions and have to be carefully chosen for each system of interest. This requirement is generally not a significant burden for detailed studies of individual systems, in particular if these are prototypical cases for which then typically a plethora of high-level or experimental data is available that can be used for the parametrization. It becomes critical though, if fast estimates are needed, for instance to assess the  catalytic activity of a large variety of electrode materials, morphologies, active sites or electrolyte components, or if unknown and complex electrochemical reactions are studied for which no reference data is available. For such cases and for potential further increases in efficiency, an even higher level of coarse graining of the liquid phase becomes appealing, in which all solvent DOFs are altogether merely described via a polarizable continuum, cf.~Fig.~\ref{fig:coarse_grain}.

Following the concurrent multiscale modeling philosophy of QM/MM or QM/RISM, such implicit solvation schemes are in the SLI context predominantly employed to describe the equilibrium solvent response on a (metallic) electrode computed at a first-principles level of theory. Again, flexibility exists whether to replace the entire electrolyte in a so-called fully implicit approach, or to retain an explicit quantum or molecular mechanical description of (parts of) the inner DL, with latter models referred to as hybrid explicit/implicit models. Reduced to a continuum, the implicitly treated electrolyte is then just a polarizable medium with a dielectric permittivity. While an isotropic, constant tensor in the bulk of the electrolyte, this permittivity can in principle vary closer to the symmetry-breaking SLI. Additionally, it needs to be artificially reduced to vacuum permittivity inside the explicitly treated region of the simulation cell so as to not introduce spurious polarizability on top of the one intrinsically provided by the quantum or molecular mechanical description of the corresponding atoms or molecules. This region of vacuum permittivity inside the overall simulation cell is commonly referred to as solvation cavity, a word coined within the traditional field of implicit solvation of finite moleculear solutes. As discussed in detail in Section~\ref{sec:Elstat}, different classes of implicit solvation schemes are categorized by the functional form employed to describe these spatial variations of the dielectric permittivity tensor. This form determines the electrostatic solvent response and could in principle be chosen to be non-local to reach similar levels of accuracy as RISM models.\cite{sundararaman2015spicing}
However, the use of such functional forms would unavoidably require the introduction of a multitude of system-specific parameters, thereby nullifying the original motivation for this effective methodology. 

For planar electrodes (typically described by crystalline slabs with low-index surfaces in the corresponding first-principles supercell calculations), it is therefore common to only consider a local and stationary dielectric tensor with components that vary exclusively as a function of the vertical distance $z$ to the surface.\cite{bonthuis2012profile} In fact, typically even the tensorial nature of the permittivity is neglected and a simple functional form for the scalar permittivity $\epsilon(z)$ is employed. As this omits all structure in the liquid and especially any kind of directed interactions with the surface, such effects are instead considered by additional effective non-Coulomb energy functionals as discussed in more detail in Section~\ref{sec:ImpSolvNonelstat} below. This particular strategy then allows to employ a minimum number of parameters for the dielectric modeling function and these non-Coulomb energy corrections as further discussed in Section~\ref{sec:param}. 

We also discuss prevalent fitting strategies for these parameters in Section~\ref{sec:param}, but note already here that the simplicity of this prevalent approach does not only reflect the objective of creating a computationally most effective, transferable solvation approach. To some extent and as mentioned before it is also dictated by the present scarcity of interface-sensitive experimental or high-level theoretical reference data that does not warrant a more detailed (physical) modeling with a concomitantly increased number of parameters. This aspect notably also concerns the powerful possibility of extending implicit solvation schemes from pure liquids to electrolytes by additionally modeling the ionic charge distribution as discussed more in Section~\ref{sec:ImpSolvElectrolyteModels}. Most of these models rely on the traditional diffuse DL theory, providing a functional form between ion distributions and electrostatic potential as developed by Gouy, Chapman and Debye in the beginning of the last century.\cite{gouy1910constitution,chapman1913theory,gouy1916sur,debye1923theory,borukhov1997steric} Since this original approach, many corrections regarding e.g.~non-mean-field ionic correlation effects, steric size corrections or ion-surface interactions have been made. While physically clearly motivated, each of these corrections necessarily gives rise to further parameters. Even though it is in particular this capability to account for the ionic counter charges that is presently predominantly exploited for the SLI context, it is thus again a specific issue of this application field in how much these more advanced electrolyte models can be parametrized or are in fact really necessary for the specific counter charge modeling aspect.

\subsection{Separation of the grand potential energy functional}
\label{sec:ImpSolvGeneralFormulation}

As apparent from the discussion in the last section, different levels of theory ranging from high-level quantum chemistry to force fields or interatomic potentials may generally also be chosen for the description of the solid electrode (and an explicitly treated part of the inner DL). In the remainder of this review we will nevertheless focus on the use of DFT for this task, as this is the predominantly taken approach in implicit solvation works on SLIs and electrocatalysis at metallic electrodes to day.\cite{trindell2019well} With minor modifications, many of the concepts and discussions are readily adapted to the other levels of theory though.

As described in the introduction around Fig.~\ref{fig:overview}, in the SLI context, the employed DFT supercell at volume $V$ generally only represents a grand-canonical sub-system, which is connected to bulk reservoirs of species that represent the rest of the (macroscopic) system. For the electrochemical environment, these would naturally include an electrochemical potential $\tilde{\mu}_{\rm el}$ for the electrons, electrochemical potentials $\tilde{\mu}_{{\rm ion},i}$ for different ionic electrolyte species $i$, and chemical potentials $\mu_{{\rm solv},j}$ for different neutral solvent species $j$. In Chapter \ref{sec:echemSLI} we will detail how these potentials are set for the SLI context, but for the time being they are simply given constants. For such given constants, the true equilibrium structure and composition of the electrified interface would result from an exhaustive grand-canonical sampling and thermodynamic averaging of all nuclei and electronic DOFs inside the supercell---with nuclei DOFs here and henceforth denoting the detailed geometric structure and chemical composition of the system and electronic DOFs referring to those of the DFT-part of the system. In the coarse-grained solvation modeling reviewed here, this typically infeasible task is separated into two stages. First, solvation effects are evaluated for an individual interface configuration characterized by say a given electrode geometry and chemical composition with specifically adsorbed reaction intermediates at its active sites. The chemical composition $N_\alpha$ of chemical species $\alpha$ in this explicitly and DFT-described part of the system is thus fixed, and under an unanimous Born-Oppenheimer approximation the thermodynamic sampling and averaging is restricted to the remaining (canonical) electronic and (grand-canonical) nuclei DOFs of the electrolyte. In other words, one thus evaluates the thermodynamic stability of the electronic ground-state configuration for the given static nuclei charge density $\rho_{\rm nuc,QM}=\rho_{\rm nuc,QM}({\bf r})$ and in contact with a fully equilibrated electrolyte. In a subsequent step detailed in Section~\ref{sec:aitd}, an {\em ab initio} thermodynamics framework is then employed to compare the stability of different such explicit interface configurations and compositions, and the one exhibiting the highest stability is identified as the closest approximant to the true grand-canonical equilibrium SLI structure within the tested space of configurations.

In this and the remaining sections of this chapter we will concentrate on the first of these two stages. In this stage, there is thus one defined chemical composition $N_\alpha$ of the DFT-described part of the system, and in this respect this stage then 
encompasses the more traditional use of implicit solvation schemes in the molecular DFT context with finite solutes. The central ansatz taken to accomplish the thermodynamic evaluation at this stage is to partition the overall system's energy, and establish a grand potential energy functional of the charge density distribution $\rho_{\rm is} = \rho_{\rm is}({\bf r})$ of the classical electrolyte and the electron density $\rho_{\rm el,QM}=\rho_{\rm el,QM}({\bf r})$ of the DFT-described part
\begin{subequations}
\begin{align}
\Omega^{N_\alpha}[\rho_\mathrm{el,QM},\rho_{\rm is}] &= \left. F_\mathrm{QM}\left[\rho_\mathrm{el,QM}\right] \right|_{\rho_{\rm nuc,QM}} \;+\; \Omega_\mathrm{is}[\rho_\mathrm{el,QM}, \rho_{\rm is}] \quad ,
\label{eq:refinal_free}
\end{align}
which is minimized by the equilibrated charge density distribution $\rho^\circ_{\rm is}$ and the ground-state electron density $\rho^\circ_{\rm el,QM}$. Here, $F_\mathrm{QM}$ is the free energy functional of the pure quantum system and $\Omega_\mathrm{is}$ is the grand potential of the surrounding electrolyte. For simplicity of notation, we drop in the following the subscript ``QM'' (e.g.~$F_\mathrm{QM}\rightarrow F$), and consistently denote all properties related to the electrolyte with the subscript ``is'' (for implicit solvent). Within the employed Born-Oppenheimer approximation, we also henceforth refrain from explicitly stating the only parametric dependence of $F$ on the nuclei charge density $\rho_{\rm nuc}$. Within Kohn-Sham (KS) DFT, $F$ is commonly expressed as
\begin{align}
F[\rho_\mathrm{el}]&=\underbrace{\overbrace{
\mathcal{T}^\mathrm{S}_\mathrm{el}[\rho_\mathrm{el}]  + \mathcal{V}^\mathrm{mf}[\rho_\mathrm{el}] + E^\mathrm{xc}[\rho_\mathrm{el}]}^{E^\mathrm{KS}[\rho_\mathrm{el}]}  + \mathcal{T}_\mathrm{nuc}[\rho_\mathrm{el}]}_{U[\rho_\mathrm{el}]} - TS[\rho_\mathrm{el}]\quad.
\label{eq:omega_ks}
\end{align}
Here, $\mathcal{T}_\mathrm{el}^\mathrm{S}$ is the kinetic energy functional of the non-interacting electrons and $\mathcal{T}_\mathrm{nuc}$ represents the kinetic energy functional of the nuclei (usually evaluated only as a post-correction at $\rho^\circ_{\rm el}$). The Coulomb energy functional $\mathcal{V}^\mathrm{mf}$ contains both nuclei-nuclei interactions described explicitly and electronic interactions described on the mean-field level, while additional electronic interactions are accounted for through the DFT exchange-correlation functional $E^\mathrm{xc}$. $E^\mathrm{KS}$ is generally referred to as the KS energy functional, and finally, $TS$ represents entropic corrections at the given temperature $T$. As indicated, all terms in $F$ with the exception of the last one are often summarized under the header of the internal energy functional $U$.

Importantly, $F[\rho_\mathrm{el}]$ with all its terms is exactly the functional also underlying regular DFT calculations and does thus not depend on the electrolyte distribution $\rho_{\rm is}$. We correspondingly refer to a multitude of excellent accounts on KS DFT for further details on this functional.\cite{becke2014perspective} All electrolyte-induced changes of the ground-state electron density arise instead from the optimization of the grand potential $\Omega^{N_\alpha}[\rho_\mathrm{el},\rho_{\rm is}]$ in eq.~(\ref{eq:refinal_free}) and not $F[\rho_\mathrm{el}]$ alone. In contrast, $\Omega_\mathrm{is}[\rho_\mathrm{el},\rho_{\rm is}]$ as the second part of this grand potential refers to the electrolyte in its equilibrium distribution, and this does depend on the detailed charge distribution of the DFT-described solute and thus its electron density $\rho_{\rm el}$. Conceptually, in order to determine $\Omega_{\rm is}[\rho_\mathrm{el},\rho_{\rm is}]$ all electrolyte DOFs would therefore have to be sampled in the presence of a given $\rho_{\rm el}$, and then the interdependence of electrolyte and DFT system charge densities would require an iterative cycle or generally a numerical optimizer to minimize the functional $\Omega^{N_\alpha}[\rho_\mathrm{el},\rho_{\rm is}]$ with respect to the electron density at a corresponding equilibrium charge density distribution of the electrolyte. This has e.g.~been realized in electrostatic QM/MM embedding,\cite{lim2016seamless,stecher2016first} where molecular dynamics simulations are used to sample the equilibrium distributions of the electrolyte DOFs. Similarly QM/RISM simulations have been employed, in which a more coarse-grained model is used to derive the electrolyte equilibrium distribution corresponding to a given electron density of the QM system\cite{nishihara2017hybrid} as already mentioned in Section~\ref{sec:FundCont}.

The great advantage of implicit solvation schemes over these less coarse-grained approaches is that there a model solvation grand potential $\Omega_{\rm is}[\rho_\mathrm{el}]$ is derived solely as an explicit functional of the electron density $\rho_\mathrm{el}$. This leads to a dramatic reduction of computational effort, as then the evaluation of the resulting closed form of $\Omega^{N_\alpha}[\rho_\mathrm{el}]$ can be achieved for a given $\rho_\mathrm{el}$ in one go. In fact, corresponding schemes are often directly integrated into the DFT program packages by simply adding routines that evaluate and add the $\Omega_{\rm is}[\rho_\mathrm{el}]$ contribution within the regular KS DFT minimization procedure. For this, it seems at first natural to separate the model grand potential functional into formal terms analogous to the quantum free energy functional $F[\rho_\mathrm{el}]$,
\begin{align}
\Omega_\mathrm{is}[\rho_\mathrm{el}] =\underbrace{
\mathcal{T}_\mathrm{is}[\rho_\mathrm{el}] + \mathcal{V}_\mathrm{is}[\rho_\mathrm{el}]
}_{U_\mathrm{is}[\rho_\mathrm{el}]} - TS_\mathrm{is}[\rho_\mathrm{el}] -\sum_i \tilde{\mu}_{\mathrm{ion},i} \langle N_{\mathrm{is,ion},i}\rangle[\rho_\mathrm{el}] - \sum_j \mu_{\mathrm{solv},j} \langle N_{\mathrm{is,solv},j} \rangle[\rho_\mathrm{el}]\quad ,
\label{eq:omegais}
\end{align}
\label{eq:omegafull}
\end{subequations}
with the respective kinetic, potential and internal energy functionals $\mathcal{T}_\mathrm{is}$, $\mathcal{V}_\mathrm{is}$  and $U_\mathrm{is}$, and the entropic contribution denoted by $S_\mathrm{is}$. Note that as a grand potential, $\Omega_\mathrm{is}$ formally also contains contributions due to the electrochemical potentials of the ionic ($\tilde{\mu}_{\mathrm{ion},i}$) and chemical potentials of the neutral solvent species ($\mu_{\mathrm{solv},j}$). The inclusion of these terms---and especially their average particle numbers $\langle N_{\mathrm{is,ion},i}\rangle$ and $\langle N_{\mathrm{is,solv},j} \rangle$ in the implicit electrolyte---does at first seem counter-intuitive given that all explicit solvent and ion degrees of freedom have been coarse-grained out. Yet, as we will see below, even implicit ionic and solvent molecule concentrations in the simulation box depend on the electrochemical environment (e.g.~the applied potential). Therefore the exchange of both kinds of particles with the extended electrolyte (as represented by the (electro)chemical potentials) needs to be accounted for, at least approximately.

In general, and as further elaborated on in Section~\ref{sec:aitd}, one is actually rarely interested in the absolute grand potential of eq.~(\ref{eq:refinal_free}). Instead it is differences in free energies, and thus differences in grand potentials at their respective optimal electronic densities, that are the main descriptors of chemical reactions. Similarly, comparisons with experiment---which are generally used for model parametrization---are also most easily done on the level of solvation free energies,\cite{hille2019generalized} which in turn are differences between the grand potential at optimized densities in solvent and in vacuum. For this purpose and considering the strong approximations to be made anyway, the fine separation into the various formal terms in eq.~(\ref{eq:omegais}) is not ideal. With the aim to later on exploit partial cancellations and to ultimately create computationally most tractable terms, it has instead proven more convenient to group the different contributions by their physical origin\cite{tomasi2005quantum}
\begin{equation}
    \Omega_{\rm is}[\rho_\mathrm{el}] = \mathcal{V}_\mathrm{is}^\mathrm{mf}[\rho_\mathrm{el}] + \Omega_\mathrm{is}^\mathrm{non-el}[\rho_\mathrm{el}] + \Omega_\mathrm{is}^{\rm ion}
		[\rho_\mathrm{el}] \quad .
		\label{eq:Gsolv}
\end{equation}
Here, $\mathcal{V}_\mathrm{is}^\mathrm{mf}$ is the mean-field contribution due to the electrostatic response of the continuous polarizable medium describing the pure liquid. Interactions with the pure liquid beyond this mean-field electrostatics are accounted for by the second term, which summarizes a number of so-called non-electrostatic contributions
\begin{equation}
    \Omega_\mathrm{is}^\mathrm{non-el}[\rho_\mathrm{el}]\;=\; \Omega_\mathrm{is}^{\rm cav}[\rho_\mathrm{el}] + G_\mathrm{is}^{\rm rep}[\rho_\mathrm{el}] + G_\mathrm{is}^{\rm dis}[\rho_\mathrm{el}] + G_\mathrm{is}^{\rm tm}[\rho_\mathrm{el}] \quad ,
  \label{eq:dGnonel}  
\end{equation}
while the last term $\Omega_{\rm is}^{\rm ion}[\rho_\mathrm{el}]$ in eq.~(\ref{eq:Gsolv}) describes all additional effects introduced by ions in the electrolyte. Even though in practice often further lumped together, cf.~Section~\ref{sec:param}, we here distinguish four non-electrostatic contributions. $\Omega_\mathrm{is}^{\rm cav}[\rho_\mathrm{el}]$ denotes the grand potential cost of forming a cavity in the solvent for the solute to be placed in. Making this space for the solute necessarily changes the particle numbers of solvent molecules of the implicitly-described liquid in the supercell and thus involves particle exchange with the reservoirs with a concomitant dependence on the chemical potentials of the solvent components. We accordingly denote this term here as a grand potential, even though most available literature refers to it as a cavitation free energy functional. $G_\mathrm{is}^{\rm rep}[\rho_\mathrm{el}]$ commonly represents the contribution due to exchange or Pauli repulsion interactions, effectively also including an entropic contribution due to the resulting changes to the potential energy surface (PES). The third term, $G_\mathrm{is}^{\rm dis}[\rho_\mathrm{el}]$, similarly represents dispersion or van der Waals interactions. Finally, $G_\mathrm{is}^{\rm tm}[\rho_\mathrm{el}]$ is the free energy functional accounting for changes in the thermal motion of the solute. 
Note that all of the non-electrostatic terms and $\Omega_{\rm is}^{\rm ion}[\rho_\mathrm{el}]$ thus contain potential, kinetic and entropic contributions. Nevertheless, each of these terms has been proven to be computationally accessible and in the following sections, we will now further elaborate on these various contributions to $\Omega_{\rm is}[\rho_\mathrm{el}]$, starting first with a pure solvent and the discussion of the dominant electrostatic $\mathcal{V}_\mathrm{is}^\mathrm{mf}[\rho_\mathrm{el}]$ term in Section~\ref{sec:Elstat} and the non-electrostatic $\Omega_\mathrm{is}^\mathrm{non-el}[\rho_\mathrm{el}]$ in Section~\ref{sec:ImpSolvNonelstat}. In Section~\ref{sec:ImpSolvElectrolyteModels}, ions are then added on top of that to arrive at full implicit electrolyte models that also include a model $\Omega_{\rm is}^{\rm ion}[\rho_\mathrm{el}]$ grand potential. The general objective in all of these sections is to derive (closed) expressions for these functionals of the electron density, which then allows to (straightforwardly) add these contributions into the KS DFT minimization process. As noted before, the true free energy is then formally given by the grand potential $\Omega^{N_\alpha}[\rho^{\circ}_\mathrm{el}]$ evaluated at the resulting optimized density $\rho^{\circ}_\mathrm{el}$, cf.~eq.~(\ref{eq:omegafull}a). However, it is important to note that, to this end, the practical implementations acknowledge the aforementioned fact that predominantly only grand potential energy differences are required. In such differences of say systems $A$ and $B$, 
\begin{align}
    \Delta \Omega(A,B) \;=\; \Omega^A[\rho^\circ_\mathrm{el}(A)] - \Omega^B[\rho^\circ_\mathrm{el}(B)] \quad,
    \label{eq:omegadiff}
\end{align}
contributions to $\Omega_{\rm is}[\rho_\mathrm{el}]$ that are not particularly sensitive to the detailed form of the optimized densities $\rho^\circ_\mathrm{el}(A)$ and $\rho^\circ_\mathrm{el}(B)$ will largely cancel. From this perspective, no efforts are therefore made to account for such contributions in the derived functional expressions in the first place. While formally describing the absolute solvation grand potential of eq.~(\ref{eq:Gsolv}), we thus emphasize that in practice many of the expressions discussed in the next sections only work for free energy differences. In fact, not least for reasons of computational efficiency the practical implementations often also consider only some terms within $\Omega_{\rm is}[\rho_\mathrm{el}]$ in the functional minimization. One justification for this is an assumed negligible impact of the omitted terms on the final optimized electron density. Another pragmatic one is that any error incurred through the omission is effectively compensated in the fitting of the model parameters to reference data.\cite{ringe2017transferable} A prominent example for this is to only consider the electrostatic $\mathcal{V}_\mathrm{is}^\mathrm{mf}[\rho_\mathrm{el}]$ in the minimization, and evaluate all non-electrostatic free energy contributions only as a post-correction on the basis of the resulting electron density that was thus exclusively optimized with respect to the dominant mean-field polarization effect of the surrounding liquid.

\subsection{Electrostatics of solvation}
\label{sec:Elstat}
\subsubsection{Potential energy and polarization models}
\label{sec:PotenPolMod}

The mean-field electrostatic $\mathcal{V}_\mathrm{is}^\mathrm{mf}$ is the contribution to the solvation grand potential most intuitively associated with the response of a solvent to a solute. Considering it jointly with the Coulomb energy functional $\mathcal{V}^{\rm mf}$ in the minimization of the grand potential energy functional of eq.~(\ref{eq:refinal_free}) accounts for the polarization response of the continuum solvent to the net charge distribution of the solute $\rho$ (resulting from the electron $\rho_\mathrm{el}$ and nuclei $\rho_\mathrm{nuc}$ charge densities of the DFT part of the system) and vice versa. To derive this contribution, we consider the static displacement field $\boldsymbol{D}$ which arises from the collection of these explicit charges in the system and is screened by the surrounding medium. $\boldsymbol{D}$ is given by the generalized Poisson equation (GPE)\cite{jackson1998classical}
\begin{equation}
   \nabla \boldsymbol{D}=\left( \rho_\mathrm{el}+\rho_\mathrm{nuc}\right) \quad.
    \label{eq:GPE}
\end{equation}
The displacement field is related to the electric field $\boldsymbol{E}$ of the explicit charge distribution via the polarization vector $\boldsymbol{P}$, representing permanent and induced dipoles in the system. The functional form of $\boldsymbol{P}$ is generally quite complicated, but depends on
the relative and generally non-local dielectric permittivity tensor $\boldsymbol{\varepsilon}=\boldsymbol{\varepsilon}_\mathrm{tot}/\varepsilon_0$ (with the absolute $\boldsymbol{\varepsilon}_\mathrm{tot}$ and vacuum permittivity $\varepsilon_0$), 
\begin{equation}
    \boldsymbol{D}=\varepsilon_0\boldsymbol{E}+\boldsymbol{P}[\boldsymbol{E},\boldsymbol{\varepsilon}] \quad.
    \label{eq:displacement}
\end{equation}
Technically, this makes $\boldsymbol{D}$ a functional of the electric field $\boldsymbol{E}$, which itself is an implicit functional of the net charge density ($\rho_\mathrm{el}+\rho_\mathrm{nuc}$) via eq.~(\ref{eq:GPE}). For reasons of legibility we dropped these dependencies though.
In this definition of $\boldsymbol{D}$, the permittivity tensor is assumed to be static---i.e.~time independent, but may still vary in space, e.g.~to account for the symmetry breaking through a finite solute or an extended interface. Note that eq.~(\ref{eq:displacement}) also omits higher-order multipolar terms that might arise in the medium. For water, this approximation is generally well justified because the solvent molecules' electric field is dominated by its dipole moment. Higher-order terms can, however,\cite{slavchov2014quadrupole} be important in non-aqueous solvents with sizeable higher-order multipole moments, but to our knowledge no DFT program package yet supports an implicit solvent parametrization including such higher-order terms. 

The GPE of eqs.~(\ref{eq:GPE}) and (\ref{eq:displacement}) provides a direct relation of electric field and charge density which is generally valid, with and without a polarizable medium. It can be used to find an analytic expression for the electrostatic Coulomb potential energy contribution of an arbitrary embedded charge distribution
\begin{align}
    \left(\mathcal{V}^\mathrm{mf}+\mathcal{V}_\mathrm{is}^{\rm mf}\right)[\rho_\mathrm{el}]=\frac{1}{2}\int \left(\rho_\mathrm{el}+\rho_\mathrm{nuc}\right) \phi =\frac{1}{2}\int \boldsymbol{E}\boldsymbol{D} \mathrm{d} \boldsymbol{r} \quad,
    \label{eq:potengeneral}
\end{align}
where the last equation can be obtained from inserting eq.~(\ref{eq:displacement}) into eq.~(\ref{eq:GPE}), using the divergence theorem, neglecting the surface terms and finally substituting $\boldsymbol{E}=-\nabla \phi$, with $\phi$ the electrostatic potential.

\begin{figure}[htb]
\includegraphics[width=0.9\textwidth]{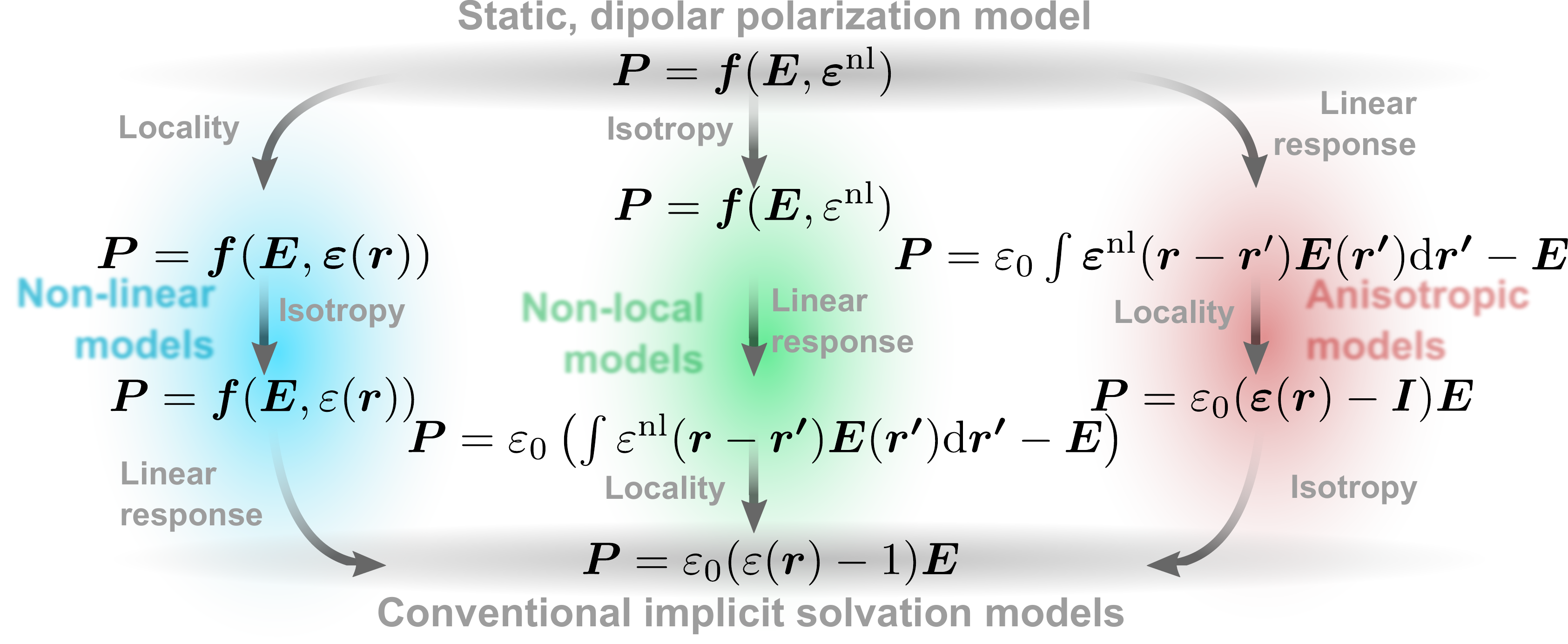} 
\caption{{\bf Categorization of different electrostatic solvation models}. From the general starting point of a static non-linear, non-local and anisotropic model (top) several approximations can be made to ultimately arrive at the linear, local and isotropic polarization model most commonly applied in present-day DFT codes.}
\label{fig:dielectric_categories}
\end{figure}

The assumption of a static, i.e.~frequency independent, dielectric permittivity implies that the solvent adapts instantaneously to the electron and nuclei charge distribution of the solute. While this is generally a good approximation for the solvent response on thermodynamic equilibrium and potentially even for transition states of chemical reactions, it will over-screen fast molecular dynamics, such as vibrations or charge-transfer processes.\cite{gauthier2017solvation} On top of that, the simulation of electronically excited states has been shown to generally also necessitate a frequency-dependent dielectric response.\cite{baumeier2014electronic,duchemin2016combining} For most other cases, however, the static, dipolar response model is a good starting point for further approximations. As compiled in Fig.~\ref{fig:dielectric_categories}, these lead to three main categories of dielectric models, namely non-linear, non-local and anisotropic ones. Non-linearity in the solvent response can be important in cases where the electric field is large, which actually can be the case inside the electric DL.\cite{schwarz2020electrochemical,gunceler2013importance} Notwithstanding, mostly a Taylor expansion of $\boldsymbol{P}$ as a function of $\boldsymbol{E}$ around $\boldsymbol{E}=0$ can be truncated after the linear term (linear-response approximation), i.e.
\begin{align}
    \boldsymbol{P}\approx 0+\varepsilon_0\left({\boldsymbol \varepsilon}-\boldsymbol{I}\right) \boldsymbol{E} \quad,
    \label{eq:linearresponse}
\end{align}
with the medium's electric susceptibility directly expressed as $({\boldsymbol \varepsilon}-\boldsymbol{I})$.\cite{jackson1998classical} Next, non-locality in the solvent response is important, whenever solvent molecule correlations occur, e.g.~close to charged solutes. The spherically averaged  liquid  susceptibility (SaLSA) model represents one example that accounts for non-locality.\cite{sundararaman2015spicing} SaLSA has been also coarse-grained into a computationally more feasible local version (charge-asymmetric nonlocally determined local-electric, CANDLE), with the dielectric permittivity being derived from the non-local response.\cite{sundararaman2015charge} Non-locality may also be employed to account for an effective size of solvent molecules, since the electric field at a certain position then affects the solvent density in a finite solvent radius around it.\cite{andreussi2019solvent} This can be relevant to prevent solvent from penetrating into small pockets formed by the solute, see the discussion on the dielectric function below. Finally, anisotropic properties of the dielectric permittivity are, of course, generally important in systems with reduced symmetry. This is notably the case at electrified SLIs where even at a planar interface the dielectric tensor would at least feature two independent dielectric tensor components, parallel $\varepsilon_{\vert\vert}$ and vertical $\varepsilon_{\perp}$ to the surface.\cite{bonthuis2013beyond}  

While non-linearity, non-locality and anisotropy could thus well be of relevance for SLIs, most implicit solvation models that have been implemented into DFT program packages to date neglect all three of them and are based on the most simple case of a linear, local and iosotropic dielectric model $\varepsilon({\bf r})$. For this case, the GPE becomes
\begin{equation}
    \nabla \boldsymbol{D}=\varepsilon_0\nabla\left[ \varepsilon(\boldsymbol{r})\boldsymbol{E}\right]=-\varepsilon_0\nabla\left[ \varepsilon(\boldsymbol{r})\nabla\phi\right]=\left( \rho_\mathrm{el}+\rho_\mathrm{nuc}\right)
    \label{eq:GPE_linearresponse}
\end{equation}
and the electrostatic Coulomb potential energy of eq.~(\ref{eq:potengeneral}) can be further simplified. Using eq. (\ref{eq:linearresponse}), it then features separately the electrostatic energy functional contributions of the DFT part and of the implicit solvent
\begin{align}
    \left(\mathcal{V}^\mathrm{mf}+\mathcal{V}^{\rm mf}_\mathrm{is}\right)[\rho_\mathrm{el}]=\frac{1}{2}\int \left(\rho_\mathrm{el}+\rho_\mathrm{nuc}\right) \phi=\underbrace{\frac{\varepsilon_0}{2}\int \boldsymbol{E}^2 \mathrm{d} \boldsymbol{r}}_{\mathcal{V}^\mathrm{mf}}+\underbrace{\frac{1}{2}\int \varepsilon_0\left(\varepsilon({\boldsymbol r})-1\right)\boldsymbol{E}^2 \mathrm{d} \boldsymbol{r}}_{\mathcal{V}^{\rm mf}_\mathrm{is}} \quad,
\label{eq:elstat_energy}
\end{align}
with $\phi$ an implicit functional of $\rho_\mathrm{el}$ via eq.~(\ref{eq:GPE_linearresponse}). Since the latter GPE cannot be solved analytically for most dielectric functions, a closed form is typically not attainable and a numerical solution is required. Common methods for this include fixed point iterations or the conjugate gradient technique employing the analytically known Green's function of the Poisson equation in vacuum.\cite{andreussi2012revised,ringe2016function} Alternatively, for certain functional forms of the dielectric function multi-center multipole expansions have been shown valuable,\cite{sinstein2017efficient} or mappings onto a finite grid and solution via standard finite difference or finite element techniques. Regardless of this technical realization, the conceptual changes to a DFT code to incorporate the Coulomb electrostatic contribution at this level of dielectric model are nevertheless minimal. In fact, while the entire self-consistency cycle around the KS equations is untouched, the only change is that the electrostatic potential does no longer satisfy the normal Poisson equation, but is instead given by the GPE of eq.~(\ref{eq:GPE_linearresponse}).\cite{ringe2017first}

\subsubsection{Dielectric function}
\label{sec:DielecFunc}

For the linear, local and isotropic case, the dielectric permittivity $\varepsilon({\bf r})$ may generally still vary in space. As already introduced in Section~\ref{sec:FundCont}, in present-day implicit solvation schemes this is primarily reduced to modeling a transition from the bulk solvent permittivity $\varepsilon_0 \varepsilon_\infty$ (with the relative solvent permittivity $\varepsilon_\infty$) deep inside the electrolyte to the vacuum permittivity $\varepsilon_0$ inside the DFT-described part of the supercell. The optimum location and form of this transition is generally system specific. Optimum refers hereby to the best possible reproduction of the true solvation effects within the confines of the chosen dielectric continuum model, and---in particular in the widespread approach to even include the inner DL fully into the implicit model---system specific includes an actual dependency on the electrode structure and chemical composition. In principle, this optimum location and form for a specific system could be determined from high-level explicit simulations.\cite{bonthuis2013beyond} However, this would negate the original motivation to use an implicit solvent model for its efficiency gain and to e.g.~screen a large number of different SLIs. Implicit solvation schemes rely therefore typically on a sufficiently simple functional form of $\varepsilon({\bf r})$ which includes as much system-relevant physics as possible while maintaining an optimum transferability. Obviously, this implies a trade-off between a more physically accurate description for particular systems (then typically involving a larger number of parameters that need to be determined) and a more simplified model with as little parameters as possible to describe qualitative trends over a wide range of systems.

\begin{figure}[ht]
\includegraphics[width=0.8\textwidth]{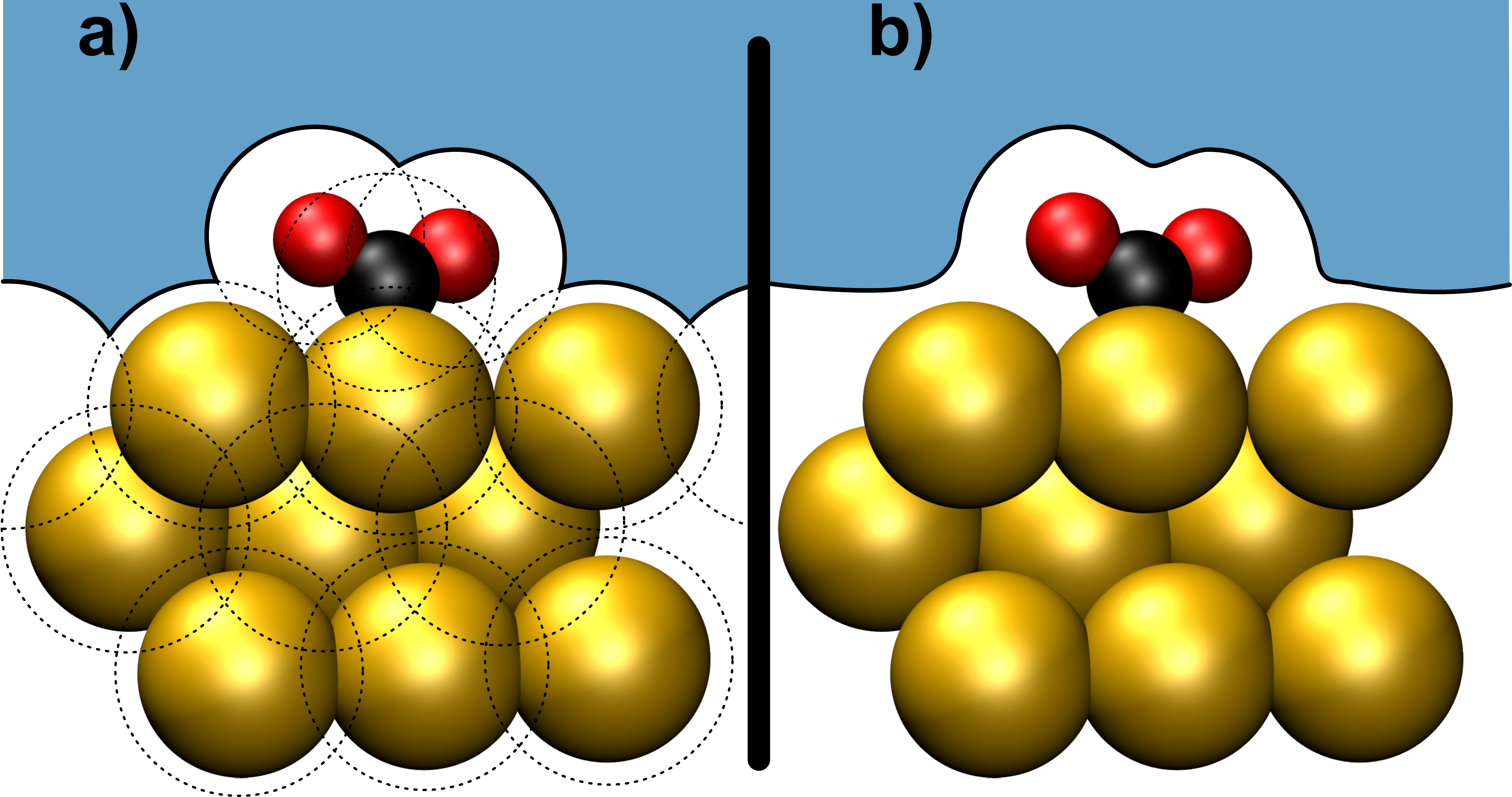}
\caption{{\bf Illustration of different types of dielectric transition between solute and solvent}. For the example of an adsorbed CO$_2$ molecule at a single-crystal surface, a) shows the solvation cavity resulting from the superposition of atom-centered spheres based on eq.~(\ref{eq:trans-atr}), while b) shows the solvation cavity as defined by an isosurface of the electron density.}
\label{fig:solv_cavity}
\end{figure}

Favoring higher transferability, the dielectric transition is often approximated by a mere switching function between bulk solvent and vacuum, resulting in the formation of a solvation cavity. The location of the dielectric transition thereby has to be expressed in an appropriate molecular descriptor that is readily available in any DFT calculation. For this and as illustrated schematically in Fig.~\ref{fig:solv_cavity}, traditional implicit solvation techniques as dominantly used in molecular chemistry often rely on defining a solvation cavity by summing up atom-centered shape functions $s({\bf r})$, so that
\begin{equation}
\varepsilon(s(\boldsymbol{r}))=\left(\varepsilon_\infty-1\right)\left\{\prod_\alpha s_{\boldsymbol{r}}(|\boldsymbol{r}-\boldsymbol{R}_\alpha|,\boldsymbol{\mathcal{P}}_\alpha)\right\}+1\quad.
\label{eq:trans-atr}
\end{equation}
Here $s_{\boldsymbol{r}}$ is a shape function going from 0 in the solute region to 1 in the bulk electrolyte, $\{\boldsymbol{R}_\alpha\}$ are the positions of the nuclei, and $\boldsymbol{\mathcal{P}}_\alpha=(\{r_\alpha\},\dots)$ is a vector of parameters, containing e.g.~the exclusion radius $r_\alpha$ for each atom and the transition smoothness of the shape function. The simplest shape function is just a single Heaviside function $s_{\boldsymbol{r}}=\theta(|\boldsymbol{r}-\boldsymbol{R}_\alpha|-r_\alpha)$, with the atomic radii $r_\alpha$ as the only parameters. These radii are usually either taken as tabulated van der Waals radii for each chemical element or fitted to reproduce some experimental data as discussed in more detail in Section~\ref{sec:param}. One advantage of using such a sharp step function, also sometimes referred to as apparent surface charge approach, is the efficiency with which the GPE can be solved using boundary element methods. Yet, in most cases\cite{sinstein2017efficient} this comes at the expense of additional approximate corrections for errors due to parts of the QM charge density lying beyond the transition. Corrections for this outlying-charge error are correspondingly integral parts of well-known implicit solvation approaches like the polarizable continuum model (PCM),\cite{tomasi2005quantum} the solvent model (SMx),\cite{marenich2007self} or the conductor-like screening model (COSMO)\cite{klamt1993cosmo} that rely on such sharp step functions. As an alternative, recently also smoothed step functions were proposed and adapted specifically for SLI simulations (soft-sphere continuum solvation - SSCS model\cite{fisicaro2017soft}), then, however, requiring additional parameters for the functional form of this transition.

In general, defining the cavity based on atom-centered shape functions has the advantage of easily being able to implement dielectric regions, e.g.~at dielectric interfaces, by assigning different values to the local dielectric permittivity. Additionally, solvation radii can be assigned separately to each atom based on their chemical environment. This allows for great flexibility in the definition of the dielectric function and, potentially, a more accurate prediction of solvation energies. Unfortunately, such a treatment also results in a larger parameter space, risking overfitting\cite{hille2019generalized} with the generally rather small available training sets as further discussed in Section~\ref{sec:param}. Furthermore, the reliance on atom-centered shapes may lead to the formation of encapsulated solvent pockets in lower-density parts of the solute.\cite{andreussi2019solvent} In particular in the context of extended metallic electrodes, filling such pockets with solvent unlikely reflects the correct physics.

Both of these limitations may be overcome in a different, equally popular approach. It recognizes that the presence of electron density---readily available in a DFT calculation---naturally separates explicitly treated regions from the rest of the supercell. The solvation cavity can thus be defined by an iso-surface of the electron density. Regions of lower $\rho_{\rm el}$ than the chosen iso-value are then classified as the solvent, while regions of higher $\rho_{\rm el}$ obviously represent the DFT-treated part of the system. In practice, smoothed shape functions are employed,
\begin{equation}
\varepsilon(\rho_\mathrm{el}(\boldsymbol{r}))=(\varepsilon_\infty-1) s_{\rho_\mathrm{el}}\left(\rho_\mathrm{el}(\boldsymbol{r}),(\rho_\mathrm{el,min},\rho_\mathrm{el,max})\right)+1 \quad ,
\end{equation}
where $\rho_\mathrm{el,min}$ and $\rho_\mathrm{el,max}$ are the minimal and maximum electron density between which the shape function $s_{\rho_\mathrm{el}}$ switches from bulk solvent to vacuum. This kind of parametrization has for instance been employed in the self-consistent continuum solvation (SCCS) model by Andreussi \textit{et al.}\cite{andreussi2012revised}. Equivalently, also the iso-value itself could be used as parameter, with the transition width then as corresponding second parameter.\cite{fattebert2003first,petrosyan2005joint} Various smooth shape functions have been proposed in the literature,\cite{andreussi2012revised,sundararaman2014weighted,fisicaro2017soft} resulting, however, in quite similar predictive accuracy of molecular solvation energies. While this suggests the actual shape to be less influential for the model performance, some functions like the one proposed in the SCCS model are constructed to have an exactly zero gradient outside the transition region, which is beneficial for the numerical solution.\cite{andreussi2012revised,ringe2016function} The advantage of the electron density based approach in general is that the solvation cavity adapts self-consistently to the electron density and exhibits thus a more physically reasonable and smooth shape.\cite{andreussi2012revised} From a technical standpoint, though, dielectric functions based on the electron density are slightly more involved to implement due to additional Pulay forces arising there.\cite{ringe2017first}

Both, atom-centered shape function and electron density based approaches are generally challenged in the description of solutes at different charge states. In the molecular context, different parameter sets defining the solvation cavity are often required for anions on the one hand, and cations and neutral molecules on the other.\cite{dupont2013self} To overcome this limitation, Sundararaman \textit{et al.}~proposed an extended form of the dielectric function,\cite{sundararaman2014weighted} that in addition to defining the transition region via the electron density allowed for a correction based on the locally averaged outward electric field. This field has inverse signs for cation- and anion-like regions, and thus provides the model with the fundamental capability to shift the dielectric transition region accordingly without the need to invoke different parameters. A similar approach has recently been followed by Truscott and Andreussi,\cite{truscott2019field} who utilized the SSCS atom-centered shape function model and allowed the atomic spheres to relax their radius depending on the value of the electric field flux through their surface. 

\subsection{Non-electrostatics of solvation}
\label{sec:ImpSolvNonelstat}

The interaction between solute and solvent is not solely restricted to the electrostatic mean-field treatment described in the last section, even though especially for the study of electrified interfaces changes in the electrostatic potential can be expected to be dominant.\cite{tomasi2005quantum} Nevertheless, it is often minute changes to free energy profiles of reactions at these interfaces that can result in crucial changes of the catalytic activity or in particular of catalytic selectivities---and for such minute changes the additional beyond mean-field and non-electrostatic interactions could prove decisive. In this section we discuss the corresponding terms in the solvation grand potential, cf.~eq.~(\ref{eq:dGnonel}), the physical background for them and how they are commonly treated. As will become apparent, this treatment is generally highly effective and thus incurs in principle multiple additional parameters. Not least from a parametrization point of view, but also for reasons of computational efficiency and to exploit potential error cancellation, modern implementations in DFT packages therefore rarely calculate these terms individually.\cite{tomasi2005quantum} Instead, some or all of these terms are instead lumped together into empirical functions with a minimum number of parameters. Highly successful examples for this are the SMx\cite{cramer1999implicit} family of methods or the SCCS approach.\cite{andreussi2012revised}
As it is important to understand the physical backgrounds of these terms to appreciate the origin of the added free parameters and the lumping strategies, we will nevertheless  discuss each term in more detail in the following. The parametrization done in practice is then covered in Section \ref{sec:param}, while a more complete overview of non-electrostatic treatments in other (not necessarily implicit) solvation models can for example be found in the recent review by Schwarz and Sundararaman\cite{schwarz2020electrochemical} or the exhaustive review by Tomasi, Menucci, and Cammi.\cite{tomasi2005quantum}

\subsubsection{Cavitation grand potential, $\Omega_{\rm is}^\mathrm{cav}$}
\label{sec:ImpSolvCavity}

The placement of a solute, be it a single molecule, a cluster, or an extended electrode surface always leads to the displacement of solvent molecules to form the solvation cavity. The work necessary for this displacement is commonly referred to as the cavity formation energy. It can, in principle, be calculated from explicit solvent simulations, e.g.~employing Monte Carlo or molecular dynamics\cite{postma1982thermodynamics,pohorille1990cavities,floris1997free,hofinger2005simple}, or information-theoretic maximum-entropy simulations \cite{hummer1996an,pratt2002molecular}. Yet, such a costly treatment is obviously not a desirable basis for the development of a simple cavitation grand potential functional within the context of implicit solvation models.

Instead, such development relies to a large extent on scaled particle theory, which essentially employs a hard-sphere representation of solvent and solute.\cite{reiss1959statistical} In this case, the formed cavity is simply the excluded volume around a solute given in terms of the hard spheres of solute and solvent molecules. For such a simplified model, $\Omega_{\rm is}^\mathrm{cav}[\rho_{\rm el}]$ can then be established analytically to yield an explicit expression that depends only on molecular parameters of solute and solvent.\cite{reiss1960aspects} One example is the solution of Pierotti,\cite{pierotti1976scaled} which is e.g.~implemented in the popular PCM solvation model, and reads up to third order in the hard-sphere radius $r_{\rm hs}$ of a given solute\cite{tomasi2005quantum}
\begin{equation}
     \Omega_{\rm is}^\mathrm{cav}=k_B T\left\lbrace -\ln(\zeta)+ \xi\left(\frac{r_{\rm hs}}{r_{\rm hs,solv}}\right)+\left[\xi+\frac{\xi^2}{2}\left(\frac{r_{\rm hs}}{r_{\rm hs,solv}}\right)^2\right]
\right\rbrace \quad .
\label{eq:gcavP}
\end{equation}
Here, $k_B$ is Boltzmann's constant, and both $\zeta=\zeta(r_{\rm hs,solv})$ and $\xi=\xi(r_{\rm hs,solv})$ are unit-less auxiliary functions of $r_{\rm hs}$ and the solvent hard-sphere radius $r_{\rm hs,solv}$. Note that this formulation only accounts for a single sphere type each for all solute and for all solvent species, and thus does not necessarily reflect the actual shape of the cavity very well. As a remedy, extensions to multiple different radii have e.g.~been proposed by Claverie {\em et al.}\cite{langlet1977studies} Nevertheless, the accuracy of such scaled particle theory based approaches still rests fully on the choice of solute and solvent radii. Many approaches have correspondingly been taken to fit such radii to various experimental properties\cite{cole1957induced,mayer1963molecular,wilhelm1971estimation} and at various experimental conditions\cite{ben-amotz1993molecular,tang2000excluded} (thereby implicitly including the grand-canonical dependence of the cavity formation on the electrochemical environment). For a comprehensive discussion of all these approaches we refer the reader to the excellent review by Tomasi and co-workers.\cite{tomasi2005quantum} Here we only note, that typically the cavity used to establish the expression for $\Omega_{\rm is}^\mathrm{cav}[\rho_{\rm el}]$ does not resemble the solvation cavity used in the mean-field electrostatic $\mathcal{V}_{\rm is}^{\rm mf}[\rho_{\rm el}]$. Given the effective nature of implicit solvation models, this is not {\em per se} a problem. It does, however, potentially add more and unnecessary parameters.

A different approach, based on the seminal work of Uhlig,\cite{uhlig1937solubilities} instead tries to link $\Omega_{\rm is}^\mathrm{cav}[\rho_{\rm el}]$ to the solvent's macroscopic surface tension, thereby eliminating the need to define species-specific parameters altogether. Where this original formulation assumed a spherical cavity of size $r_{\rm cav}$ around the entire solute and independence of solvent parameters beyond the surface tension, more recent formulations account for geometric properties and density of the solvent,\cite{tunon1993continuum} or for deviations from the spherical shape.\cite{tolman1949effect} Especially the latter correction by Tolman\cite{tolman1949effect} proved popular and reads
\begin{equation}
    \Omega_{\rm is}^\mathrm{cav}=4\pi r_{\rm cav}^2 \overline{\gamma}\left(1-\frac{2\delta}{r_{\rm cav}}\right)
    \label{eq:tolman}
\end{equation}
with an effective surface tension $\overline{\gamma}$ and a parameter $\delta$ accounting for deviations from the spherical form. In principle, a direct connection between cavitation energy and surface tension seems obvious, considering that a cavity is essentially an internal interface between solvent and vacuum. Yet, it is not at all clear, that such a relation also has to hold on the microscopic level where cavities are not significantly bigger than solvent molecules, or at least that $\overline{\gamma}$ is in any sense connected to the macroscopic surface tension. Yet, a number of works\cite{floris1997free,huang2001scaling} have shown the Tolman equation, eq.~(\ref{eq:tolman}), to hold and $\overline{\gamma}$ to be near indistinguishable from the macroscopic surface tension. 

The fact remains, though, that also this approach needs parameters describing the shape of the cavity on top of those already used in the mean-field electrostatic model. This can be avoided by recognizing that the term $4\pi r_{\rm cav}^2 \left(1-\frac{2\delta}{r_{\rm cav}}\right)$ in eq.~(\ref{eq:tolman}) essentially just describes the surface area of the cavity, per definition of the surface tension as free energy per area. Based on this, Scherlis and co-workers suggested\cite{scherlis2006unified} that a most straightforward expression for the cavitation grand potential functional could be, 
\begin{equation}
    \Omega_{\rm is}^\mathrm{cav}=\gamma A_{\rm cav} \quad ,
    \label{eq:scherlis1}
\end{equation}
with $\gamma$ the macroscopic surface tension of the solvent and $A_{\rm cav}$ now the surface area of the solvation cavity employed in the electrostatic model. In electrostatic models where the cavity is defined through a step function in the dielectric permittivity, such an area can be calculated quite straightforwardly through some form of tessellation of the surface.\cite{sinstein2017efficient} In models that rely on a continuous dielectric function with a smoothed transition, the surface area of the cavity seems less obvious. To this end, Scherlis {\em et al.}\cite{scherlis2006unified} employed the concept of a quantum surface. Introduced by Cococcioni and co-workers\cite{cococcioni2005electronic} and refined by Andreussi {\em et al.}\cite{andreussi2012revised}, this is essentially a continuous integral over the points in space, which are part of the finite transition region of the shape function $s_{\rho_\mathrm{el}}$, where $\nabla s_{\rho_\mathrm{el}}\ne0$. Numerically, the integral over the gradient is solved by rewriting the gradient as derivative of the electron density by employing the chain rule and differentiation using a finite difference
\begin{equation}
    A_{\rm cav} = \int {\rm d}\boldsymbol{r} \nabla s_{\rho_\mathrm{el}} \approx \int {\rm d}\boldsymbol{r}\left\lbrace\left(s_{\rho_\mathrm{el}}\left[\rho_{\rm el}(\boldsymbol{r})-\frac{\Delta}{2}\right]-s_{\rho_\mathrm{el}}\left[\rho_{\rm el}(\boldsymbol{r})+\frac{\Delta}{2}\right]\right)\times
\frac{|\nabla \rho_{\rm el}(\boldsymbol{r})|}{\Delta}\right\rbrace \quad.
\label{eq:scherlis2}
\end{equation}
This describes a thin film between two density iso-surfaces with a thickness $\Delta$. The exact value of $\Delta$ thereby proved to be unimportant as long as it is large enough to avoid numerical noise due to the real-space integration grid of the specific DFT code, and small enough to still follow the contours of the cavity.\cite{scherlis2006unified} 

On the plus side, based on eqs. (\ref{eq:scherlis1}) and (\ref{eq:scherlis2}), $\Omega_{\rm is}^\mathrm{cav}[\rho_{\rm el}]$ may then straightforwardly be determined without adding any free parameters beyond those already necessary for the electrostatic part---if indeed the macroscopic surface tension $\gamma$ is employed. As discussed in Section~\ref{sec:param}, $\gamma$ may also be seen as an empirical parameter, in which case at least still only one additional parameter would be required. This more effective view is also more consistent with a downside of the cavity definition through the quantum surface concept of eq.~(\ref{eq:scherlis2}). Since the latter depends on $\rho_{\rm el}$ and its gradient, additional terms arise when explicitly including a corresponding cavitation functional term in the KS DFT minimimization. For this reason, the free energy contribution due to a $\Omega_{\rm is}^{\rm cav}[\rho_{\rm el}]$ based on eqs. (\ref{eq:scherlis1}) and (\ref{eq:scherlis2}) is typically only considered as a post-correction for an electrostatically optimized electron density as already discussed at the end of Section~\ref{sec:ImpSolvGeneralFormulation}. 

Finally, a conceptually related approach to this is the weighted-density cavity formation model by Sundararaman and co-workers.\cite{sundararaman2014weighted} There, instead of a cavity composed of overlapping spheres, one formulates a solvent-center cavity, where the tails of the electron density are expanded by the van der Waals radius of the solvent molecules to gain a more physical representation of the solvent accessible area of a solute. Based on this approach one can then derive an expression for $\Omega_{\rm is}^\mathrm{cav}[\rho_{\rm el}]$ that fulfills known physical limits for very small cavities or on the opposite end for droplets of solvent in vacuum.

\subsubsection{Exchange repulsion, $G_{\rm is}^\mathrm{rep}$}
\label{sec:g-rep}

While $\Omega_{\rm is}^\mathrm{cav}[\rho_{\rm el}]$ represents the thermodynamic cost of creating a cavity in the solvent for the solute to fit in, it does not include actual interactions between solute and solvent that are lost in the coarse graining of the solvent DOFs. The free energy functional $G_{\rm is}^\mathrm{rep}[\rho_{\rm el}]$ is supposed to account for repulsive such interactions, predominantly arising from Pauli exchange. While there is a whole hierarchy of methods developed to treat this term,\cite{tomasi1994molecular} modern implicit solvation models generally employ only either of two routes, a more quantum-mechanically inspired one and a more empirical one.\cite{tomasi2005quantum} Recognizing that exchange repulsion originates fundamentally from the overlap of the electron densities of solute and solvent,\cite{margenau2013theory} $G_{\rm is}^\mathrm{rep}[\rho_{\rm el}]$ is in the former approximated from the explicitly available electron density lying outside of the cavity\cite{amovilli1997self} or in the latter through a Lennard-Jones based metric of how close the various solute atoms could approach the cavity.

In the former more quantum-mechanically inspired ansatz, the exchange repulsion functional is specifically given as an overlap integral over the explicit DFT electron density outside the cavity with a model solvent electron density approximated as a simple Gaussian with a width $\xi_{\rm G}$,
\begin{equation}
    G_{\rm is}^\mathrm{rep} =  \frac{4\pi}{\xi_{\rm G}} n^{\rm val}_\mathrm{solv}c_{\rm solv}\int_{\rm >cavity} {\rm d}\boldsymbol{r}\; \rho_\mathrm{el}(\boldsymbol{r}) \quad .
    \label{eq:grep-qm}
\end{equation}
Here, $c_{\rm solv}$ is the constant solvent concentration and $n^{\rm val}_\mathrm{solv}$ the number of valence electrons of the solvent species. The advantage of this approach is that the functional expression can be straightforwardly inserted into the KS DFT Hamiltonian. To this end, the integral over all external space of eq.~(\ref{eq:grep-qm}) is transformed into a 2D integral over the cavity surface $A_{\rm cav}$, which is numerically solved via tesselation. The price for this simplicity is a parameter $\xi_{\rm G}$ which largely lacks any physical motivation and with which the repulsion free energy contribution resulting from this model functional can be scaled to any desired value.

A corresponding tesselation is also the basis for the second, more empirical scheme, which essentially approximates a possible electron density overlap of solute and solvent by how close individual solute atoms come to the cavity surface. With the tesselation yielding units labelled by $k$ with surface area $A_{{\rm cav},k}$ and surface normal $\mathbf{n}_k$, the exchange repulsion functional is then given as\cite{floris1991dispersion}
\begin{subequations}
\begin{equation}
    G_{\rm is}^\mathrm{rep} = n^{\rm val}_\mathrm{solv}c_{\rm solv}
    \sum_{j \in {\rm solv}} N_j \sum_{\alpha \in {\rm solute}} \sum_{k \in {\rm cavity}} A_{{\rm cav},k} \, \mathbf{X}_{\alpha,j}(\boldsymbol{r}_{\alpha k}) \cdot \mathbf{n}_k \quad .
    \label{eq:grep-class}
\end{equation}
Next to the sum over surface tesserae, the other two sums range over all explicitly treated atoms $\alpha$ in the solute and all chemically unique atomic species $j$ in the solvent. $N_j$ denotes the number of times the species $j$ is contained in a solvent molecule, and the auxiliary distance vector
\begin{equation}
    \mathbf{X}_{\alpha,j}(\boldsymbol{r}_{\alpha k})
    =-\frac{1}{9}\frac{d^{(12)}_{\alpha j}}{\vert\boldsymbol{r}_{\alpha k}\vert^{12}}\boldsymbol{r}_{\alpha k}
\end{equation}
\end{subequations}
encodes a Lennard-Jones type repulsive interaction between solute atom $\alpha$ and solvent molecules, with the latter represented by the cavity units and thus at a distance $\boldsymbol{r}_{\alpha k}\cdot \mathbf{n}_k$ apart. In the form of the Lennard-Jones $d^{(12)}_{\alpha j}$ for each pair of solute and solvent species, this approach adds multiple additional parameters, which need to be determined, e.g.~via fitting to experimental reference values.\cite{tomasi2005quantum} On the other hand, the computational overhead of this approach is negligible given that most of the other contributions to the solvation free energy demand such a surface tesselation anyway.

Importantly, both methods reduce in fact again to integrals over the surface area of the cavity. This observation inspired Andreussi and co-workers\cite{andreussi2012revised} to simplify the calculation of the repulsion energy even further.
Making again use of Cococcioni and co-workers' quantum surface concept,\cite{cococcioni2005electronic} they simply formulated $G_{\rm is}^\mathrm{rep}[\rho_{\rm el}]$ (actually only in sum together with $G_{\rm is}^\mathrm{dis}[\rho_{\rm el}]$ as discussed below) as linearly dependent on the electrostatic cavity surface area $A_{\rm cav}$ and potentially also its volume $V_{\rm cav}$,
\begin{equation}
    G_{\rm is}^\mathrm{rep} + G_{\rm is}^\mathrm{dis} = \alpha A_{\rm cav} +\beta V_{\rm cav} \quad .
    \label{eq:grep_Andreussi}
\end{equation}
The advantage of this approach over eqs.~(\ref{eq:grep-qm}) and (\ref{eq:grep-class}) is its unparalleled computational efficiency (when again only evaluating it as a post-correction) and the fact that it adds only two adjustable parameters\cite{sinstein2017efficient}, as further discussed in Section~\ref{sec:param}.

\subsubsection{Dispersion interactions, $G_{\rm is}^\mathrm{dis}$}

Similar to $G_{\rm is}^\mathrm{rep}[\rho_{\rm el}]$ and indeed often treated in a very similar fashion or grouped together with it, $G_{\rm is}^\mathrm{dis}[\rho_{\rm el}]$ is supposed to account for another type of intermolecular interaction between solute and solvent molecules that is lost in the coarse graining process, namely attractive dispersion. With the relevance of solute-solvent dispersion on solute structure\cite{piana2015water} and energetics\cite{hwang2015how} well documented, a great number of methods have been devised to derive approximate expressions for $G_{\rm is}^\mathrm{dis}[\rho_{\rm el}]$.\cite{tomasi1994molecular}  Again, these approaches can be roughly categorized into more quantum mechanically inspired and more empirical approaches. Of the former, a popular approach, implemented e.g.~in the PCM model,\cite{amovilli1994calculation} is based on the theory of McWeeny.\cite{mcweeny1992methods} Without going into too much detail---see for example ref.~\citenum{tomasi2005quantum} for a full description---and similarly to the quantum mechanical treatment of the repulsion energy, also this approach can be boiled down to an integral over the cavity surface, yet this time over the electrostatic potential and the normal component of the electrostatic field. Both are represented in the basis functions of the underlying DFT method, which, at least in localized basis function codes, tend to be not very dense near the cavity surface.\cite{ringe2016function} Therefore, the accuracy of the quantum mechanical calculation of $G_{\rm is}^\mathrm{dis}[\rho_{\rm el}]$ tends to strongly depend on the chosen basis set. Properties of the solvent and solute enter this approach in the form of a multiplicative factor that depends among others on the first ionization energy of the solvent or average electronic transition energies. In particular also the complex integrals involved in the calculation, render the overall computational cost of this approach significantly higher than that of the other non-electrostatic contributions.

For this reason, a lot of implementations opt for a more empirical approach instead.\cite{tomasi1994molecular} An ansatz analogous to eq.~(\ref{eq:grep-class}) leads then to
\begin{subequations}
\begin{equation}
    G_{\rm is}^\mathrm{dis} = n^{\rm val}_\mathrm{solv}c_{\rm solv}
    \sum_{j \in {\rm solv}} N_j \sum_{\alpha \in {\rm solute}} \sum_{k \in {\rm cavity}} A_{{\rm cav},k} \, \mathbf{X'}_{\alpha,j}(\boldsymbol{r}_{\alpha k}) \cdot \mathbf{n}_k \quad ,
    \label{eq:gdis-class}
\end{equation}
only now with an auxiliary distance vector that encodes a London type attractive dispersion,
\begin{equation}
    \mathbf{X'}_{\alpha,j}(\boldsymbol{r}_{\alpha k})
    =-\frac{1}{3}\frac{d^{(6)}_{\alpha j}}{\vert\boldsymbol{r}_{\alpha k}\vert^{12}}\boldsymbol{r}_{\alpha k} \quad .
\end{equation}
\end{subequations}
Obviously, this approach thus incurs again a set of parameters ($d_{\alpha j}^{(6)}$) which need to be determined. 

Finally and also in exact analogy to exchange repulsion, each of these approaches to estimating $G_{\rm is}^\mathrm{dis}[\rho_{\rm el}]$ boils numerically down to a tessellation of the cavity surface. We note that instead of the here described geometric surface tessellation, one could in principle also integrate over any suitable cavity shape function, such as the aforementioned weighted density solvent-center cavity.\cite{sundararaman2014weighted}
In any case, based on the observation that $G_{\rm is}^\mathrm{dis} $ is just an integral over the cavity surface, Still and co-workers\cite{still1990semianalytical} proposed a simple description as a function of solvent accessible area, or indeed, the surface area of the solvation cavity. As noted above, this idea was later expanded upon in the work of Andreussi {\em et al.}\cite{andreussi2012revised}~where the dispersion functional is then described together with the exchange repulsion functional through eq.~(\ref{eq:grep_Andreussi}).

\subsubsection{Thermal motion, $G_{\rm is}^\mathrm{tm}$}

As discussed above, solvation of any solute generally alters that solute's PES. Foremost, one pictures this in form of an altered equilibrium structure of the solute compared to the vacuum one, such that e.g.~hydrophobic groups avoid exposure to the solvent, zwitterionic structures are stabilized by polar solvents or the internal hydrogen-bond network is rearranged.\cite{chowdhry2008vibrational} 
However, the altered PES could in principle also lead to changes in the vibrational modes of the solute that would correspondingly need to be accounted for through another free energy functional term $G_{\rm is}^{\rm tm}[\rho_{\rm el}]$.\cite{chowdhry2008vibrational} For molecular solutes, this would then additionally cover changes of the solute's rotational and translational entropy.\cite{ben-naim2013solvation} 
The latter do not play a role at extended SLIs, and 
solvent-induced changes to the vibrational modes of an adsorbate are likely small compared to those arising from the adsorption itself or from ongoing chemical reactions. Therefore, to our knowledge no implicit solvation implementations for the SLI context have hitherto explicitly considered a $G_{\rm is}^\mathrm{tm}[\rho_{\rm el}]$ term.

\subsection{Electrolyte models}
\label{sec:ImpSolvElectrolyteModels}

The theories introduced in the last two sections yield expressions for the mean-field electrostatic $\mathcal{V}_\mathrm{is}^\mathrm{mf}[\rho_\mathrm{el}]$ and non-electrostatic $\Omega_\mathrm{is}^\mathrm{non-el}[\rho_\mathrm{el}]$ terms in the model grand solvation potential, cf.~eq.~(\ref{eq:Gsolv}). These expressions are already sufficient to establish implicit solvation models for pure liquids. However, real electrochemistry or electrocatalysis almost invariably works with electrolytes with finite salt concentrations. Indeed, the presence of salt can actually even be substantial for the chemical reactions and the way they proceed. As already discussed, at SLIs ions act as counter charges to compensate the surface charge of the electrode. They are thus potentially strongly enriched particularly in the inner DL close to the electrode, and their presence may not least crucially impact the stabilities of reaction intermediates.\cite{ringe2019understanding} In this section, we therefore continue with the extension of implicit solvation models to electrolytes and notably Poisson-Boltzmann (PB) theory, which forms the unanimous basis for most of these extensions to day. Practically, this proceeds again by deriving computationally tractable or closed expressions for the contributions grouped into the ion grand potential term $\Omega_{\rm is}^{\rm ion}[\rho_\mathrm{el}]$ of eq.~(\ref{eq:Gsolv}).

\subsubsection{Poisson-Boltzmann theory}

In a dilute electrolyte solution, one may reasonably assume that the solvent dielectric response is not (significantly) modified by the small ion concentrations, and interactions between the generally quite distant ions can be well described on a mean-field level. In the DFT supercell, one realization of such a dilute electrolyte could be to simply place a small number of point-like ions at fixed and not too close positions to each other inside the implicit solvent part. For a corresponding static ion charge distribution $\rho_{\rm ion}=\rho_{\rm ion}({\boldsymbol r})$, as well as under the mentioned assumption of unmodified solvent dielectric response and only mean-field ion-ion interactions, then the only term that we would have to consider in $\Omega_{\rm is}^{\rm ion}[\rho_\mathrm{el}]$ is a straightforward potential energy functional $\mathcal{V}^\mathrm{mf,ion}_{\rm is}$. Together with the analogous mean-field potential energy functionals of the DFT part and the pure implicit liquid it would be given as
\begin{align}
\left(\mathcal{V}^\mathrm{mf}+\mathcal{V}^{\rm mf}_\mathrm{is}+\mathcal{V}^\mathrm{mf,ion}_{\rm is}\right)[\rho_\mathrm{el}]=\frac{1}{2}\int \left(\rho_\mathrm{el}+\rho_\mathrm{nuc}+\rho_\mathrm{ion} \right) \phi \quad .
\label{eq:ionpotfunct}
\end{align}

\begin{figure}
    \centering
    \includegraphics[width=\textwidth]{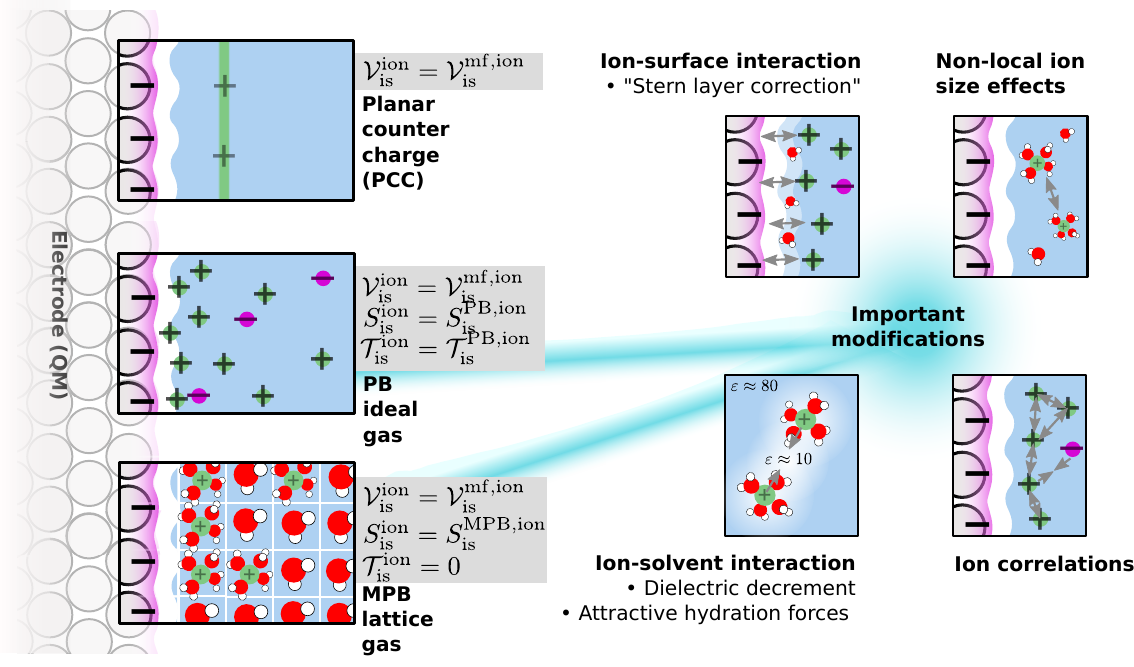}
    \caption{{\bf Schematic representation of various electrolyte models currently used for the description of SLIs.} Planar counter charge (PCC) models place rigid ions in a Helmholtz layer like arrangement, while Poisson Boltzmann (PB) models determine the ionic distribution self-consistently in the total electrostatic potential. Various important modifications of PB theory are highlighted and discussed in the text.}
    \label{fig:ion_models}
\end{figure}

As visualized in Fig.~\ref{fig:ion_models} such static ion distributions are indeed employed in so-called planar counter charge (PCC) models that are specifically developed for the description of planar SLIs\cite{andreussi2019continuum} and that we will further motivate in Section~\ref{sec:pcc}.
However, for the general objective of deriving functional expressions for an electrolyte that is equilibrated in its response to the given electrode or solute with its (DFT) net charge density, it makes no sense to manually ascribe fixed ion positions. Indeed, the equilibrated ion density should be a result of the theory, and not an input. Therefore, this equilibrated density will have to adapt to the electrostatic potential, to which the ions though actually contribute themselves. This already shows that in such a case self-consistency between $\rho_{\rm ion}$ and $\phi$ in eq.~(\ref{eq:ionpotfunct}) has to be reached. Most commonly, a corresponding self-consistent description of the ion distribution is achieved within the famous PB theory,\cite{gouy1910constitution,gouy1916sur,chapman1913theory,debye1923theory} which treats the ions as a gas that interacts only via mean-field electrostatic interactions within the continuum dielectric. Referring to dedicated accounts on PB theory\cite{gray2018nonlinear} for full derivations and a full appraisal, we here only compile the resulting expressions for the ion grand potential functional. For simplicity, we furthermore focus here and in the remainder of this electrolyte section also on an electrolyte with a cationic concentration $c_+=c_+({\boldsymbol r})$ due to only one cation species of mass $m_+$ and charge $+z$, and an anionic concentration $c_-=c_-({\boldsymbol r})$ due to only one anion species of mass $m_-$ and charge $-z$. Reflecting the additionally considered ion dynamics, the corresponding PB ion grand potential functional
\begin{subequations}
\label{eq:PB-functional}
\begin{align}
    \Omega_{\rm is}^{\rm ion}[\rho_\mathrm{el}] = 
    \mathcal{T}_\mathrm{is}^\mathrm{PB,ion}[\rho_\mathrm{el}] +
    \mathcal{V}^\mathrm{mf,ion}_{\rm is}[\rho_\mathrm{el}] - TS_\mathrm{is}^\mathrm{PB,ion}[\rho_\mathrm{el}]
    \quad ,
\end{align}
now includes kinetic and, applying the famous Sackur-Tetrode equation,\cite{tetrode1912chemische,sackur1911anwendung,sackur1913universelle} also entropic contributions. For the $+z/-z$ electrolyte they read
\begin{align}
    &\mathcal{T}_\mathrm{is}^\mathrm{PB,ion}=\frac{3}{2} k_\mathrm{B}\int \left\{c_++c_-\right\}\mathrm{d}\boldsymbol{r}\\
    &S_\mathrm{is}^\mathrm{PB,ion}=-k_\mathrm{B}\int \biggl\{c_{+}\left[\ln \left(c_{+} \lambda_+^{3}\right)-\frac{5}{2}\right]+ c_{-}\left[\ln \left(c_{-} \lambda_-^{3}\right)-\frac{5}{2}\right]\biggr\} \mathrm{d}\boldsymbol{r} \quad ,
\label{eq:PBtheory_func}
\end{align}
\end{subequations}
with $\lambda_\pm=\frac{h}{\sqrt{2\pi m_\pm k_\mathrm{B}T}}$ the thermal wave length.
The potential energy functional still holds as before in eq.~(\ref{eq:ionpotfunct}), of course, now with the ion charge density given as $\rho_{\rm ion}= z(c_+ - c_-)$.

The starting point to obtain the self-consistent ion concentrations and electrostatic potential to evaluate these functional expressions is as before the GPE. Within the prevalent isotropic, linear and local dielectric model, and under the already mentioned assumption that the dielectric response of the solvent is not changed by the ion density, the electrolyte charge distribution can straightforwardly be added to the GPE of eq.~(\ref{eq:GPE_linearresponse}), simply by extending the source terms on the right hand side
\begin{align}
    - \varepsilon_0 \nabla \left[\varepsilon({\boldsymbol r}) \nabla \phi\right]&= (\rho_\mathrm{el}+\rho_\mathrm{nuc}+\rho_\mathrm{ion}) \quad .
\label{eq:gpe_ions}
\end{align}
PB theory then additionally makes the assumption that in the equilibrated electrolyte the ions are Boltzmann-distributed in the electrostatic potential
\begin{align}
&c_\pm=c_{\infty,{\rm ion}} \exp\left(\mp\frac{z \phi}{k_\mathrm{B}T}\right)\quad,
\label{eq:pb_concentrations1}
\end{align}
with $c_{\infty,{\rm ion}}$ the constant equal concentration of cations and anions in the bulk of the electrolyte. Within the mean-field PB gas ansatz, $c_{\infty,{\rm ion}}$ is in turn readily related to the bulk ion electrochemical potential via $\tilde{\mu}_{\mathrm{ion},\pm} =k_\mathrm{B}T\ln\left(\lambda_\pm^3c_{\infty,{\rm ion}}\right)$.
Inserting eq.~(\ref{eq:pb_concentrations1}) into the ion-including GPE of eq.~(\ref{eq:gpe_ions}), leads finally to the famous PB equation (PBE) itself
\begin{equation}
    \varepsilon_0\nabla \left[\varepsilon \nabla \phi\right]=- \left(\rho_\mathrm{el}+\rho_\mathrm{nuc}+2z c_{\infty,{\rm ion}} \sinh\left(-\frac{z \phi[\rho_\mathrm{el}(\boldsymbol{r})]}{k_\mathrm{B}T}\right)\right) \quad,
    \label{eq:PBE-full}
\end{equation}
which does not explicitly contain the spatially varying ionic concentrations anymore. In practice, this PBE can thus be implemented into the KS-DFT minimization in a way completely analogous to the GPE of the ion-free case, and then be solved at each electron density optimization step. However, due to the complicated non-linear nature of the PBE, and the associated computational cost of solving it, it is popular to instead solve a simplified linearized version of it.\cite{chipman2004solution,mathew2019implicit} This linearized Poisson-Boltzmann equation (LPBE) can be obtained by truncating a Taylor expansion of the sinh-term in eq.~(\ref{eq:PBE-full}) around $\phi=0$ (here assumed to be the bulk electrolyte potential) after the linear term\cite{debye1923theory}
\begin{equation}
        \varepsilon_0\nabla \left[\varepsilon \nabla \phi\right]=- \left(\rho_\mathrm{el}+\rho_\mathrm{nuc}- \frac{2z^2 c_{\infty,{\rm ion}} }{k_\mathrm{B}T} \phi[\rho_\mathrm{el}(\boldsymbol{r})]\right) \quad.
    \label{eq:LPBE-full}
\end{equation}
The corresponding LPBE grand potential functional terms can then be derived analogously from a Taylor expansion of eq.~(\ref{eq:PB-functional}).\cite{ringe2016function}

As mentioned at the beginning of this sub-section, the assumptions underlying PB theory restrict its formal range of applicability to dilute electrolytes. Close to electrified SLIs, however, high ion concentrations may accumulate even for electrolytes that are indeed dilute in the bulk.\cite{borukhov1997steric,kilic2007steric}
This motivates corrections to PB theory that account for then increased ion-ion and ion-solvent correlations. Despite the non-locality of these interactions, a series of local ion density approximation models have been proposed to keep the simplicity of the PB model intact. They are summarized in Fig.~\ref{fig:ion_models} above and will be briefly outlined in the following.

\subsubsection{Finite ion size corrections}
\label{sec:mpb-corr}
In the original formulation of PB theory, ions are point-like. This means, that for stronger fields local ion concentrations could in principle reach unphysically high values. An immediate fix to this problem is to simply give ions a finite size, which then leads to size-modified PB (MPB) theory.\cite{bikerman1942structure} While MPB can be derived in various ways,\cite{wang2013simulations} the most physically intuitive derivation is based on a lattice model with a uniform cell size $a$ for (solvated) ions and solvent molecules, where each lattice site can at max hold only one particle, cf.~Fig.~\ref{fig:ion_models}. This way, the lattice mimics short-range ion-ion repulsion and by construction does not allow unphysically high local ion concentrations. For this model, an ion grand potential functional can be developed using the configurational partition function of solvent molecules and ions, and then applying a mean-field approximation.\cite{borukhov2000adsorption,borukhov1997steric,kralj-igliv1996simple,ringe2017first} The model thus  corrects in a mean-field way for ion repulsions, and the kinetic energy and entropy functionals are modified as
\begin{align}
    &\mathcal{T}_\mathrm{is}^\mathrm{MPB,ion}=0
    \label{eq:T_MPB}
    \\
    &S_\mathrm{is}^\mathrm{MPB,ion}=-k_\mathrm{B}\int \biggl\{c_{+}\left[\ln \left(c_{+} a^{3}\right)-1\right]+ c_{-}\left[\ln \left(c_{-} a^{3}\right)-1\right]\nonumber\\
    &\quad \quad \quad \quad +\left(\frac{1}{a^3}-c_+-c_-\right)\ln\left(1-c_+a^3-c_-a^3\right)+c_++c_-\biggr\} \mathrm{d}\boldsymbol{r} \quad .
\label{eq:MPB_functional}
\end{align}
The ion concentrations now effectively follow a Fermi-Dirac-like statistics due to the maximum occupancy of the lattice cells
\begin{align}
    &c_\pm=c_{\infty,{\rm ion}} \frac{\exp\left(\mp\frac{z \phi}{k_\mathrm{B}T}\right)}{1-2c_{\infty,{\rm ion}} a^3+2c_{\infty,{\rm ion}} a^3 \cosh\left(\frac{z\phi}{k_\mathrm{B}T}\right)} \quad, 
\label{eq:c_MPB}
\end{align}
with the relation between the bulk concentration and ion electrochemical potential modified to $\tilde{\mu}_{\mathrm{ion},\pm}= k_\mathrm{B}T\ln\left(\frac{c_{\infty,{\rm ion}} a^3}{1-2c_{\infty,{\rm ion}} a^3}\right)$. In direct analogy to the unmodified PB case, inserting these concentrations into the GPE of eq.~(\ref{eq:gpe_ions}) leads then to the so-called MPB equation, which in turn gives the desired functional relation between  electrostatic potential and electron density and which can be solved within the KS DFT minimization as before. 

The MPB ion concentration profiles converge to the PB profiles for $a\rightarrow 0$, if the same bulk electrochemical potential reference is used. One can easily see that this then implies $a=\lambda_+=\lambda_-$, i.e.~within PB theory the thermal wavelength plays the same role as the ion size in MPB theory. The two theories thus have a common algebraic origin, but a different physical one. Indeed, in contrast to PB theory and as seen in eq.~(\ref{eq:T_MPB}), the MPB model lacks an ion kinetic energy functional. Yet, since it equally lacks a corresponding entropy contribution from the ionic motion and with the two terms canceling each other in PB theory, the same functional form is nevertheless recovered in both theories for the small ion size limit. Note also that the original MPB theory, and the corresponding equations above, were developed for equally sized cations and anions. It has since been extended to asymmetric electrolytes, e.g.~by extending the statistical lattice model with sublattices,\cite{han2014mean,mceldrew2018theory} by introducing potential-dependent ion sizes\cite{kornyshev2007double}, or by other means.\cite{popovic2013lattice,maggs2016general} There are also efforts to go beyond the lattice approximation,\cite{may2019differential} deriving functional expressions from experimental equation of state data or from equations defining atomic or molecular interactions such as closure relations to the Ornstein-Zernicke equation, cf.~Section~\ref{sec:FundCont}.\cite{antypov2005incorporation,maggs2016general} Notwithstanding, the resulting energy functionals are generally still based on a local approximation for the ion density. Such local approaches to ion-ion interactions offer generally a simple correction to PB theory for those situations where ions are crowded, e.g.~due to strong electric fields. However, in case of strong variations of the ion concentration profiles the description of $\Omega_{\rm is}^{\rm ion}[\rho_{\rm el}]$ as a local functional of ion concentrations may break down altogether. Such cases may then necessitate a more involved non-local treatment.\cite{antypov2005incorporation} 

\subsubsection{Ion-induced solvent structuring}
\label{sec:solventstruct}

One important physical effect of the ions completely omitted so far is simply the fact that in regions with high ion concentrations few solvent molecules may reside, and if they do they are likely highly structured around the ions. In such situations the dielectric continuum approximation for the solvent likely breaks down and ion interactions become much more specific than the the hitherto included mean-field electrostatics.

Accounting for this effect, Burak and Andelman\cite{burak2000hydration} derived a corrective short-range ionic interaction potential contribution to the mean-field electrostatic potential from Monte Carlo simulations. By truncating the virial expansion of the PB partition function after second order, they were then able to derive a simple analytic expression for the free energy. Although this direct expansion approach is of great interest for the development of improved PB-based theories, it is less practical due to the required knowledge of the system-dependent fluctuating short-range potential. An approach that is in this respect more in the spirit of effective parametrized continuum models has been put forward by Bohinc, Shrestha and May.\cite{bohinc2012poisson,bohinc2011poisson,bohinc2017incorporation} There, they represented the additional short-range forces by a parametrized Yukawa potential, arriving at a simple correction to PB theory. Next to this, several other approaches have been developed in the past for which we refer the interested reader to an extensive review on this topic.\cite{ben-yaakov2011ion}

All of the above corrections share the fact that the resulting corrections are non-local in the sense that solvent structure at a point in space is also influenced by the ion concentration in its vicinity (e.g.~via the aforementioned short-range potentials).
A much simpler and local variant to correct for ion-induced solvent structuring is the dielectric decrement approach, cf.~Fig.~\ref{fig:ion_models}. From simulations\cite{chandra2000static} and various experimental works,\cite{glueckauf1964bulk,wei1990dielectric,wei1992ion,buchner1999dielectric} it has been found that the isotropic dielectric permittivity of water varies linearly with the salt concentration at small to medium (1.5\,M) salt concentrations,\cite{ben-yaakov2011dielectric}
\begin{equation}
    \varepsilon_\infty(c_{\infty,{\rm ion}})=\varepsilon_\infty(c_{\infty,{\rm ion}}=0)+\beta c_{\infty,{\rm ion}} \quad,
\label{eq:dielectric_decrement}
\end{equation}
where $\beta$ is a generally negative, ion-specific dielectric decrement coefficient indicating how easy the water structure can be polarized by the presence of salt. The equation can be also written as a function of the local salt concentration, but importantly anionic and cationic contributions cannot be separated without making further assumptions, due to a lack of experimental data.\cite{ben-yaakov2011dielectric} In any case, based on this evidence for water one could 
simply consider $\beta$ as a further variable parameter and modify the dielectric function employed for the mean-field electrostatics, cf.~Section~\ref{sec:DielecFunc}, to additionally depend (linearly) on the local ion concentration. In the SLI context with aqueous electrolytes, this effective dielectric decrement approach also enjoys recent popularity to model a stronger (sometimes ice-like) water structuring in the inner DL through an accordingly reduced dielectric permittivity in that region.\cite{melander2019grand} In summary, there are thus a number of methods of varying degrees of complexity that allow to mimic an ion-induced local structuring of the solvent around a solute. Thereby, they extend the validity of PB or MPB approaches to higher ion concentrations, yet often at the price of additional parameters that need to be determined.

\subsubsection{Coulombic ion correlations}

In the case of small ion concentrations with thus effectively large ion separations and in strongly screening solvents like water, the mean-field interaction between the dissolved ions assumed in PB theory is generally a good approximation. However, it may quickly break down for solvents with smaller dielectric permittivity, for higher ion concentrations (such as in ionic liquids\cite{skinner2010capacitance,mezger2008molecular}), or for multi-valent ions with stronger Coulomb forces.\cite{burak2000hydration,kornyshev2007double,bazant2011double} In these cases, the electrostatic force is much more non-local and fluctuating, leading for example to the effect of overscreening at charged interfaces.\cite{levin2002electrostatic} Overscreening in the context of electrified SLIs refers to the presence of higher amounts of counter charge in the electrolyte close to the electrode than needed to compensate the surface charge, followed by a smaller net charge of opposite sign to satisfy overall electroneutrality. 

An account for the corresponding ion correlations requires in general field theoretical approaches, using loop expansions, which lead to substantially more complicated expressions than in PB theory.\cite{sahin2014nonlocal,buyukdagli2016beyond} A promising, more approximate approach by Bazant \textit{et al.}~instead leads to a simple correction of the mean-field electrostatic potential energy, cf.~eq.~(\ref{eq:ionpotfunct}), in form of one added term\cite{bazant2011double}
\begin{align}
   \mathcal{V}_{\rm is}^\mathrm{non-mf,ion}[\rho_\mathrm{el},\phi]=-\int \frac{\varepsilon_0\varepsilon({\bf r})}{2}l_c^2(\boldsymbol{\nabla}^2\phi)^2\mathrm{d} \boldsymbol{r}\quad,
\end{align}
where the parameter $l_c$ represents an electrostatic correlation length. This demonstrates nicely that the electrostatic energy is lowered due to overscreening by enhancing the curvature of $\phi$. The theory was shown to give overscreened ion distribution profiles in close agreement with molecular dynamics simulations resulting in realistic estimates of the potential-dependent capacitance compared to experimental reference data.\cite{bazant2011double}

\subsubsection{Ion-solute interaction and Stern layer formation}
\label{sec:ionsol-corr}

PB and MPB theories as well as their extensions are usually derived without the actual presence of the solute. Therefore, the only coupling between solute and ions is the hitherto discussed mean-field electrostatic coupling. Just as highlighted in Section~\ref{sec:ImpSolvNonelstat} for the pure liquid, this neglects additional interactions between the ions and the solute that were for the solvent summarized in the $\Omega_\mathrm{is}^\mathrm{non-el}[\rho_\mathrm{el}]$ term in eq.~(\ref{eq:Gsolv}). A prominent such non-electrostatic correction for the ions would be an additional repulsive contribution which prevents ions to approach the solute too closely. In protic solvents, the formation of a corresponding ion-free solvent region called Stern layer is for instance a consequence of the large size of the hydrated cations.

\begin{figure}
    \centering
    \includegraphics[width=0.8\textwidth]{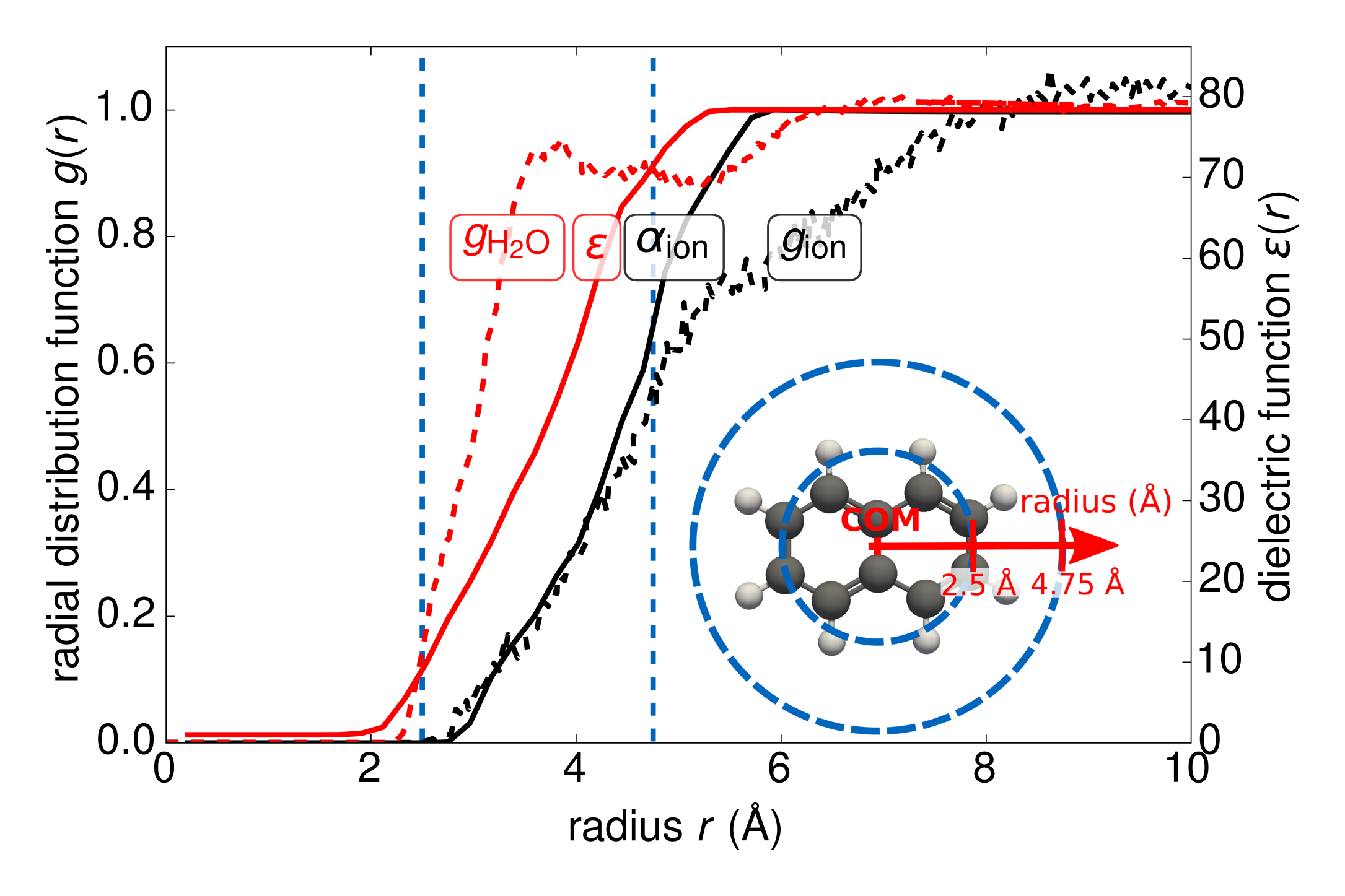}
    \caption{{\bf Creation of an ion-free Stern layer around a molecular solute.} Compared is the solvation environment around the center of mass (COM) of naphthalene in a 2.18M NaCl solution as obtained from explicit molecular dynamics simulations \cite{li2014small} (dashed lines) and with a Stern-layer corrected implicit MPB model (solid lines). Data in red represents the spherically-averaged radial distribution function (RDF) of the oxygen atoms in the
    explicit water solvent ($g_{\rm H_2O}$) and the 
    corresponding spherically-averaged dielectric
    function $\varepsilon$ for the implicit model. Data in black are the spherically-averaged RDF for the ions and the corresponding ion-exclusion function $\alpha_{\rm ion}$. Both the onset of the solute solvation shell and the radial Stern layer shift of the ionic distribution are rather
    well reproduced. To better grasp the involved scales, two dotted vertical lines illustrate the radial distance to the molecule COM as shown in the top view in the inset. Adapted with permission from ref.~\citenum{ringe2017transferable}, American Institute of Physics (AIP).}
    \label{fig:rdf_plot}
\end{figure}

Most straightforwardly, this kind of physics can be implemented by an additional repulsion potential added on top of the mean-field potential. Alternatively, the repulsion potential can simply be expressed as an exclusion function $\alpha_{\mathrm{ion},\pm}=e^{-\frac{\phi^\mathrm{rep}_\pm}{k_\mathrm{B}T}}$ for cations and anions, respectively. The exclusion function prevents the ions from approaching the solute to a certain distance,\cite{ringe2017first} similar to the dielectric shape function controlling the solvent's place of closest approach (see ref.\citenum{ringe2017first} for a full derivation following the statistical lattice approach). This leads to a modified set of ion concentration functions, which for the case of a +z/-z electrolyte read
\begin{align}
    &c_\pm=c_{\infty,{\rm ion}} \alpha_{\mathrm{ion},\pm}[\rho_\mathrm{el}]\frac{\exp\left(\mp\frac{z \phi}{k_\mathrm{B}T}\right)}{1-2c_{\infty,{\rm ion}} a^3+2c_{\infty,{\rm ion}} a^3 \alpha_{\mathrm{ion},\pm}[\rho_\mathrm{el}] \cosh\left(\frac{z\phi}{k_\mathrm{B}T}\right)} \quad.
\end{align}
For the sake of convenience the functional form for $\alpha_{\mathrm{ion},\pm}$ can be chosen identical to the dielectric shape function $s_{\rho_\mathrm{el}}$, to vary in between 0 in the ion-free region and 1 in the ion-contained electrolyte region, yet with a different cutoff parameter that can be individually tuned.\cite{ringe2017transferable} This simplified model was shown to be able to account for short ranged ion-solute interactions, at the price of the additional cutoff and shape parameters. As detailed in Section~\ref{sec:param}, a careful tuning of these parameters to reproduce molecular experimental reference data yields a plot like in Fig.~\ref{fig:rdf_plot}, which nicely illustrates the achieved creation of an ion-free Stern layer close to the solute with an extent and location that agrees well with the results of molecular dynamics simulations with explicit solvent. Below, we will refer to a MPB model that additionally provides such a Stern-layer functionality as S-MPB model.

\subsubsection{Planar counter charge models}
\label{sec:pcc}

All the ion models presented so far are based on diffuse layer theory, essentially assuming mobile gas like ions that migrate and equilibrate in a mean-field potential. While these more physical approaches are most valuable for a general treatment of solvation, the aforementioned, much simpler PCC approach to place rigid ions into the supercell, cf.~Fig.~\ref{fig:ion_models}, is of particular interest and convenience for the context of planar SLIs. As further discussed in Section~\ref{sec:surfcharge} below, its primary purpose is to introduce a counter charge distribution that exactly compensates a net charge of the electrode to achieve an overall charge-neutral supercell. As the name says, this ionic counter charge distribution is simply modeled as a smoothed out Gaussian charge plane.\cite{andreussi2019continuum} The advantage of this method is that the ionic charges can be freely shifted in space, thereby providing some flexibility, e.g.~in modeling asymmetric DFT supercells with only one slab side exposed to the electrolyte. Physically, the PCC model resembles if at all the situation in the inner DL, where the ions are assumed to be highly crowded. In dominantly studied aqueous electrolytes, the obvious crudeness of this approach is fortunately to some extent remedied by the strong screening capabilities of the polar solvent, which renders the DL potential drop less sensitive to the exact location of the ions.

\subsection{Parametrization}
\label{sec:param}

As apparent from the presentation so far, implicit solvation methodologies come invariably with a set of (in principle system-dependent) parameters, and it is these parameters that crucially determine the accuracy of this highly effective approach to solvation. To recap the previous sections, parameters arise generally in the functional expressions accounting for electrostatic, non-electrostatic and ionic contributions. In the standard linear, local and isotropic dielectric formulation of the electrostatic contribution, parameters are needed to define the location of the solvation cavity, or additionally the dielectric transition region. As discussed in Section~\ref{sec:DielecFunc}, these can be atomic radii in the case of spatially parametrized dielectric functions, or iso-values of the electron density in the density-dependent case. 
The non-electrostatic energy functional also gives rise to a varying number of parameters that depends strongly on the models of choice. If the models separately account for cavitation, dispersion or repulsion, quite a large number of parameters can quickly arise. In contrast, the simplified SCCS model of Andreussi \textit{et al.}\cite{andreussi2012revised}~lumps all of this into just two parameters that scale the solvation cavity volume and surface. Lastly, the ionic energy functional comes with its own number of parameters to express deviations from the PB theoretical description. These can either be the introduction of a finite ion size parameter, a parameter to describe the Stern layer and thus solute-ion interactions, or parameters to describe ion-solvent interactions in form of a dielectric decrement. 

Depending on the model complexity and the way it considers these various contributions, a largely different total number of parameters can result. This number can range from just four in the minimal SCCS model\cite{andreussi2012revised} to 64 parameters in the popular SMD model.\cite{marenich2009universal} These parameters may then be determined to give an optimum account of maybe only a single solvent/solute combination, maybe a particular solvent (e.g.~water in the case of the SCCS model), or aiming at maximum transferability for a whole range of solvents (as in the SMD case). At the same time, due to the fitting procedure, there is always the possibility of some degree of error cancellation, when for instance the non-electrostatic contributions compensate for some of the shortcomings of the DFT functional itself.\cite{sinstein2017efficient}
A key question is thus, to which degree the use of parameters can improve the physics and transferability of the implicit solvation model. This question relates directly to the size, quality and information content of the available training data to which parameters can be fitted. 

In principle, training data should be selected that is as close as possible to the intended application, in this case SLIs, or if possible even electrified SLIs. Unfortunately and as further discussed in Chapter 3, experimental reference data is very rarely available for these systems, and if it is, it is often not suited for the parametrization of a microscopic solvation model. The main reason for this is that at SLIs various other effects overlap with pure solvation contributions as we will see in the next chapter below. In contrast, molecular solvation data is much more widely available, at least for water as a solvent. Therefore, many implicit solvation studies on SLIs have adapted parameters that have originally been derived from a parametrization to such molecular data. This is not only critical from the viewpoint of the largely different chemistries, involving solutes composed of light organic elements in one case and extended electrodes composed of heavy transition metals in the other. There are also practical problems that arise not least from the different dimensionality of the problem. For instance and as outlined in Section~\ref{sec:g-rep} above, non-electrostatic contributions can be expressed as a function of the volume of the solvation cavity, which in turn is not defined for extended interfaces. Such issues have led to the formulation of functional expressions that only involve quantities compatible with SLIs. In the mentioned example, this is a non-electrostatic model that only considers the cavity surface and not the volume.\cite{andreussi2012revised,sinstein2017efficient} Nevertheless, also such models are then re-parametrized using the existing molecular databases. The performance of the cavity-surface model was there found to be similar to the original one including the cavity volume. While this suggests that a parametrization of SLI-compatible models is possible, it still does not tell how well the molecular parameters will transfer to the SLI context and we will come back to this issue in the next chapter.

\begin{table}
\begin{tabular}{c|c|c|c|c}
    Database & \# Solvents & \# Solutes & \# Hydration  & \# NAQ solv.   \\
     &   &  &   energies &   energies  \\
     \hline
    FreeSolv (v0.51)\cite{mobley2014freesolv} & 1 (W) & 643N & 643 & 0\\
    Wang\cite{wang2016automatic} & 1 (W) & 668N & 668 & 0\\
    Rizzo-DGHYD\cite{rizzo2006estimation} & 1 (W) & 538N/52C & 603 & 0\\
    Kelly\cite{kelly2006aqueous} & 1 (W) & 106C & 106 & 0\\
    MNSol (v2.0)\cite{marenich2012minnesota} & 106 (W,NAQ) & 662N & 389 & 2648 \\
    Solv@TUM (v1.0)\cite{SOLVATUMmediatum,hille2019generalized,SOLVATUMgithub} &145 (NAQ) & 658N &0& 5952\\
    CompSol\cite{moine2017estimation} & 732 (W,NAQ,IL,T) & 863N & 397 (581*) & 3786 (13386*) \\
\end{tabular}
\caption{{\bf Databases containing experimentally measured solvation energies of molecular solutes at room temperature.} C = charged, N = neutral, W = water, NAQ = no-aqueous solvents, IL = ionic liquids, T = temperature dependence. The numbers were extracted from the respective databases directly. The Wang database is to a large part constructed from the FreeSolv database. *Only solvation energies evaluated at $25\pm2^\circ$C have been considered, while the number in brackets refers to the complete number of solvent-solute combinations for which at least one temperature data is available.}
\label{tab:molecular_solvation_databases}
\end{table}

If one accepts that the parametrization is done with molecular data, the next obvious questions are which and how much of such data is available, how diverse the database is in terms of a wide range of molecular chemistries and whether the tabulated quantities are in fact really suited for the parametrization at hand. In the long, independent history of molecular solvation modeling, these questions have been satisfactorily addressed through the built-up of databases of primarily experimental solvation free energies. As apparent from Table~\ref{tab:molecular_solvation_databases}, these databases are indeed sizable and partly contain data for a wide variety of solvents.
Out of these, the Minnesota solvation (MNSol) database\cite{marenich2012minnesota} was among the first to provide experimental solvation energies of a wide range of over 600 neutral organic molecules in over 100 different solvents. Over the years, this database has been extended and various other databases have appeared. The FreeSolv\cite{mobley2014freesolv} database is currently the largest collection of neutral molecule solvation free energies in water (then called hydration energies). It consists of over 600 entries and should thus allow a meaningful parametrization even of more complex models.

Within the implicit solvation framework defined in this review, a molecular solvation free energy is calculated as 
\begin{equation}
    \Delta G_\mathrm{solv}=\Omega^{N_\alpha}_{\varepsilon_\infty={\rm solv}}[\rho_\mathrm{el}^\circ]-\Omega^{N_\alpha}_{\varepsilon_\infty=1}[\rho_\mathrm{el}^\circ] \quad ,
\end{equation}
where the two grand potential terms correspond to the solute in the solvent and the solute in vacuum, at their respective (generally different) ground state electronic densities $\rho_\mathrm{el}^\circ$. Note that it is awkward to see a free energy on the left hand side of the equation, and a difference of grand potential energies on the right hand side. We here simply attest to the fact that (measurable) solvation free energies are generally seen as a property of the full (macroscopic) system and not of the grand-canonical sub-system technically employed in the calculations. Starting our survey of model performance with the ubiquitous solvent water, implicit solvation models trained by these molecular databases can typically predict hydration energies with a mean absolute error (MAE) between 0.6~kcal/mol (large parameter space models like SMD\cite{marenich2009universal} or SM8\cite{marenich2007self}) up to 1.2~kcal/mol (small parameter space models like the SCCS model\cite{andreussi2012revised}). In general, though, these numbers are difficult to compare, since rarely the same set of training and test molecules have been used, see e.g.~ref.\citenum{ou2021assessing}
for a notable exception. Nevertheless, it generally seems that standard implicit solvation models have had a hard time to decrease the accuracy below about ~0.5~kcal/mol, which could thus somehow mark what can realistically be expected at such high level of coarse graining. Recent reports of ground-breaking 0.14~kcal/mol MAEs\cite{alibakhshi2021improved} with new machine-learned implicit solvation models have thus also to be seen in light of the actual accuracy of the underlying experimental data. Different solvation databases have been found to have an error of up to 0.25~kcal/mol relative to each other,\cite{hille2019generalized,moine2017estimation} which agrees with the experimental error that is estimated for solvation energies of neutral solutes based on measured partition coefficients.\cite{hille2019generalized,thompson2004new} Too highly parametrized models could therefore run the risk of overfitting of experimental errors. A connected problem is the occurrence of solutes in the training set which are reactive in solution.
The optimized SCCS model was, for example, found to perform well for most molecular components, apart from carbonic acids and amines. These are precisely those compounds which are mostly present in solution in their dissociated form at associated vastly different solvation energy. This highlights the importance of a careful curation of the reference databases.

\begin{figure}
    \centering
    \includegraphics[width=0.8\textwidth]{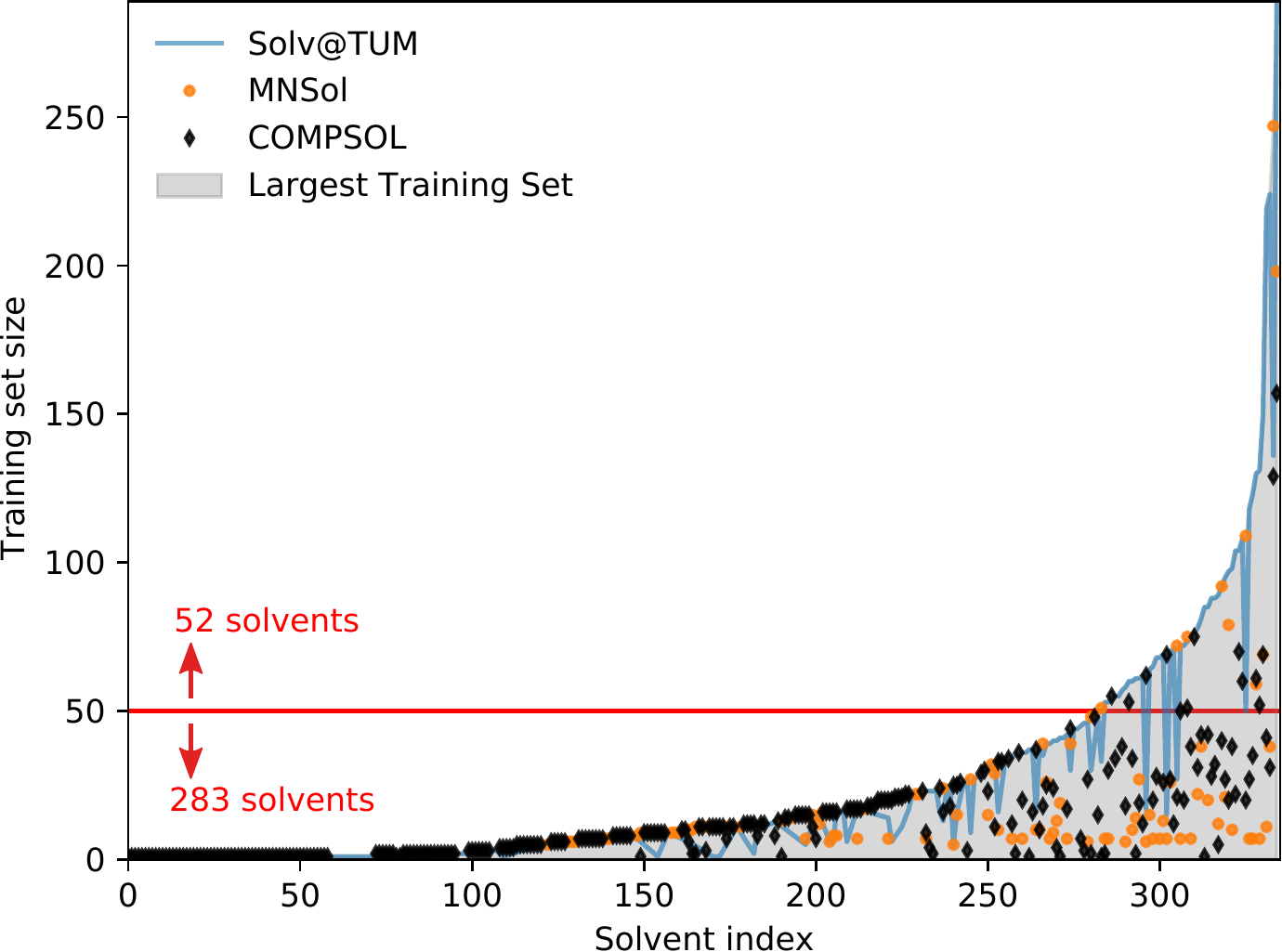}
    \caption{{\bf Number of solvation energy entries (``training set size'') per non-aqueous solvent in the three largest corresponding experimental databases.} The solvents are sorted according to their largest training set size in all of the three databases. Training set sizes below 50 are prone to significant overfitting errors, in particular if the solutes are not homogeneously distributed over the chemical space.}
    \label{fig:solvation_databases_nonaq}
\end{figure}

Next to the simulation of aqueous solvation, also non-aqueous solvents are of high importance for electrochemistry, such as e.g.~for lithium ion batteries\cite{xu2004nonaqueous} or the electrocatalytic reduction of CO$_2$.\cite{konig2019solvents} As shown in Table \ref{tab:molecular_solvation_databases}, the Solv@TUM database\cite{SOLVATUMmediatum,hille2019generalized,SOLVATUMgithub} currently has the largest total collection of non-aqueous solvation energies of neutral organic molecules, but this does not tell how large a training set is available for each individual non-aqueous solvent. Figure~\ref{fig:solvation_databases_nonaq} correspondingly compares the MNSol, Solv@TUM and COMPSOL databases regarding this amount of available solvation energies for each non-aqueous solvent. It is apparent that for most of these solvents the databases actually contain less than 50 solvation free energy entries. As pointed out by Hille \textit{et al.},\cite{hille2019generalized} such small training set sizes for the implicit solvation model can result in significant overfitting of the data. This would likely reduce the transferability even for those implicit solvation models that have only a few fitting parameters. Unfortunately, the situation is even worsened by an often low chemical diversity of the organic solutes contained in these small test sets, which may further lead to bias in the achieved parametrization.\cite{hille2019generalized} These issues provide a motivation especially for smallest parameter space implicit solvation models that rather trade quantitative accuracy with a somewhat robust extrapolation outside of the small training regime. As we will elaborate further in Chapter 3 below, this objective fits actually very well with the realization that the primary value of implicit solvation modeling at SLIs is presently more the provision of a counter-charge model than the actual account of solvation effects. Within this perspective, a recent reformulation was able to reduce the parameter space of the SCCS model to a single non-electrostatic parameter, while still resulting in a reasonably accurate prediction of solvation free energies for most solvents.\cite{hille2019generalized} This one parameter can furthermore be estimated from the solvent bulk dielectric permittivity, enabling the prediction of solvation free energies for arbitrary solvents with known permittivity.

The problem of small training set sizes becomes even more critical when transitioning to solvation free energies of charged solutes. Estimating the solvation energy of ions is a key challenge of high priority, as charged systems appear constantly as reactants or reaction intermediates in electrochemistry,\cite{koper2013theory} but also e.g.~in bio-\cite{cardona2012charge} or organic chemistry.\cite{jones2007reviews} The solvation energy of dissociated acids for example is important for the estimation of the acid dissociation constant,\cite{mcgrath2013calculation} or the solvation energy of charged redox species for the calculation of redox potentials.\cite{jinnouchi2008aqueous} Due to their large electrostatic stabilization, charged solutes exhibit solvation energies that are an order of magnitude larger than the ones of neutral solutes. However, the experimental measurement of single molecule ionic solvation energies requires thermodynamic cycles and the knowledge of the absolute solvation free energy of an arbitrary reference ion, usually a proton.\cite{lewis1961thermodynamics,pliego2002gibbs,kelly2006aqueous} Especially, the latter has been found to be prone to errors up to 2~kcal/mol.\cite{kelly2006aqueous} Nevertheless, various attempts have been made to parameterize implicit solvation models to ionic solvation data, as e.g.~the Rizzo-DGHYD database\cite{rizzo2006estimation} containing 52 solvation energies of cations and anions. In using this database for parametrizing the SCCS model, Dupont \textit{et al.}~found substantially different cavity parameters for anions, cations and neutral molecules,\cite{dupont2013self} which one would generally avoid for the modeling of SLIs with a varying charge state depending on the applied potential. This issue may be due to two drawbacks of the original SCCS approach. First, relying on the electron density to define the dielectric function, cf.~Section \ref{sec:DielecFunc}, means that anions show significantly larger cavities compared to cations and neutral molecules at comparable density iso-values.\cite{fisicaro2017soft,sinstein2017efficient} Furthermore, SCCS does not account for explicit solvent-solute correlation, which becomes significant in the case of high fields near localized charges. As noted above, electric field corrected dielectric functions have shown promise to go beyond this limitation.\cite{sundararaman2014weighted,truscott2019field}

Next, in an attempt to increase the training data for implicit solvation models, the CompSol database has recently been published adding further quantities beyond the traditional solvation free  energies at standard state. These are mainly temperature-dependent solvation free energies and solvation energies in ionic liquids. In terms of solvation energy entries, this database is now by far the largest. In order to make full use of it though, the implicit solvation model actually has to be able to somehow account for the additional physics in this data. Indeed, temperature-dependent solvation data could provide an additional constraint on the functional form of electrostatic, cavity or dispersion energy contributions, all of which could in principle depend on the temperature.\cite{gupta2012modeling,gupta2013modeling,liu2015order} While some implicit solvation models incorporating temperature effects have been communicated,\cite{chamberlin2008extension} these have not found their way into widespread use to date.

As a final point, we note that all of the databases discussed above focus on solvation free energies for vanishing ionic concentrations in the solvent. They are thus not suited for the determination of the ionic parameters appearing in electrolyte models that go beyond the plain PB approach. Ionic parameters suffer therefore presently from the highest scarcity of reference data. One remedy is to realize that finite salt concentrations are known to alter the solvation free energy of neutral solutes in aqueous solution nearly linearly. This is described by the so-called Setchenow equation\cite{Sechenow1892-AnnChimPhys,long1952activity}
\begin{equation}
\Delta G_{\rm solv}(c_{\infty,{\rm ion}}) - \Delta G_{\rm solv}(0) = k_{\mathrm{s}} \frac{k_{\mathrm{B}} T}{\log _{10}(\mathrm{e})} c_{\infty,\mathrm{ion}} \quad ,
\end{equation}
with the Setchenow coefficient $k_{\mathrm{s}}$ and $e$ the electronic charge as a positive value. From tabulated Setchenow coefficients, one can estimate that a 1~M ion concentration in the electrolyte decreases solvation free energies by 0.1-0.3~kcal/mol, where the dominating effect is the energy penalty to create the ionic cavity.\cite{ringe2017transferable} Such apparently small changes to the solvation free energies can still have critical consequences, at least if one thinks of biochemistry where they are known to induce protein folding. Accounting for these changes may also be key in fitting implicit models to experimental solvation data with possibly finite salt concentrations. In this respect, tabulated Setchenow coefficients actually represent a direct way to determine the ionic parameters of the implicit solvation model. Ringe \textit{et al.}~used such a database of experimentally tabulated Setchenow coefficients to optimize the electron density cutoff that controls the ionic cutoff function $\alpha_\mathrm{ion}^\pm$ using an SCCS/S-MPB model.\cite{ringe2017transferable} This density cutoff was found to be correlated with the hydration number of the ions in the solution, showing that the parametrized model was able to predict ion-specific hydration effects for neutral molecules. 

To recapitulate, most contemporary implicit solvation models applied to the simulation of SLIs tend to use parameters derived from molecular solvation databases as a basis. While the actual transferability of these parameters to the SLI context is still unclear, there is generally at least sufficient molecular experimental reference data available to achieve a coherent parametrization for water as a predominant solvent also in interfacial electrocatalysis applications. The situation worsens quickly for non-aqueous solvents and is critical for ionic parameters. While this sets a perspective for the complexity of implicit solvation modeling one can aspire to, it clearly shows that even in the context of molecular solvation there is still room for improvement through the establishment of larger and chemically more diverse reference data bases including entries beyond standard state solvation free energies.

\subsection{Implicit solvation implementations in DFT program packages}

As we have shown in the previous sections, a wide variety of implicit solvation methodologies exist and it is their recent implementation into DFT program packages that can also deal with extended surfaces (e.g.~through the use of periodic boundary condition supercells) that has enabled such kind of modeling for the SLI context at all.

Table~\ref{tab:implicit_solvation_codes} shows a compilation of the implicit solvation methods and features that have been implemented into various state-of-the-art, periodic and non-periodic DFT program packages at the time of submission of this article. It clearly shows a tendency of predominant use of local, linear and isotropic dielectric models. All-electron DFT packages traditionally use sharp apparent surface charge (ASC) models, cf.~Section 2.3.2, instead of smooth dielectric models. This is partly due their  logarithmic integration grid structure to resolve localized core basis functions, which makes the solution of the GPE or PBE over the whole computational domain numerically difficult. ASC models have been implemented also in connection with the LPB equation to simulate simplified SLIs.\cite{mennucci1997evaluation} Among all these realizations of implicit solvation models, FHI-aims has been the first all-electron DFT package implementing a smooth dielectric response model (SCCS\cite{andreussi2012revised}), extended by an advanced Stern-layer and ionic size (lattice model) corrected PB (S-MPB) ion representation.\cite{ringe2016function,ringe2017transferable,ringe2017first} It also introduced an efficient Newton solver linearizing the S-MPB equation\cite{ringe2016function}, which has recently also been adapted by other DFT packages.\cite{nattino2019continuum} Q-Chem is an interesting alternative, in particular for non-equilibrium (frequency-dependent dielectric function) solvation calulations or heterogeneously structured dielectrica (e.g.~solvation at vacuum-water or liquid-liquid interfaces) via ASC methods. Q-Chem has also recently implemented the smooth dielectric S-MPB model to support modeling of electrolytes.\cite{stein2019poisson} 

The so far discussed program packages are ideal for the simulation of electrochemistry of finite-size nanoparticles. The implicit solvation simulation of extended metallic electrodes, realized by supercells with periodic boundary conditions, has been made available in several ASC schemes, implemented in CRYSTAL,\cite{labat2018implicit} GAUSSIAN\cite{cossi2004continuum} and Dmol$^3$.\cite{delley2006conductor} However, this domain is clearly dominated by pseudo-potential and then mostly plane-wave based DFT program packages, with their efficient Fourier-transform algorithms for periodic systems. Here, QUANTUM ESPRESSO\cite{andreussi2019continuum} and JDFTx\cite{schwarz2020electrochemical} are presently the clear leaders with most advanced implementations of solvation and ion models. In addition, QUANTUM ESPRESSO provides most sophisticated correction schemes for removing periodic boundary condition in the normal direction of the surface slab to avoid artificial slab-slab interactions.\cite{dabo2008electrostatics,andreussi2014electrostatics} VASP, arguably the most popular DFT code in the theoretical electrochemistry community, provides so far only basic implicit solvation functionality, but at least also a LPB solver which provides counter charges and then allows simulations of charged interfaces.\cite{mathew2019implicit} As pointed out in recent works, the use of VASP requires special care though due to the not self-explanatory shifting of the electrostatic potential and also problems with the dipolar slab correction which is supposed to correct slab-slab interactions across periodic boundaries.\cite{gauthier2019unified,gauthier2019practical,gauthier2019challenges} Other packages, such as GPAW,\cite{melander2019grand} ONETEP\cite{womack2018dl_mg} and BIGDFT\cite{fisicaro2016generalized} have recently also reported the required implementations for S-MPB based implicit solvation models and should thus also be valid options for future modeling of electrified interfaces.

\begin{table}[!ht]
\scriptsize
\begin{tabular}{l|l|
>{\columncolor{Gray}}l|
>{\columncolor{Gray}}c| 
>{\columncolor{Gray}}c|
>{\columncolor{Gray}}c| l}
\multicolumn{2}{c}{} & \multicolumn{4}{c}{\textcolor{gray}{\textbf{Dielectric model ($\varepsilon$)}}}&\\\hline
                   &   &  \textbf{Lin. model}/  &    &     \textbf{Non-lin.}                &      \textbf{Het.}           &  \\
                   \cline{5-6} 
\multirow{-2}{*}{ \textbf{DFT package}} & \multirow{-2}{*}{  \textbf{BCs}}& \textbf{Shape $s$} &                     \multirow{-2}{*}{\textbf{Non-loc}}  &         \textbf{Aniso}              &        \textbf{Non-eq.}             & \multirow{-2}{*}{\textbf{Salt model}} \\ \hline
%
%%%%%%%Q_CHEM%%%%%%%%%%
%
\multicolumn{7}{c}{\textbf{---full potential / all-electron---}}\\\hline
                   &  &  $s_{\rho_\mathrm{el}}$/$s_{\boldsymbol{r}}$,\cite{fisicaro2017soft,stein2019poisson}  &   &    $\times$                &      $\checkmark$\cite{coons2016hydrated,coons2018quantum}           & S-MPB\cite{stein2019poisson},\\
                    \cline{5-6}
                   &&$s_{\rho_\mathrm{el}}$\cite{claudino2019automatic}/$s_{\boldsymbol{r}}$&&&$\checkmark$\cite{cammi1999linear,cossi2001time,you2015comparison,cammi1995nonequilibrium,cossi2000separation}& LPB\\
                   &&& & &\cite{improta2006state,you2015comparison,tomasi1994molecular} &\\
\multirow{-4}{*}{Q-Chem\cite{shao2015advances}} & \multirow{-4}{*}{F}&(ASC\cite{chipman2002implementation})&   \multirow{-4}{*}{$\times$} &         \multirow{-3}{*}{$\times$}              &       \cite{mewes2017on,coons2016hydrated,coons2018quantum}             &   (ASC\cite{lange2011simple})  \\ \hline  
%                  
%%%%%%%FHI-aims%%%%%%%%%%
%
                   &&      $s_{\rho_\mathrm{el}}$\cite{andreussi2012revised,ringe2016function},      &         &    $\times$                &      $\times$           & S-MPB\\
                    \cline{5-6}
\multirow{-2}{*}{FHI-aims\cite{blum2009ab}} &  \multirow{-2}{*}{F}&$s_{\rho_\mathrm{el}}$(ASC\cite{sinstein2017efficient}) &                      \multirow{-2}{*}{$\times$}  &         $\times$              &        $\times$             & \cite{ringe2016function,ringe2017transferable,ringe2017first} \\ \hline
%                  
%%%%%%%CRYSTAL%%%%%%%%%%
%
%\cite{labat2018implicit}
                   &&      $s_{\boldsymbol{r}}$       &         &    $\times$                &      $\times$          & \\
                    \cline{5-6}
\multirow{-2}{*}{CRYSTAL\cite{dovesi2018quantum,r.2017crystal17}} &  \multirow{-2}{*}{P\cite{labat2018implicit}/F}&(ASC\cite{labat2018implicit,vassetti2021evaluation}) &                      \multirow{-2}{*}{$\times$}  &        $\times$              &      $\times$           & \multirow{-2}{*}{$\times$} \\ \hline
%                  
%%%%%%%Jaguar%%%%%%%%%%
%
                   &  &    $s_{\boldsymbol{r}}$          &       &   $\times$                &      $\times$           &LPB  \\
                   \cline{5-6}
\multirow{-2}{*}{Jaguar\cite{bochevarov2013jaguar}} &  \multirow{-2}{*}{F} & (ASC\cite{nicholls1991rapid,tannor1994accurate,marten1996new,marenich2007self,kelly2005sm6})&                      \multirow{-2}{*}{$\times$}  &         $\times$              &       $\times$             &  (ASC\cite{nicholls1991rapid,tannor1994accurate,marten1996new}) \\ \hline
%                  
%%%%%%%GAMESS%%%%%%%%%%
%
                   &  & $s_{\rho_\mathrm{el}}$/$s_{\boldsymbol{r}}$ &                                 &      $\times$           & $\checkmark$\cite{si2009heterogeneous}&LPB\\
                   \cline{5-6}
\multirow{-2}{*}{GAMESS\cite{barca2020recent}} &  \multirow{-2}{*}{F} & (ASC\cite{tomasi2005quantum}) &\multirow{-2}{*}{$\times$}         &       $\checkmark$\cite{mennucci1997evaluation,cances1997new,cances1998new}                &    $\checkmark$\cite{minezawa2011implementation,si2010analytic,wang2010excited,yoo2008solvent}                 & (ASC\cite{cossi1998ab})\\ \hline
%                  
%%%%%%%GAUSSIAN%%%%%%%%%%
%
     &   &    $s_{\rho_\mathrm{el}}$/$s_{\boldsymbol{r}}$    &                    &      $\times$ & $\times$          &  LPB \\
     \cline{5-6}
\multirow{-2}{*}{GAUSSIAN\cite{frisch2016gaussian16}} &  \multirow{-2}{*}{  P\cite{cossi2004continuum}/F}&   (ASC\cite{tomasi2005quantum})  &   \multirow{-2}{*}{$\times$} &        $\checkmark$\cite{mennucci2003ab}              &       $\checkmark$\cite{fukuda2011nonequilibrium,cammi2000fast,cossi2001time,scalmani2006geometries,improta2006state}             &  (ASC\cite{cossi1998ab})\\ \hline
%                  
%%%%%%%Dmol3%%%%%%%%%%
%
     &   &    $s_{\boldsymbol{r}}$    &                    &      $\times$ & $\times$         &   \\
     \cline{5-6}
\multirow{-2}{*}{ Dmol$^3$} &  \multirow{-2}{*}{  P\cite{delley2006conductor}/F}& (ASC\cite{klamt1993cosmo,delley2006conductor})   &   \multirow{-2}{*}{$\times$} &        $\times$              &       $\times$             &  \multirow{-2}{*}{$\times$}\\ \hline
%                  
%%%%%%%TURBOMOLE%%%%%%%%%%
%
     &   &     $s_{\rho_\mathrm{el}}$/$s_{\boldsymbol{r}}$      &                    &      $\times$  & $\times$          &   \\
     \cline{5-6}
\multirow{-2}{*}{ TURBOMOLE\cite{TURBOMOLE2017}} &  \multirow{-2}{*}{F}&   (ASC\cite{klamt1993cosmo,klamt2015comprehensive,schafer2000cosmo})  &   \multirow{-2}{*}{  $\times$ } &         $\times$              &        $\checkmark$\cite{klamt1996calculation,scalmani2006geometries}              &  \multirow{-2}{*}{$\times$ }\\ \hline
%             
%%%%%%%NWChem%%%%%%%%%%
%
     &   &         &                    &     $\times$ & $\times$           &   \\
     \cline{5-6}
\multirow{-2}{*}{ NWChem\cite{apra2020nwchem}} &  \multirow{-2}{*}{  F}& \multirow{-2}{*}{ ASC\cite{klamt1993cosmo,york1999smooth}}  &   \multirow{-2}{*}{  $\times$} &        $\times$              &        $\times$             &  \multirow{-2}{*}{$\times$}\\ \hline
\multicolumn{7}{c}{\textbf{---pseudopotential---}}\\\hline

%                  
%%%%%%%VASP%%%%%%%%%%
%
                   &   &     $s_{\rho_\mathrm{el}}$\cite{mathew2014implicit,petrosyan2005joint}     &   &   $\times$                &      $\times$           &  \\ 
                   \cline{5-6}
\multirow{-2}{*}{ VASP\cite{kresse1996efficient}} &  \multirow{-2}{*}{P}& &                       \multirow{-2}{*}{$\times$}  &          $\times$              &         $\times$            & \multirow{-2}{*}{LPB\cite{mathew2019implicit}}  \\ \hline
%                  
%%%%%%%QE%%%%%%%%%%
%
                   &  &      $s_{\boldsymbol{r}}$\cite{fisicaro2017soft,andreussi2019solvent},     &&     $\times$                &      $\times$           &  LPB, S-MPB, \\
                    \cline{5-6}
\multirow{-2}{*}{ QE\cite{garrity2014pseudopotentials,giannozzi2009quantum,giannozzi2020quantum,giannozzi2017advanced}} & \multirow{-2}{*}{P/F} & $s_{\rho_\mathrm{el}}$\cite{andreussi2012revised}&                    \multirow{-2}{*}{ $\checkmark$\cite{andreussi2019solvent}}         & $\checkmark$              &        $\times$             &PCC\cite{fisicaro2016generalized,nattino2019continuum}\\ \hline
%                  
%%%%%%%BigDFT%%%%%%%%%%
%
                   &   &    $s_{\boldsymbol{r}}$\cite{fisicaro2017soft}&     &     $\times$                &      $\times$           &  \\
                    \cline{5-6}
\multirow{-2}{*}{ BigDFT\cite{dawson2020complexity,mohr2015accurate,mohr2014daubechies,genovese2008daubechies,ratcliff2020flexibilities}} &   \multirow{-2}{*}{F/P\cite{fisicaro2020wet}}&  \multirow{-2}{*}{ $s_{\rho_\mathrm{el}}$\cite{andreussi2012revised}}&               \multirow{-2}{*}{    $\times$}  &         $\times$              &       $\times$             &  \multirow{-2}{*}{MPB\cite{fisicaro2016generalized}} \\ \hline
%                  
%%%%%%%GPAW%%%%%%%%%%
%
                   &  P/ &  $s_{\boldsymbol{r}}$,\cite{held2014simplified}       &&    $\times$                &      $\times$          & DD-S-MPB, \\     
                   \cline{5-6}          
\multirow{-2}{*}{ GPAW\cite{mortensen2005real,enkovaara2010electronic}} &F\cite{held2014simplified,otani2006first,melander2019grand}& $s_{\rho_\mathrm{el}}$,\cite{andreussi2012revised,fattebert2003first} &                      \multirow{-2}{*}{$\times$}  &         $\times$              &        $\times$            & LPB, PCC\cite{melander2019grand}  \\ \hline
%                  
%%%%%%%PWMat%%%%%%%%%%
%
                   &   &         &     &$\times$                &      $\times$          &  \\
                    \cline{5-6}   
\multirow{-2}{*}{ PWMat\cite{jia2013analysis,jia2013fast}} & \multirow{-2}{*}{P} &  \multirow{-2}{*}{ $s_{\rho_\mathrm{el}}$\cite{andreussi2012revised,gao2020substantial}}&                      \multirow{-2}{*}{$\times$}  &         $\times$              &       $\times$             &  \multirow{-2}{*}{S-LPB\cite{fisicaro2016generalized,gao2020substantial}}\\ \hline
%                  
%%%%%%%JDFTx%%%%%%%%%%
%
                  &   &     $s_{\boldsymbol{r}}$,\cite{fisicaro2017soft}   &$\checkmark$  &     $\checkmark$\cite{gunceler2013importance,sundararaman2017evaluating,sundararaman2018improving}                &                & CS-MPB,\cite{gunceler2013importance}   \\
          &&$s_{\rho_\mathrm{el}}$\cite{petrosyan2005joint,gunceler2013importance,sundararaman2017evaluating,sundararaman2015charge}&\cite{petrosyan2007joint,sundararaman2015spicing,sundararaman2014weighted}  &\cite{sundararaman2015spicing,sundararaman2014weighted} &\multirow{-2}{*}{$\times$}&S-CS-MPB,\cite{sundararaman2018improving}\\ 
          		\cline{5-6}
\multirow{-3}{*}{ JDFTx\cite{sundararaman2017jdftx}} &   \multirow{-3}{*}{P/F}& &   &        $\checkmark$\cite{JDFTx_input_tensor}              &        $\times$             & LPB\cite{gunceler2013importance,letchworth-weaver2012joint} \\ \hline     
%                  
%%%%%%%CP2K%%%%%%%%%%
%
     &   &     $s_{\rho_\mathrm{el}}$\cite{andreussi2012revised,fattebert2003first,yin2017periodic}    &                    &     $\times$ & $\times$          &   \\
     \cline{5-6}
\multirow{-2}{*}{ CP2K\cite{kuhne2020cp2k}} &  \multirow{-2}{*}{P}&     &   \multirow{-2}{*}{  $\times$} &         $\times$              &        $\times$             &  \multirow{-2}{*}{$\times$}\\ \hline
%                  
%%%%%%%ONETEP%%%%%%%%%%
%
                   &  &     $s_{\boldsymbol{r}}$,\cite{fisicaro2017soft,ARCHERsolv,womack2018dl_mg}   &  &     $\times$               &     \checkmark$^a$\cite{DziedzicONETEP}           & S-MPB\cite{womack2018dl_mg} \\               
                    \cline{5-6}   
\multirow{-2}{*}{ ONETEP\cite{prentice2020onetep}} &\multirow{-2}{*}{P/F\cite{DziedzicONETEP}} &   $s_{\rho_\mathrm{el}}$\cite{dziedzic2011minimal,fattebert2003first,ARCHERsolv,womack2018dl_mg}&                     \multirow{-2}{*}{$\times$}  &         $\times$              &        $\times$             & \cite{bhandari2020electronic,dziedzic2020practical,ARCHERsolv,womack2018dl_mg} \\ \hline
%                  
%%%%%%%CASTEP%%%%%%%%%%
%
     &   &         &                    &      $\times$ & $\times$          &   \\
     \cline{5-6}
\multirow{-2}{*}{ CASTEP\cite{clark2005first}} &  \multirow{-2}{*}{F}&  \multirow{-2}{*}{$s_{\rho_\mathrm{el}}$\cite{fattebert2003first,ARCHERsolv,womack2018dl_mg}}  &   \multirow{-2}{*}{  $\times$} &         $\times$              &        $\times$             &  \multirow{-2}{*}{$\times$}\\ \hline
\end{tabular}
    \caption{{\bf Overview over published implementations of implicit solvation models in various DFT program packages}. This compilation is to provide a rough picture of all the implemented features and corresponding references, with a focus on the electrostatic and ionic part of the grand potential functional. For the shape function, if not ASC is specified, a smooth dielectric step function is used. Legend: QE = QUANTUM ESPRESSO, BC = Boundary condition (referring to solvation model implementation, P=periodic, F=free), ASC = Apparent Surface Charge, Non-loc.~= Non-local, Non-lin.~= Non-linear, Aniso = Anisotropic (dielectric tensor), Het.~= Heterogeonous (different bulk dielectric permittivities in different regions, to model e.g.~systems at the air-water interface, or liquid-liquid interfaces), Non-eq.~= Non-equilibrium/frequency-dependent, PB = Poisson-Boltzmann, MPB = Lattice Size-Modified Poisson-Boltzmann, S = Stern correction, CS = hard sphere crowding effects based on Carnahan-Starling equation of state\cite{carnahan1969equation}, PCC = planar counter charge, DD = dielectric decrement. Lastly, GAMESS and Gaussian support a variety of ASC models and the user is referred to the documentation and the Tomasi paper for further review of the available methods.\cite{tomasi2005quantum} $a$: Only supports the use of regions with vacuum permittivity, not a different permittivity.}
    \label{tab:implicit_solvation_codes}
\end{table}

\newpage

\section{Implicit solvation models applied to electrified SLIs}
\label{sec:echemSLI}

\subsection{\emph{Ab initio} thermodynamics framework}
\label{sec:aitd}

Having established all methodological ingredients to implicit solvation schemes in Chapter~\ref{sec:solvtheory}, we now proceed to discuss their application in the context of electrified SLIs, and in particular at metal electrodes. Already in the introduction we had motivated that SLI applications presently focus predominantly on thermodynamic quantities, but that the actual DFT supercell typically only represents a grand-canonical sub-system in equilibrium with the general and electrochemical environment.  In order to evaluate the true thermodynamics in  SLI applications, it is therefore generally not sufficient to consider the hitherto discussed grand potential functional $\Omega^{N_\alpha}[\rho_{\rm el}] = F[\rho_{\rm el}] + \Omega_{\rm is}[\rho_{\rm el}]$, cf.~eq.~(\ref{eq:refinal_free}). This grand potential accounts for the exchange of all implicitly treated solvent particles and ions with their reservoirs and does therefore already depend on the electrochemical environment. However, this is only for one fixed chemical composition $N_\alpha$ of the explicitly, and thus DFT-described part of the system. To capture the full thermodynamics appropriately, we therefore would formally need to extend this to a modified total grand potential functional
\begin{align}
\tilde{\Omega}[\rho_\mathrm{el}]=\Omega^{\langle N_{\alpha}\rangle}[\rho_\mathrm{el}] - \sum_\alpha\tilde{\mu}_{\alpha}\langle N_{\alpha}\rangle \quad ,
\label{eq:omega-echem}
\end{align}
which additionally accounts for the possible exchange of all explicitly treated chemical species $\alpha$ of charge $q_\alpha$ with their corresponding reservoirs described through their electrochemical potentials\cite{schmickler2010interfacial}
\begin{eqnarray}
\tilde{\mu}_\alpha = \mu_\alpha +q_\alpha\phi \quad . \label{eq:echempot}
\end{eqnarray}
Full minimization of this total grand potential functional would then yield the average particle number $\langle N_{\alpha}\rangle$ of each explicitly described species at equilibrium.

Depending on the application at hand, it is typically convenient to distinguish sub-groups among these explicit chemical species. Frequently, one considers substrate atoms of the (metal) electrode with chemical potentials $\mu_{\rm sub}$, neutral solvent species $j$ with chemical potentials $\mu_{{\rm solv},j}$ (e.g.~water molecules in aqueous solvents), ions $i$ of the electrolyte with electrochemical potential $\tilde{\mu}_{{\rm ions},i}$ and electrons with electrochemical potential $\tilde{\mu}_{\rm el}$, with the latter indeed also just another chemical species in the thermodynamic sense. To this end, the electrocatalysis context and the use of an implicit solvation model dictate utmost care and add severe challenges in establishing a fully consistent set of corresponding electrochemical potentials. For one, the same chemical species might exist in both explicit and implicit parts of the system. This applies notably to mixed explicit/implicit models, where the inner DL is (partly) included in the DFT-treated part of the system. A common example are ice-like rigid water layers \cite{gauthier2019unified,choi2021understanding,van2019assessment,ping2015solvation,blumenthal2017energy,hormann2019absolute} to approximate the Helmholtz layer structure at metal electrodes in aqueous solutions.\cite{parsons1997metal,fumagalli2018anomalously,kornyshev2007electrochemical,nihonyanagi2013structure} For such systems, inconsistencies between the $\mu_{{\rm solv},j}$ or $\tilde{\mu}_{{\rm ion},i}$ employed in the explicit minimization of eq.~(\ref{eq:omega-echem}) and the one of the implicit electrolyte model could connect an erroneous free energy gain to the exchange of in principle equivalent explicit particles with implicitly described ones, or vice versa. If one indeed performed the full formal minimization of eq.~(\ref{eq:omega-echem}), this would then for instance spuriously favor to either describe the entire solvent in the DFT supercell explicitly or implicitly. A further challenge comes from surface chemical reactions, which can again not only convert explicitly and implicitly described species into another, but which in the electrocatalysis context in fact often involve the interconversion of species commonly assigned to different sub-groups. A prominent example would be protonation reaction steps, where a (charged) proton from the electrolyte and an electron end up forming part of a (neutral) reaction intermediate specifically adsorbed at the electrode.

The ongoing struggle to achieve such consistent sets of electrochemical potentials, in particular within the confines of present-day implicit solvation models, is one central reason, why contemporary works dodge the formal full minimization of the total grand potential of eq.~(\ref{eq:omega-echem}). A second crucial one concerns the intractability of the concomitant configurational sampling and thermodynamic averaging. Such sampling obviously would have to include all possible structures and chemical compositions of the SLI, a task that in particular due to the possibility of strong {\em operando} changes of working (electro)catalysts is generally as unfeasible as it is in thermal surface catalysis.\cite{reuter2016ab} In fact, electrocatalysis adds an additional level of complexity by the existence of charged species $\alpha$ (ions or electrons). If the SLI model contained in the DFT supercell does not account for the contribution of the diffuse DL to the 
full compensating counter charge, then as highlighted in Chapter \ref{sec:intro} this would imply the necessity to extend the sampling also over different overall charge states of the DFT supercell. However, in order to achieve an appropriate description of the extended interface and the metallic band structure of the electrode, DFT implementations predominantly employ periodic boundary conditions. This, for technical reasons, generally restricts such calculations to overall charge-neutral supercells and would thus prevent any such sampling of different supercell charge states. As already alluded to at several occasions, it is specifically the versatility with which implicit electrolyte models allow to include ionic counter charges into the DFT supercell that addresses this problem and we will discuss this in more detail in Section 3.4. Even if the grand-canonical sampling involved in the formal minimization of eq.~(\ref{eq:omega-echem}) can then be restricted to overall charge-neutral supercells, it is still generally impractical to perform this sampling simultaneously for the number of electrons and explicit particle species. This has to do with the predominantly canonical ansatz for the electron DOFs of major DFT packages (JDFTx\cite{sundararaman2017grand} and very recently also ONETEP\cite{bhandari2021electrochemistry} forming rare exceptions). Rather than adjusting the electron number as to grand canonically equilibrate with an imposed electron electrochemical potential (in electrochemistry given by the applied electrode potential), this ansatz imposes a fixed electron number $N_{\rm el}$---with the $\tilde{\mu}_{\rm el}$ to which this refers to then an outcome of the calculation. While not least the coupling to a potentiostat can still allow to indirectly determine the electron number that matches an applied electrode potential, cf. Section~\ref{sec:constantpotconstantcharge}, this is in general technically better not mixed with a simultaneous adaption of the chemical composition $N_\alpha$ of the DFT calculation.

For these multiple reasons, the prevalent approach in first-principles based SLI works with implicit solvation is to use an {\em ab initio} thermodynamics framework as also widespread in thermal surface catalysis research.\cite{reuter2016ab} Instead of the full minimization of the total grand potential functional of eq.~(\ref{eq:omega-echem}), such a framework considers a total grand potential functional at a fixed chemical composition $N_\alpha$ of the DFT-described part, 
\begin{align}
\tilde{\Omega}^{N_\alpha} [\rho_\mathrm{el}]=\Omega^{N_\alpha}[\rho_\mathrm{el}] - \sum_\alpha\tilde{\mu}_{\alpha} N_{\alpha} \quad.
\label{eq:omega-excess}
\end{align}
This $\tilde{\Omega}^{N_\alpha}[\rho_{\rm el}]$ is then evaluated and minimized individually for different trial chemical compositions $N_\alpha$ and geometric structures of the SLI, where we recall that the latter is included through the Born-Oppenheimer parametric dependence of the involved quantum-mechanical free energy functional $F[\rho_\mathrm{el}]$ on the positions $\{{\boldsymbol R}_\alpha\}$ of the explicitly treated species, cf.~eq.~(\ref{eq:refinal_free}). Subsequently, comparison of the resulting free energies for the individual candidate structures and compositions allows to conclude on their relative thermodynamic stability. In fact, the true equilibrium SLI structure and composition will yield a minimum such free energy, and with the exception of the neglected fluctuations around this equilibrium (contained in $\langle N_\alpha \rangle$ in eq.~(\ref{eq:omega-echem})) this free energy will be the same as the one one would also obtain from the full formal minimization of $\tilde{\Omega}[\rho_{\rm el}]$.

This {\em ab initio} thermodynamics approach is highly convenient. Not least, as there are no additional force terms beyond those already arising in the consideration of $\Omega^{N_\alpha}[\rho_{\rm el}]$. This allows to straightforwardly perform geometry optimizations or structural sampling through molecular dynamics, if only the employed DFT code has an implemented implicit solvent model to evaluate $\Omega^{N_\alpha}[\rho_{\rm el}]$ and associated force terms. This ease is treacherous though, as the obtained relaxed structure and sampled ensemble is, of course, restricted to the once fixed chemical composition. Likely even more consequential, all chemical sampling is now outsourced to the trial $N_\alpha$ explicitly tested. In other words, while the full formal minimization of $\tilde{\Omega}[\rho_{\rm el}]$ will yield the true equilibrium SLI structure and composition by construction, any evaluation of $\tilde{\Omega}^{N_\alpha} [\rho_\mathrm{el}]$ will only allow to conclude that among all compositions $N_\alpha$ (and corresponding structures $\{{\boldsymbol R}_\alpha\}$) explicitly tested, the one that yields the minimum free energy is the closest approximant to the true SLI structure and composition within the configurational space spanned by the trial structures and compositions. 

This kind of "poor man's sampling" is not only critical because of the human bias possibly introduced in the selection of trial structures and compositions. This is a problem that is generic to the described {\em ab initio} thermodynamics framework and we will not further discuss it here. More specific to the electrified SLIs and implicit solvation context is instead that also the fixed-composition total grand potential functional $\tilde{\Omega}^{N_\alpha} [\rho_\mathrm{el}]$ still depends on the bulk electrochemical potentials. Inconsistencies in these references will therefore also in general sensitively affect the relative stabilities of trial structures and compositions, and the corresponding conclusion on the closest equilibrium approximant. However, evaluation of $\tilde{\Omega}^{N_\alpha} [\rho_\mathrm{el}]$ in mindfully chosen configurational sub-spaces, e.g.~in the simplest case just different structures of the same chemical composition, can ease or even entirely lift these dependencies. Furthermore, with the focus typically on free energy differences as discussed in Section~\ref{sec:ImpSolvGeneralFormulation}, further error cancellation might be exploited by taking these differences already for the individual trial candidates, rather than only after performing the corresponding two full minimizations. As we will see in Section \ref{sec:constantpotconstantcharge}, 
this is prominently exploited in performing so-called constant-charge rather than constant-potential calculations. In this respect, the pragmatic focus on the fixed-composition total grand potential $\tilde{\Omega}^{N_\alpha}[\rho_\mathrm{el}]$ can provide highly useful insight and can circumvent issues that within the context of present-day implicit solvation models and DFT calculations would render a formal full minimization of $\tilde{\Omega}[\rho_\mathrm{el}]$ largely useless---even if it could be achieved practically.

This tight integration of implicit solvation models into the {\em ab initio} thermodynamics framework has advantages and disadvantages. On the negative side, it is often difficult to judge how well an employed implicit model really describes solvation effects at the electrified interface, as the computed thermodynamic quantities may also be affected by specificities of the {\em ab initio} thermodynamics ansatz. This has, not least, bearings on the parametrization issues of present-day implicit solvation schemes, as many of the quantities commonly measured in contemporary SLI electrochemistry can simply not be used to assess, advance or re-parametrize existing implicit solvation models. On the positive side, implicit solvation capabilities like the representation of counter charges within the DFT supercell are actually central to overcome some of the {\em ab initio} thermodynamics limitations---and this may in fact turn out to be even more relevant conceptually than the originally intended (and likely quite crude) account of the solvation effects {\em per se}. In the following sections we will further illustrate these various aspects. We will begin with specific thermodynamic quantities that are least affected by the {\em ab initio} thermodynamics framework (and its limitations) and thus most sensitive to the implicit solvation modeling itself, and then gradually move over to quantities where the two approaches get increasingly intertwined.

\subsection{Potential of zero charge}
\label{sec:pzc}

The potential of zero charge (PZC) and thermodynamic quantities evaluated at the PZC are a natural starting point for this survey. The PZC is generally defined as the applied electrode potential at which there is no excess charge at the metal electrode.\cite{frumkin1975potentials,schmickler2010interfacial}. Within the scope of this review, this implies that all electrolyte counter charges in the DL vanish, cf.~Fig.~\ref{fig:overview}. The DFT supercell is thus charge neutral and any corresponding restrictions in sampling the optimum charge state of $\Omega^{N_\alpha}[\rho_{\rm el}]$ do not apply. If we furthermore concentrate on the PZC of the pristine metal electrode and assume for the moment that the known structure and composition of the latter is not changed e.g.~by any specific adsorption of electrolyte species (as most likely fulfilled at unreactive coinage metals), then there is neither any chemical composition sampling issue, nor do we have to worry about any of the (electro)chemical potentials of explicit ions, solvent or electrode species. When aiming for a comparison with experimental values, we still have to define an appropriate reference for the electron electrochemical potential and in this respect even this simple application example provides already a manifestation of the subtle issues related with the definition of a consistent set of electrochemical potentials and their references when using implicit solvation models. 

Let us henceforth generally denote electrode potentials on the absolute scale (i.e.~versus vacuum reference $\Phi_{\rm E, vac}=0$) with $\Phi_{\rm E}$ and correspondingly an experimentally measured PZC on this scale with $\Phi_{\rm E,PZC}$. Then we can exploit that $e \Phi_{\rm E,PZC}$ with $e$ the (positive) elementary charge corresponds identically to the work function of the metal immersed in solution.\cite{trasatti1990absolute,trasatti1991structure} In a periodic, canonical gas-surface DFT supercell calculation with a slab model representing the surface, the work function vs. vacuum is conveniently computed as $e(\phi_{\rm F}-\phi_{\rm vac})$. Here, $\phi_{\rm F}$ is the electron Fermi level, which in the canonical calculation equals the DFT-internal electron chemical potential and is as mentioned before an outcome of the DFT calculation once self-consistency is achieved. $\phi_{\rm vac}$ is the DFT-internal vacuum potential, which one approximately obtains as the position of the average electrostatic potential in the middle of the vacuum region between the (periodically repeating) slabs. If this vacuum region in the supercell is now filled with implicit solvent, one would think that the same difference would directly yield the PZC. Unfortunately, this is not the case, as this difference only accounts for bringing the electron to the bulk of the implicit solvent. What is thus missing to be able to align to the experimental absolute scale, is the potential difference between implicit solvent and vacuum. 

The corresponding potential drop at say a water-vacuum interface can in principle be determined from higher-level explicit simulations.\cite{leung2010surface,ambrosio2018absolute,haruyama2018electrode,hormann2019absolute} However, this drop differs substantially from the required implicit water-vacuum drop,\cite{hormann2019absolute} as the average electrostatic inner potential of bulk water that dominates the prior drop\cite{leung2010surface} vanishes in implicit models. Alternatively, one might argue that due to the vanishing polarization at implicit-vacuum interfaces, the missing potential difference might actually be small\cite{leung2010surface}. While this seems indeed supported by a recent study,\cite{hormann2019absolute} it is still not a firm basis for a quantitative alignment. Similar alignment issues arise equally for other experimental referencing scales like the predominantly employed standard hydrogen electrode (SHE) for aqueous environments. By and large, this then presently prevents the desirable direct comparison of computed and measured PZC values for different electrodes as an accuracy test of the implicit solvent model (and specifically its parametrization). 

Rather than actually assessing the performance of the implicit solvation description, the comparison to measured PZCs is therefore instead employed to empirically fit the unknown implicit solvent-vacuum potential drop\cite{letchworth-weaver2012joint,gunceler2013importance} or to re-parametrize the implicit solvent model to effectively match an experimental PZC, e.g.~for Pt electrodes \cite{hormann2019grand, bramley2020reconciling}. To this end, it has to be emphasized though that in such procedures experimental data needs often to be re-referenced from a reference electrode scale to the absolute scale, e.g.~using the absolute SHE potential,\cite{trasatti1990absolute} which in itself introduces quite some uncertainties on the experimental numbers. Notwithstanding, since the implicit solvent-vacuum potential drop is solvent-specific, but electrode independent, useful insight into the performance of the implicit solvation model can still be obtained from relative PZCs, i.e.~PZC trends over different electrodes. 

\begin{figure}[!ht]
    \centering
    \includegraphics[width=0.5\textwidth]{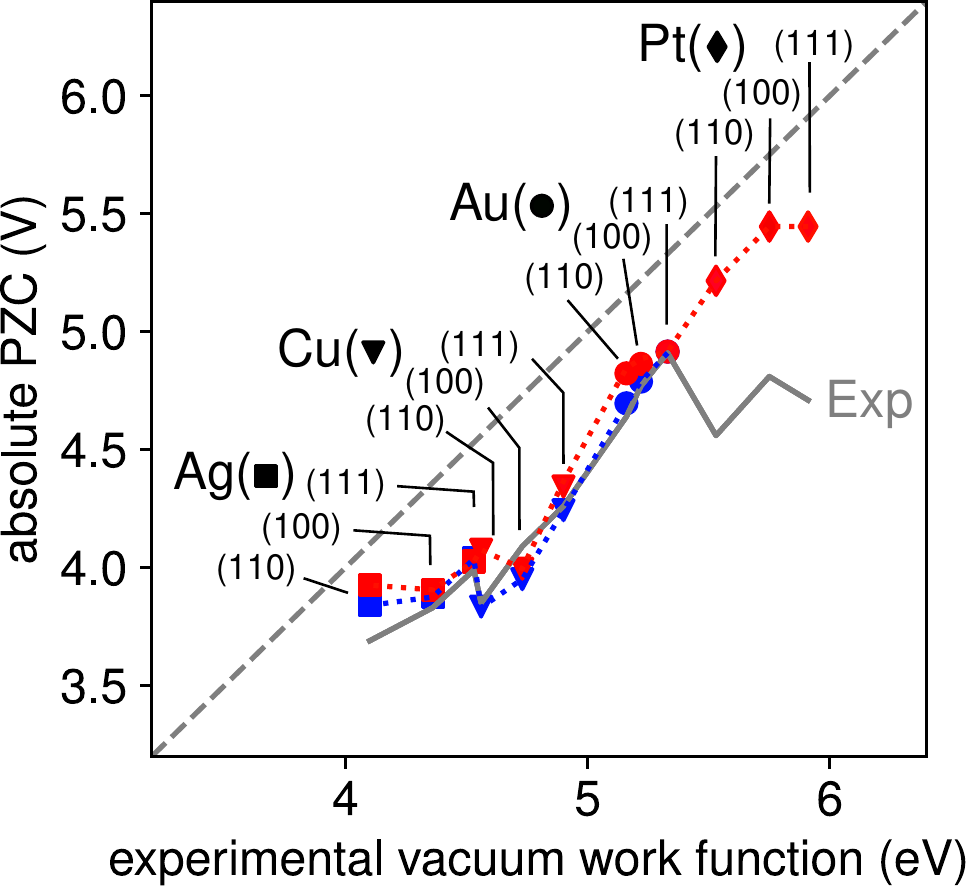}
    \caption{{\bf Relative trend of PZC values as obtained in experiments and implicit solvent calculations.} Experimental PZCs for the low-index surfaces of Ag, Cu, Au and Pt (gray line) are on the absolute scale and taken as averages of literature data compiled in the SI of ref. \citenum{hormann2019grand}. Calculated PZCs are arbitrarily aligned to the experimental PZC of Au(111) and taken from refs. \citenum{hormann2019grand} (red) and
    \citenum{letchworth-weaver2012joint} (blue).}
    \label{fig:pzc}
\end{figure}

A corresponding comparison with experimental data in water is shown in Fig.~\ref{fig:pzc} and reveals that even the fully implicit models employed in the corresponding studies capture this trend qualitatively, if not semi-quantitatively. As also apparent from Fig.~\ref{fig:pzc}, experimental absolute PZCs are consistently smaller than the corresponding vacuum work functions.\cite{bockris1969determination,trasatti1991structure,White1999modern,climent1999effect,schmickler2010interfacial} In Fig.~\ref{fig:pzc} the computed PZC values were arbitrarily aligned to the experimental PZC of Au(111) to illustrate the trend behavior, so that from there no conclusion can be drawn in how much implicit solvation models can reproduce this reduction. However, if the implicit water-vacuum drop is indeed small, then one can indeed show that they would effectively yield this reduction, albeit likely only on a quantitatively smaller scale.\cite{letchworth-weaver2012joint,fang2013theoretical,kastlunger2018controlled,hormann2019grand}. This is actually surprising, since the reduction originates in reality mainly from finite charge transfer from water molecules in the inner DL and a concomitant polarization within the first $\sim 4$\,{\AA} away from the metal surface.\cite{wasileski2008first,schnur2011challenges,le2017determining} Outliers to the captured trend in the PZCs, e.g.~the larger offset for Pt visible in Fig.~\ref{fig:pzc}, have been ascribed to an increased interfacial charge transfer that can no longer be mimicked by the implicit solvation model.\cite{le2017determining} However, for more reactive surfaces one also has to keep in mind that experimental PZCs can not least be influenced by specifically adsorbed electrolyte ions\cite{frumkin1975potentials,frumkin1980potentials}. If the calculations were repeated for a corresponding chemical composition $N_\alpha$ including such ions, the agreement might thus improve. However, in general, this uncertainty is nothing but a consequence of the "poor man's sampling" in {\em ab initio} thermodynamics, which requires the explicit testing of different such chemical compositions $N_\alpha$, rather than yielding the true equilibrium one as an outcome of the theory.

\subsection{Computational hydrogen electrode}
\label{sec:CHE}

Many of the thermodynamic quantities that are of central interest in electrocatalysis concern energetics. Take the surface free energy as a measure of the stability of the catalyst surface, or adsorption free energies as central to the surface thermochemistry (or within Br{\o}nsted-Evans-Polanyi relationships\cite{bell1936theory,bronsted1924stochiometry,evans1936further} even indicative of the reaction kinetics). Computing or evaluating such energetic quantities within the {\em ab initio} thermodynamics framework described in Section~\ref{sec:aitd} almost invariably involves comparing the relative stability of interface  configurations with different chemical composition $N_\alpha$. Next to the electron electrochemical potential referencing issues discussed in the preceding section, this then also puts the electrochemical potentials of particle species $\alpha$ like explicitly described ions, solvent molecules or electrode constituents on the agenda. In addition, the energetic quantities are typically not only required at the PZC, which could then bring up first trouble with the charge restriction of prevalent periodic boundary DFT supercell implementations. 

Let us exemplify this general problem in this section for the simple case of hydrogen adsorption (formally better proton electrosorption) in an aqueous environment. A pertinent thermodynamic quantity to compute for this case is the adsorption free energy. Within the {\em ab initio} thermodynamics ansatz this is suitably defined as an applied-potential dependent difference of fixed-composition total grand potential energies before and after the adsorption,
\begin{align}
    \Delta G^{{\rm ads},N_\alpha}_{\rm H}(\Phi_{\rm E}) =  \tilde{\Omega}^{N_\alpha, {\rm H}}[\rho^\circ_\mathrm{el,H}(\Phi_{\rm E})] - \tilde{\Omega}^{N_\alpha}[\rho^\circ_\mathrm{el}(\Phi_{\rm E})] \quad . 
\label{eq:Gads_Hproton}
\end{align}
Here, $N_\alpha$ summarizes the entire chemical composition of the electrode, which we assume to be unchanged upon adsorption apart from the additional proton (H nucleus). For simplicity of notation, we consider here only one proton per supercell, even though one would in the implicit solvation context practically prefer symmetric slab setups with adsorption of one proton per surface and thus two protons per supercell. Obviously, within the employed periodic boundary conditions adsorption of this proton per supercell corresponds as always effectively to some finite coverage, but this does not matter for our present argument. Note also that similarly as in Section~\ref{sec:param} we again attest to the fact that (measurable) adsorption free energies are generally seen as a property of the full (macroscopic) system and not of the grand-canonical sub-system technically employed in the calculations, which is why we denote them as free energies even though they are computed here as a difference of grand potential energies.

Each fixed-composition total grand potential energy in eq. (\ref{eq:Gads_Hproton}) is evaluated at its optimized equilibrium electron density. Due to the presence of the additional H nucleus, $\rho^\circ_{\rm el,H}$ will not only differ in its detailed spatial form from $\rho^\circ_{\rm el}$, but will generally also integrate up to a total number of electrons that differs by $l_\mathrm{}$---the so-called electrosorption valency \cite{lorenz1962zum,vetter1972potentialabhangigkeit,vetter1972stromfluss,schultze1973experimental,frumkin1974on,vetter1974general,lorenz1977partial,valette1978specific,foresti1998electrosorption,habib1980specific,trasatti1983interphases,de2004electrosorption,guidelli2005electrosorption,schmickler1987ionic,schmickler1988surface,schmickler2010interfacial,schmickler2014partial,hormann2020electrosorption,hormann2021thermodynamic,hormann2021thermodynamiccyclic}. With the electrode chemical composition unchanged, the dependence on all electrode chemical potentials $\mu_{\rm sub}$ cancels in the difference of eq.~(\ref{eq:Gads_Hproton}). What remains are the electrochemical potentials of the bulk reservoirs from where the proton and the additional electron density dragged to it upon adsorption came from. At the electrified interface in the aqueous environment this correspond to $\tilde{\mu}_{\rm H^+}$ of the solvated proton and the electron electrochemical potential $\tilde{\mu}_{\rm el}$ as determined by the applied potential. Using eq.~(\ref{eq:omega-excess}) we can thus rewrite eq.~(\ref{eq:Gads_Hproton}) as a difference of grand potential energies and these electrochemical potentials
\begin{align}
    \Delta G^{{\rm ads},N_\alpha}_{\rm H}(\Phi_{\rm E}) =  \left( \Omega^{N_\alpha, {\rm H}}[\rho^\circ_\mathrm{el,H}(\Phi_\mathrm{E})] - \Omega^{N_\alpha}[\rho^\circ_\mathrm{el}(\Phi_\mathrm{E})] \right) - \left( \tilde{\mu}_{\rm H^+} + l_\mathrm{} \tilde{\mu}_{\rm el} \right) \quad . 
\label{eq:Gads_Hproton2}
\end{align}
This equation now clearly carves out the entire wealth of practical problems that have to be dealt with. Under an applied potential $\Phi_{\rm E}$ away from the PZC, the electrode will generally be charged with a corresponding balancing counter charge built up in the electrolyte. Some of this counter charge will be located in the diffuse DL, which is likely outside of a practically feasible DFT supercell as discussed in Chapter \ref{sec:intro}. Unless this is suitably taken care of by an implicit electrolyte model as discussed in the next section, this would imply the computation of charged supercells. Furthermore, to evaluate eq.~(\ref{eq:Gads_Hproton2}) we also need to determine the two electrochemical potentials. While we have already seen the difficulties to align $\tilde{\mu}_{\rm el} = -e \Phi_{\rm E}$ on the absolute scale with the DFT-internal Fermi level in an implicit solvation calculation in the last section, also the explicit computation of the electrochemical potential of a solvated proton $\tilde{\mu}_{\mathrm{H}^{+}}$ is a tough endeavor\cite{cheng2009redox,cheng2010aligning,cheng2014redox}.

Intriguingly, all of these problems vanish completely with just one single and ingenious approximation. If we assume that the optimized electron density of a given interface configuration $N_\alpha$ at any applied potential $\Phi_{\rm E}$ remains the same as the one at its PZC, then there is no electrolyte counter charge as discussed in the previous section and the DFT supercell is always charge neutral. For $\rho^\circ_\mathrm{el,H}$ this implies that the adsorbed protonic charge is exactly compensated by one additional electron, i.e.~the H adsorption (proton electrosorption) process is a so-called proton-coupled electron transfer (PCET).\cite{hammes-schiffer2010introduction,sakaushi2020advances} In turn, $l_\mathrm{} = 1$ and we arrive at
\begin{align}
    \Delta G^{{\rm ads},N_\alpha}_{\rm H,CHE}(\Phi_{\rm E}) =  \left( \Omega^{N_\alpha, {\rm H}}[\rho^\circ_\mathrm{el,H}(\Phi_\mathrm{E,H,PZC})] - \Omega^{N_\alpha}[\rho^\circ_\mathrm{el}(\Phi_\mathrm{E,PZC})] \right) - \left( \tilde{\mu}_{\rm H^+} + \tilde{\mu}_{\rm el} \right) \quad . 
\label{eq:Gads_Hproton_CHE}
\end{align}
Also, the electrochemical potential calculation and alignment problem is naturally resolved, as the remaining integer sum of the two potentials in eq.~(\ref{eq:Gads_Hproton_CHE}) is simply related to the applied potential $\Phi_{\rm E}^{\rm SHE}$ on the SHE scale,\cite{trasatti1990absolute,schmickler2010interfacial, norskov2004origin}
\begin{eqnarray}
\left( \tilde{\mu}_{\rm H^+} + \tilde{\mu}_{\rm el} \right) \;=\; \tfrac{1}{2}\mu_\mathrm{H_2} -e \Phi^{\rm SHE}_{\rm E} - k_{\rm B}T\ln(10)\rm{pH}  \quad .
\label{eq:SHE}
\end{eqnarray}
Here, $\mu_\mathrm{H_2}$ is the chemical potential of hydrogen gas at standard state, which is straightforward to compute,\cite{norskov2004origin,rossmeisl2005electrolysis,rogal2007ab,todorova2014extending,reuter2016ab}
and pH is the pH value of the aqueous electrolyte. Note that the SHE scale is the predominantly employed scale in experiments anyway, which thus does allow to directly compare with experiments (as long as they are not affected by mass transport effects\cite{ringe2020double,griesser2021true}). There is correspondingly no need anymore to align the DFT-internal Fermi level to the applied potential. In fact, as the difference of grand potential energies in eq.~(\ref{eq:Gads_Hproton_CHE}) is now potential-independent, it suffices to compute it once, and the entire dependence of the adsorption free energy on the applied potential is then just analytically given by eq.~(\ref{eq:SHE}). This analytic dependence becomes even easier on the reversible hydrogen electrode (RHE) scale,
\begin{eqnarray}
\left( \tilde{\mu}_{\rm H^+} + \tilde{\mu}_{\rm el} \right) \;=\; \tfrac{1}{2}\mu_\mathrm{H_2} -e \Phi^{\rm RHE}_{\rm E} \quad ,
\label{eq:RHE}
\end{eqnarray}
i.e. all pH dependencies are in the CHE just trivially Nernstian.

\begin{figure}
    \centering
    \includegraphics[width=.8\textwidth]{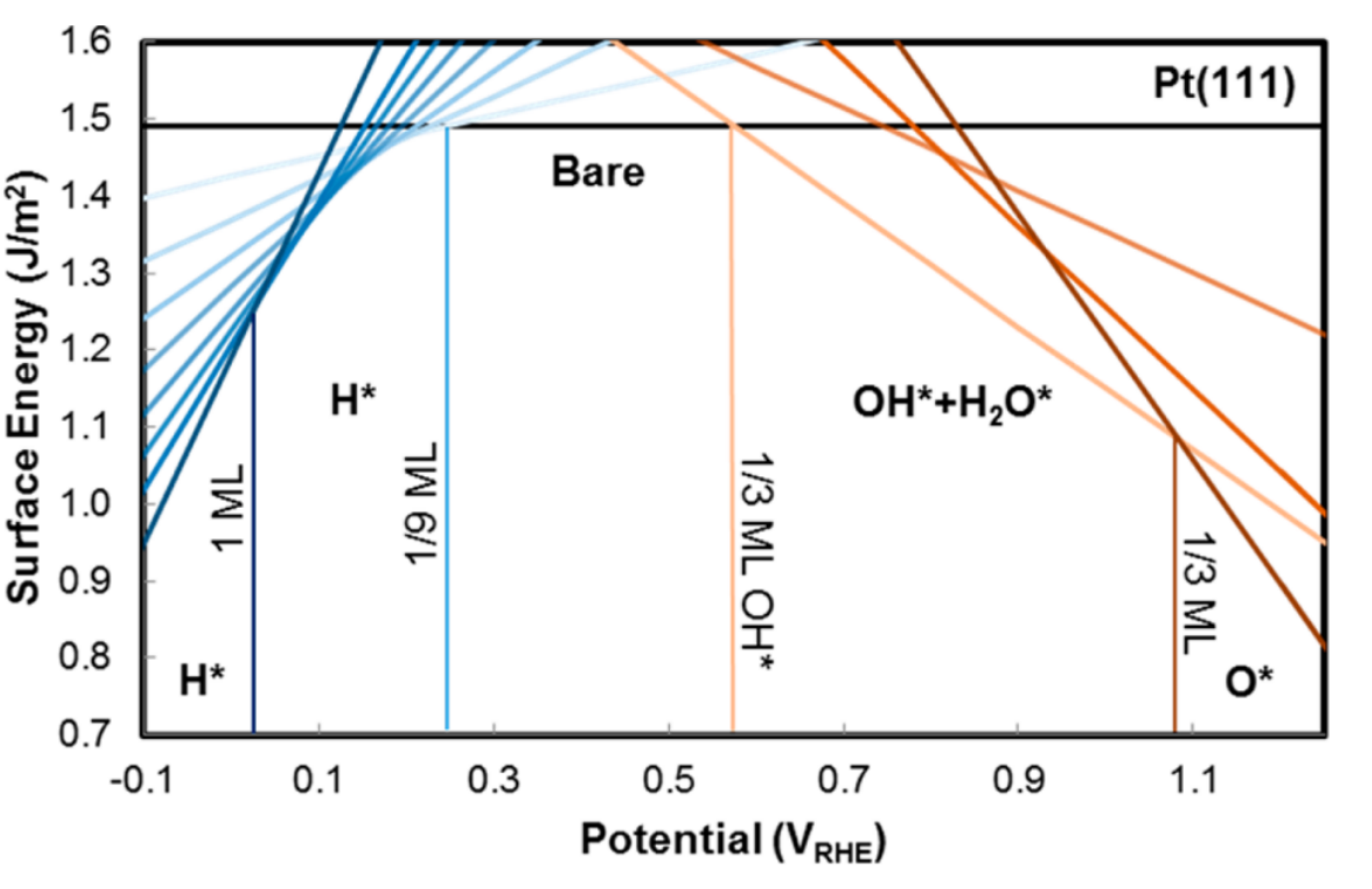}
    \caption{{\bf Surface phase diagram of Pt(111) in water as determined within the CHE approach.} Shown are computed potential-dependent surface free energies of bare Pt(111) and various H, OH and O coverages on it. Within the {\em ab initio} thermodynamics framework, surface terminations with lowest surface free energy are declared as most stable one at the corresponding potential. This yields the indicated gradual transition from H-covered over bare surface to OH- and O-covered terminations with increasingly positive potential.
    Reproduced with permission from ref.~\citenum{mccrum2017first}.}
    \label{fig:surfaceEnergyPt111}
\end{figure}

The original intention to exploit the SHE definition to circumvent the electrochemical potential referencing issues was coined computational hydrogen electrode (CHE) by Rossmeisl, N{\o}rskov and coworkers.\cite{norskov2004origin,rossmeisl2005electrolysis} Nowadays, CHE is instead essentially equated with the somewhat stronger approximation to employ PZC optimized densities as introduced in the example above. This kind of CHE approach underlies the by far dominant part of contemporary first-principles based work on electrified interfaces and electrocatalysis at them. In fact, it is fair to say that without the computational simplicity enabled by the CHE, theoretical electrocatalysis would not be where it is today.\cite{callevallejo2012first} The CHE philosophy is readily generalized to other electrodes (computational sulfur electrode, computational Li electrode,$\ldots$)\cite{mccrum2015electrochemical,hormann2015some,lespes2015using} and employed for the computation of other thermodynamic quantities. Notably, these are the aforementioned surface free energies\cite{reuter2001composition,rossmeisl2006calculated,hansen2008surface, hoermann2014stability,gossenberger2015equilibrium,mccrum2017first,opalka2019ab,griesser2021true}, i.e. the cost to create a surface with a certain structure and composition, as well as thermodynamic reaction barriers or concomitant thermodynamic overpotentials\cite{norskov2004origin,rossmeisl2005electrolysis,norskov2009towards,rossmeisl2008electrocatalysis,sakong2016importance,haobo2020active}. Using the prior applied-potential dependent surface free energies to compare the stability of a range of candidate surface structures and composition, one can readily establish surface phase diagrams, which---if the electrochemical potential dependence is resolved into a potential and pH dependence---are also known as Pourbaix diagrams. Figure~\ref{fig:surfaceEnergyPt111} illustrates this with corresponding work from McCrum~{\em et al.} for the Pt(111) surface in a water environment. Such kind of CHE surface phase diagrams are nowadays widely used to draw first conclusions on the actual surface structure and composition of electrodes under true operating conditions and we refer to excellent reviews on this topic\cite{seh2017combining, norskov2009towards,zeng2015towards,bagger2020fundamental,callevallejo2012first,abidi2021atomistic} for a  more detailed overview of the uses and merits of this kind of most popular CHE application.

If employed within the sketched CHE approach, the task of an implicit solvation model is to account for the solvation response at the PZC of the considered surface configuration. This is conceptually analogous to what was discussed in the previous section for the pristine electrode, yet with two notable, opposing differences. On the one hand, in particular for larger, more protruding, polar or hydrogen-bond affine adsorbates one can in principle expect larger solvation corrections even at the PZC.\cite{iyemperumal2017evaluating,gray2017quantifying,iyemperumal2017evaluating,zhang2019solvation,patel2018theoretical,ludwig2019solvent,gauthier2017solvation} On the other hand, as apparent from eq.~(\ref{eq:Gads_Hproton_CHE}) it is typically differences of grand potential energies that matter for the targeted thermodynamic energetic quantities and in these differences solvation corrections partly cancel. As an upshot, such corrections to adsorption free energies by fully implicit solvation models are typically small for the prototypical adsorbates of interest in aqueous environments in particular for *H and *O, while being slightly larger for *OH, *OOH or *H$_2$O, maybe reaching up to some hundred meVs for the dipolar species. By and large, this seems to agree with the results of calculations with explicit solvent,\cite{karlberg2006adsorption,rossmeisl2008electrocatalysis,zhang2019solvation,park2020elucidating} but this is most likely just due to fortuitous error cancellation in such free energy differences rather than evidence for the accuracy of present-day implicit solvation models and their existing parametrizations. More detailed analysis of the contributions points out that an arguably most important correction to existing schemes would be to account for so-called competitive solvent adsorption in the cavitation grand potential. This reflects the fact that in a thermodynamically consistent treatment of solvation at SLIs, there should be an energy cost associated with the need to first remove solvent from the pristine electrode to create space for the adsorbate. This would generally be a substrate-dependent cost,\cite{meng2004water,tonigold2012dispersive} in contrast to the existing, substrate agnostic cavitation grand potential formulations discussed in Section~\ref{sec:ImpSolvCavity}.

We think it could mainly be this missing appropriate account of competitive solvent adsorption that stands behind the (partly) dramatic discrepancies between implicit solvation results and benchmark AIMD simulations in explicit water environments.\cite{steinmann2016solvation,ludwig2019solvent,heenen2020solvation} This view would be supported by the strong correlations with the OH/H$_2$O adsorption properties of the substrate\cite{heenen2020solvation}. Note, however, that competitive solvent adsorption is also not appropriately considered in a wide range of simple explicit solvation strategies,\cite{norskov2004origin,park2020elucidating} while it is generally questionable anyway whether the limited trajectories obtained in the dynamic simulations can really faithfully mimic thermodynamic equilibrium. On the implicit solvation side, there are some hints that more substrate-specific models such as the SCSS model using soft-sphere atomic cavities might constitute a way forward while not compromising other observables.\cite{bramley2020reconciling} Nevertheless, while all of this surely indicates the need to further advance implicit solvation models (or to rather move over to mixed explicit/implicit solvation models for SLIs), the fact remains that in a CHE free energy difference like in eq.~(\ref{eq:Gads_Hproton_CHE}) for the adsorption free energy, solvation corrections evaluated at the PZCs tend to be small. One can correspondingly find multiple practitioner works in the literature, where the CHE is applied and in fact no solvation treatment is included at all, i.e. the underlying DFT calculations are actually performed for slabs in vacuum. In historical perspective, the advent of implicit solvation methodology in major periodic boundary conditions DFT packages came after the CHE approximation was firmly established and widely employed by the theoretical electrocatalysis community. The new functionality was then often employed within the prevalent CHE, rather than realizing that it could actually constitute a powerful avenue beyond it.

\subsection{Surface charging and interfacial capacitance}
\label{sec:surfcharge}

It is indeed important to realize that the typically small solvation corrections within the CHE are an invariable outcome of the PZC assumption. To assess this assumption, let us recall the physical processes actually occurring at a pristine electrode on gradual application of a potential that brings us away from its PZC. Without loss of generality, let this be a potential positive from the PZC, which will thus withdraw electrons from the electrode and lead to the formation of a positive net surface charge on the electrode surface. In order to screen the resulting electric field, a compensating counter charge will build up in the electrolyte part of the DL. Initially, this is just a capacitive charging process of the electric DL as introduced in Chapter \ref{sec:intro}. This means that the concomitant changes to the electrode electron density might induce some atomic relaxation or even stronger rearrangements in the electrode material. Also, the molecular and ionic distributions within the electrolyte will obviously change when building up the counter charge. However, at the initially small potentials there is formally no exchange of (charged) matter between these two constituents of the DL. In a fully implicit solvation model, this would thus mean that the chemical composition $N_\alpha$ of the DFT-part of the system does not change. 

Upon further increase of the potential away from the PZC, the polarization of the DL might eventually become so large, that such an exchange will occur, specifically in form of a so-called interfacial (or Faradaic) charge transfer. Here, it is now generally the transfer of ions to or from the electrolyte with a concomitant change of their charge state that reduces the electric field. 
For the considered positive potential and an aqueous electrolyte, this could for instance be the specific adsorption of anions,\cite{kasuya2016anion,pajkossy1996impedance,valette1982double,valette1981double,hu2017anion,cuesta2000adsorption,koga2001specific} or depending on the pH, the formation of hydroxyl groups at the metal electrode.\cite{berna2007new,kristoffersen2018oh,rizo2015towards} Even in a fully implicit solvation model, these new surface species would be explicitly modeled and we would correspondingly arrive at a new chemical composition $N_{\alpha'}$. This new electrode configuration then has its own new PZC,\cite{tian2009how} typically at a more positive potential as the original pristine surface (cf.~normal vs anomalous work function change\cite{gossenberger2014change}). When now further increasing the potential, we are again moving away from this PZC and the sequence of capacitive charging and interfacial charge transfer upon exceeding DL polarization continues.

What the CHE does is to approximate this sequence with a series of pure charge-transfer processes through unpolarized electrode configurations. As it only considers the electron density at the PZC of each electrode configuration, it is agnostic to capacitive charging. The increasing applied potential enters the fixed-composition total grand potential only through the changed electron electrochemical potential term as in the hydrogen adsorption eq.~(\ref{eq:Gads_Hproton_CHE}) above. Within the {\em ab initio} thermodynamics framework, any change of relative stability of different electrode configurations can thus only be captured, if the concomitant change of $N_\alpha$ includes a change in the involved number of electrons $N_{\rm el}$, as is the case for an interfacial charge transfer. By construction, the CHE approximation can therefore for instance not account for potential-induced geometric changes or stronger reconstructions of the electrode that leave the chemical composition $N_\alpha$ unchanged.

In order to overcome these limitations of the CHE it is therefore imperative to include some form of surface charging into the modeling. Remember that one of the motivations for historically introducing the CHE was the charge-neutrality restriction of prevalent periodic boundary condition DFT implementations. This restriction is elegantly addressed by the PZC assumption as the electron density then automatically integrates up to exactly match the total nuclei charge of the DFT part of the supercell. Yet, even in these codes there is in practice nothing that prevents us from adding more or less electrons into the DFT calculation to mimic surface charging. What the codes would only do (more or less unnoticed), is to introduce a homogeneous background charge that exactly compensates the net charge that would result from this varied electron number.\cite{bhandari2020electronic,womack2018dl_mg} In principle, this straightforward, so-called jellium approach can still be and is in fact largely used to model a potential-dependent electron density.\cite{lozovoi2001ab,taylor2006first,mamatkulov2011an} Obviously though, a homogeneous background charge---if applied without further corrections---is unlikely a good representation of the DL counter charge and there are several studies that highlight the nonphysical surface charging behavior obtained within this approach.\cite{kastlunger2018controlled,bhandari2020electronic,melander2019grand}

It is especially to this problem of surface charging where implicit solvation methodology adds significant flexibility. As discussed in Section \ref{sec:ImpSolvElectrolyteModels} and summarized in Fig.~\ref{fig:ion_models}, current electrolyte models offer a wide spectrum of introducing counter ions into the supercell, thereby allowing to establish overall charge neutrality without the need for a jellium background. This spectrum ranges from the simple PCC Helmholtz-layer models to the self-consistent ion distributions of PB or MPB theory, where, importantly, the ion distributions described in the latter theories include the diffuse DL. Here, it is worthwhile to emphasize the elegance with which these implicit approaches solve the problem of the wide extension of the diffuse DL part highlighted in Chapter \ref{sec:intro}. Even if this extension largely exceeds the actual dimension of the supercell employed in the practical DFT calculation, this still plays no role as it only enters the generalized Poisson equation solver of the code, cf.~Section \ref{sec:ImpSolvElectrolyteModels}. Corresponding solvers can be implemented with 
free boundary conditions in $z$-direction,\cite{andreussi2014electrostatics,dabo2008electrostatics}  i.e.~vertical to the slab surface, and are then completely independent of the finite supercell size used in the other parts of the KS DFT minimization. 
Whatever the specific electrolyte model used, its ionic counter charges thus flexibly allow to satisfy the supercell charge-neutrality condition despite a varying net surface charge on the explicitly DFT-described electrode. The implementation of corresponding implicit electrolyte models in a range of major DFT packages as summarized in Table~\ref{tab:implicit_solvation_codes} marked therefore a big conceptual step forward for the first-principles based modeling of electrified SLIs. What remains to be seen though, is what this actually brings practically for the modeling of surface charging and the truly intended computation of fixed-composition total grand potential energies $\tilde{\Omega}^{N_\alpha}[\rho_{\rm el}^\circ(\Phi_{\rm E})]$ with potential-dependent optimized electron densities $\rho_{\rm el}^\circ(\Phi_{\rm E})$, cf.~eq.~(\ref{eq:omega-excess}). Which aspects of the implicit physical counter charge model are truly important and how influential is the parametrization of the effective solvation model?

\begin{figure}[htb]
\includegraphics[width=.99\textwidth]{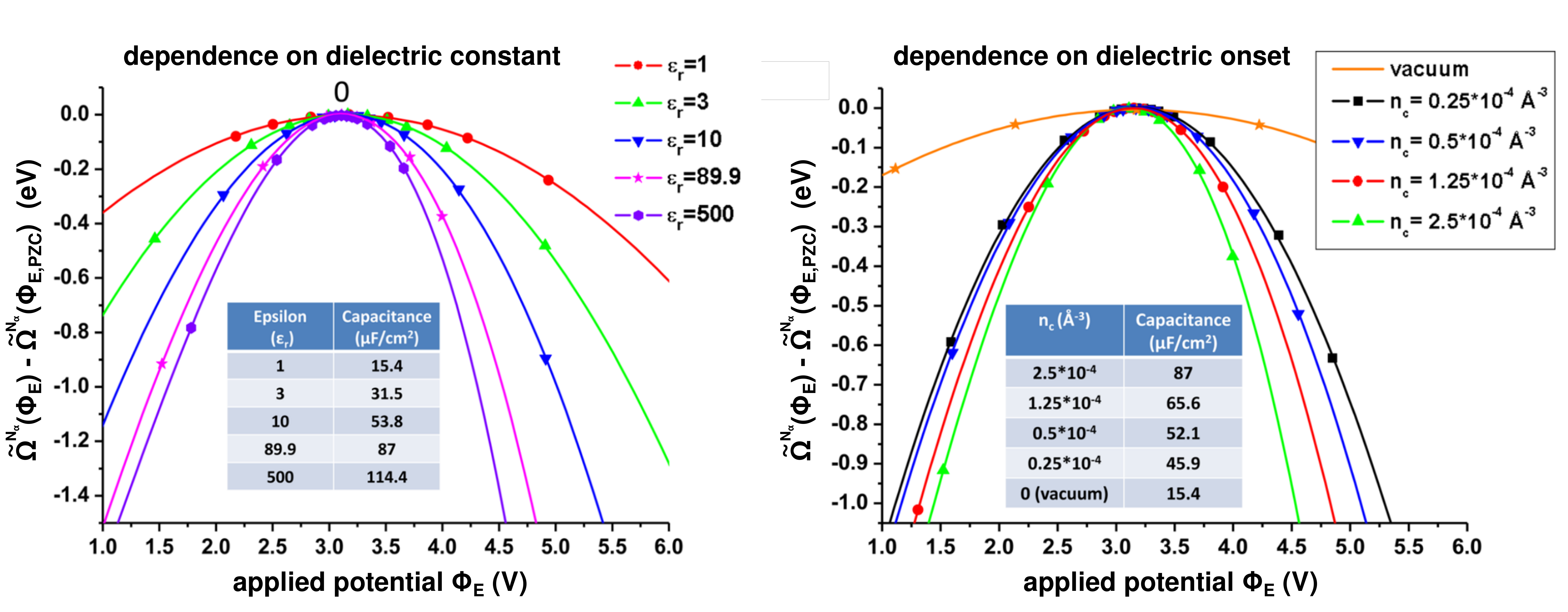}
\caption{{\bf Dependence of the interfacial capacitance on implicit solvation model parameters.} Shown is the variation of the fixed-composition total grand potential energy $\tilde{\Omega}^{N_\alpha}$ around the PZC for a model Li(110) electrode in implicit ethylene carbonate (EC) solvent ($\varepsilon(\mathrm{exp}) = 89.9$)
. The parabolic variation nicely reflects the Taylor expansion of eq.~(\ref{eq:cleanparabola}) and allows to fit the interfacial capacitance. (Left panel) Variation as a function of the bulk permittivity employed in the implicit solvation model. (Right panel) Variation as a function of the threshold charge density (called $n_{\rm c}$) employed to define the solvation cavity. Adapted from ref.~\citenum{lespes2015using}.}  
\label{fig:filhol}
\end{figure}

For a first such assessment it is instructive to perform a Taylor expansion of $\tilde{\Omega}^{N_\alpha}[\rho_{\rm el}^\circ(\Phi_{\rm E})]$ around the PZC, which can be achieved fully analytically up to second order
\cite{hormann2020electrosorption}
\begin{equation}
\tilde{\Omega}^{N_\alpha}[\rho_{\rm el}^\circ(\Phi_{\rm E})] =  \tilde{\Omega}^{N_\alpha}[\rho_{\rm el}^\circ(\Phi_{\rm E, PZC})] -
\frac{AC_\mathrm{PZC}}{2}\left( \Phi_{\rm E} -  \Phi_{\rm E, PZC} \right)^2 + 
\mathcal{O}(\left(\Phi_{\rm E} - \Phi_{\rm E,PZC}\right)^3) \quad ,
\label{eq:cleanparabola}
\end{equation}
with $C_\mathrm{PZC}$ the area normalized interfacial capacitance at the PZC, and $A$ the surface area (which in practice in a symmetric slab calculation would comprise the upper and lower side of the slab). Consistent with the above discussed physics of capacitive charging when moving away from the PZC, it is thus $C_\mathrm{PZC}$ that naturally appears in the thermodynamics as a central quantity. This analysis correspondingly suggests that next to the PZC discussed in Section~\ref{sec:pzc}, implicit electrolyte models should especially be able to appropriately describe the capacitance $C_{\rm PZC}$ at the PZC to achieve a sound potential-dependence of grand potential energies $\tilde{\Omega}^{N_\alpha}[\rho_{\rm el}^\circ(\Phi_{\rm E})]$ away from the PZC. Figure~\ref{fig:filhol} from Lespes and Filhol~\cite{lespes2015using} illustrates that in principle already fully-implicit models can offer this capability. Their data for Li(110) nicely portrays the inverted parabolas of the Taylor expansion, eq.~(\ref{eq:cleanparabola}), and the sensitive dependence of the extracted capacitance $C_{\rm PZC}$ on central parameters of the implicit solvation model, namely the bulk permittivity value and the iso-surface value of the electron density defining the solvation cavity, cf.~Section~\ref{sec:Elstat}. Note in particular the small capacitance values obtained in vacuum that can only be increased when considering the polarization response of the surrounding liquid through the implicit solvation model.

Further away from the PZC, higher-order terms in the Taylor expansion of eq.~(\ref{eq:cleanparabola}) will start to play a role, which will then feature variations of the interfacial capacitance with the applied potential. As a result, quite some work has been dedicated to construct implicit models that can reproduce experimentally observed capacitance variations with applied potential as well as electrolyte composition.\cite{andreussi2019continuum,schwarz2020electrochemical,sundararaman2018improving} As already introduced in Chapter \ref{sec:intro}, the total interfacial capacitance can be seen as arising from two capacitors in series, the inner DL and the outer DL. For high electrolyte concentrations or for potentials far away from the PZC, this total capacitance will correspondingly be dominated by the capacitance of the inner DL, where the highest potential drop occurs.\cite{parsons1997metal} Consistent with this picture, the modeling of the diffuse DL is often found to play a minor role to describe the interfacial capacitance in this limit.\cite{ringe2019understanding,ringe2020double} 
Instead, it is the appropriate parametrization of the solvation cavity boundary that critically determines the overall accuracy,\cite{sundararaman2017electrochemical} and it is within this understanding that refined models that include nonlinear electrolyte and dielectric response e.g.~in form of a dielectric decrement as discussed in Section~\ref{sec:solventstruct} are currently being pursued.\cite{sundararaman2018improving}

\subsection{Constant potential vs. constant charge calculations}
\label{sec:constantpotconstantcharge}

As discussed in Section~\ref{sec:CHE}, the capability to compute applied-potential dependent fixed-composition total grand potential energies $\tilde{\Omega}^{N_\alpha}[\rho_{\rm el}^\circ(\Phi_{\rm E})]$ is a key prerequisite to access thermodynamic energetic quantities like surface free energies or adsorption free energies, see e.g. eq.~(\ref{eq:Gads_Hproton}) for the discussed example of hydrogen adsorption. To this end, the flexibility with which implicit electrolyte models allow to consider finite surface charges in the periodic DFT supercell calculations provides primarily an opportunity to go beyond the CHE approximation. Within the prevalent canonical DFT implementations that work with a prescribed number of electrons $N_{\rm el}$, the amount of surface charge that corresponds to a particular applied potential $\Phi_{\rm E}$ can e.g. straightforwardly be obtained from the condition that the DFT-internal electron chemical potential has to equal the external electron electrochemical potential at equilibrium. As discussed before in Section~\ref{sec:pzc}, the prior is given by the Fermi level position $\phi_{\rm F}(N_{\rm el})$ with respect to the DFT-internal vacuum reference $\phi_{\rm vac}$, while the latter is imposed by the applied potential. Thus, $N_{\rm el}(\Phi_{\rm E})$ is determined by varying the number of electrons in the calculation until
\begin{align}
e( \phi_{\rm F}(N_{\rm el}) - \phi_{\rm vac} ) = \tilde{\mu}_{\rm el} = -e\Phi_{\rm E} 
 \label{eq:eq_cond_nrelectrons}
\end{align}
is fulfilled.\cite{goodpaster2016identification,kastlunger2018controlled,van2019assessment,melander2020grand} For this electron number, the desired fixed-composition total grand potential energy is then evaluated as
\begin{align}
\tilde{\Omega}^{N_\alpha}[\rho_{\rm el}^\circ(\Phi_{\rm E})] =  \Omega^{N_\alpha}[\rho^\circ_\mathrm{el}(N_\mathrm{el}(\Phi_\mathrm{E}))] +e \Phi_\mathrm{E} N_\mathrm{el}(\Phi_\mathrm{E}) \quad ,
\end{align}
which mathematically can be identified as a Legendre transformation between the variables $N_{\rm el}$ and $\Phi_{\rm E}$. Equivalent results can be achieved via the use of a potentiostat\cite{bonnet2012first} or in grand canonical DFT via an extended Hamiltonian.\cite{sundararaman2017grand} Note that, if it is of interest to understand the interface energetics for a range of applied potentials, it can be computationally more efficient to simply compute $\Omega^{N_\alpha}[\rho^\circ_\mathrm{el}(N_\mathrm{el})]$ as well as $\Phi_\mathrm{E}(N_\mathrm{el})$ ($\leftrightarrow \phi_{\rm F}(N_{\rm el})$) for a set of electron numbers $\left\{ N_\mathrm{el} \right\}$ from which the (continuous) $\tilde{\Omega}^{N_\alpha}[\rho_{\rm el}^\circ(\Phi_{\rm E})]$ can then be obtained in a straightforward way e.g.~via interpolation\cite{hormann2019grand}.

In principle, correspondingly determined $\tilde{\Omega}^{N_\alpha}[\rho_{\rm el}^\circ(\Phi_{\rm E})]$ still suffer from the difficulty of referencing to the DFT-internal vacuum reference $\phi_{\rm vac}$ in implicit solvation calculations as discussed in Section~\ref{sec:pzc}. However, fortunately only differences of such fixed-composition total grand potential energies are typically required for targeted quantities like an adsorption free energy. 
In such differences, one can consistently reference to the available bulk implicit solvent potential, which then only implies a residual constant shift of the potential dependence of a quantity like $\Delta G^{{\rm ads},N_\alpha}(\Phi_{\rm E})$ with respect to an experimental scale. Accepting such constant uncertainty, empirical values are then also conveniently taken for additionally required electrochemical potentials of those particle species that vary in these differences, like the $\tilde{\mu}_{{\rm H}^+}$ of a solvated proton in the hydrogen adsorption example.

On the other hand, the necessity to evaluate differences also creates challenges, in particular with respect to the implicit solvation modeling. To understand this, let us recap the equation with which the hydrogen adsorption free energy of Section~\ref{sec:CHE} would be determined in this constant-potential approach
\begin{align}
    \Delta G^{{\rm ads},N_\alpha}_{\rm H}(\Phi_{\rm E}) =  \tilde{\Omega}^{N_\alpha, {\rm H}}[\rho^\circ_\mathrm{el,H}(\Phi_{\rm E})] - \tilde{\Omega}^{N_\alpha}[\rho^\circ_\mathrm{el}(\Phi_{\rm E})] \quad . 
\label{eq:Gads_Hproton_recap}
\end{align}
Both fixed-composition total grand potential energies are obviously here evaluated at the same applied potential $\Phi_{\rm E}$. However, as discussed in the last section, the PZCs of the H-covered and the clean surface are generally different. This means that the evaluation of the two grand potential energies proceeds at a different relative potential with respect to the respective PZC. It could for example be that the applied potential is actually quite close to the PZC of the clean surface, but rather far away from the H-covered one. In the last section we had seen that the accuracy of implicit electrolyte models typically depends on this relative difference from the PZC. A model might be better suited to describe the potential region around the PZC, while another one is tailored just for the inner DL-dominated region far away from the PZC. In the difference of eq.~(\ref{eq:Gads_Hproton_recap}), the model can instead be required to describe quite different relative potential regions and this could introduce quite some error.

Aiming for better error cancelation, we can revisit the CHE and emphasize an aspect of it that is actually often overlooked. As discussed, the prevalent form of the CHE assumes that the optimized electron density of a given interface configuration $N_\alpha$ at any applied potential $\Phi_{\rm E}$ always remains the same as at its PZC, cf. Section~\ref{sec:CHE}. Intriguingly, this implies that in the difference required for the adsorption free energy, the two fixed-composition grand potential energies are actually evaluated at different potentials. In the hydrogen adsorption case and the corresponding eq.~(\ref{eq:Gads_Hproton_CHE}), this would namely be at the PZC of the hydrogen-covered surface $\Phi_{\rm E,H,PZC}$ and at the PZC of the clean surface $\Phi_{\rm E,PZC}$. It is now tempting to transfer this aspect to the surface charging case. If both terms in eq.~(\ref{eq:Gads_Hproton_recap}) were not evaluated at the same potential, but at the same amount of surface charging, then both fixed-composition total grand potential energies are determined at an approximately equal relative potential with respect to their PZC and one can hope for maximum error cancellation in the implicit model. Of course, on the other hand, error is introduced because at least one of the two terms is not computed at the correct applied potential---but possibly this incurred error is smaller than the error cancellation achieved. 

This is essentially the philosophy of so-called constant-charge calculations, which can in principle be carried out with explicit\cite{chan2015electrochemical,chan2016potential,sandberg2016co} or implicit\cite{ringe2019understanding,ringe2020double,vijay2020dipole,kim2021selective,lee2020electric,wu2019twodimensional} charging. In such calculations the amount of surface charge according to an applied electrode potential is determined e.g. from experimental \cite{parsons1997metal} or simulated\cite{ringe2019understanding,ringe2020double} charge-potential relations. The resulting decoupling of quantum chemistry and surface charging thus allows to control the accuracy of both scales roughly independently. While clearly inspired by the CHE, it is worthwhile to note that this approach is also closely related to traditional constant field calculations\cite{liu2003modeling,resasco2017promoter,hyman2005theoretical,lamoureux2019ph,kelly2020electric,chen2016electric,deissenbeck2021dielectric,norskov2004origin,karlberg2007estimations,panchenko2004ab,clark2019influence,che2018elucidating} as interfacial fields are naturally proportional to surface charge, and it is correspondingly not surprising that latter calculations were also taken into consideration in the early developments of the CHE approximation.\cite{liu2003modeling,panchenko2004ab,norskov2004origin,rossmeisl2005electrolysis,hyman2005theoretical,karlberg2007estimations}

Both the constant-potential\cite{goodpaster2016identification,kastlunger2018controlled,van2019assessment,hormann2019grand,melander2020grand,bonnet2012first,sundararaman2017grand,gauthier2019unified,gauthier2019practical,gauthier2019challenges,jinnouchi2017simulated,choi2021understanding} and constant-charge\cite{che2018elucidating,lamoureux2019ph,gauthier2019unified,ringe2019understanding,kelly2020electric,vijay2020dipole,ringe2020double,kim2021selective,lee2020electric,wu2019twodimensional} approach enjoy present popularity, with the former also often denoted as fully grand-canonical (FGC) approach. While they give quantitatively different results in practical supercell sizes, it is gratifying that in the thermodynamic limit of low-coverage adsorption, i.e. one adsorbate in a laterally infinitely extended supercell, both approaches will eventually coincide.\cite{gauthier2019practical} In this limit, the PZC of the clean surface will only be infinitesimally changed upon adsorption. At any applied potential, both the single-adsorbate covered and clean surface will thus be equally charged anyway, and one is in both cases also at an equal absolute and relative potential with respect to the joint PZC. For this limit, one can analogously to the constant-potential case also consider a constant-charge appropriate Taylor series expansion of the adsorption free energy, i.e. now in terms of the excess charge.\cite{gauthier2019practical,ringe2019understanding,hormann2020electrosorption} Here it has been found that in many cases, the first order term corresponding to the PZC change is dominating. In contrast, higher-order terms depending on the capacitance and thus electrolyte description are indeed less relevant, supporting the error cancelation motivation of this approach.\cite{ringe2019understanding} For the constant-potential case, a dipole-field-type first-order correction term to the CHE adsorption free energy can even analytically be derived.\cite{gauthier2019unified,hormann2020electrosorption, hormann2021thermodynamiccyclic,ringe2019understanding,ringe2020double} For the specific example of hydrogen adsorption, this term reads
\begin{align}
\Delta G^{{\rm ads},N_\alpha}_{\rm H(\theta\rightarrow0)}(\Phi_{\rm E}) -
\Delta G^{{\rm ads},N_\alpha}_{\rm H(\theta\rightarrow0),CHE}(\Phi_{\rm E}) \approx
    e\;(l -1)\;(\Phi_{\rm E} - \Phi_{\rm E,PZC}) \quad ,
\end{align}
with the already earlier introduced electrosorption valency $l$ as measure of the number of electrons dragged onto the electrode upon adsorption of the proton. In the CHE, $l_{\rm CHE}=1$, corresponding to a PCET process, while in general $l\ne 1$.

\begin{figure}[htb]
\includegraphics[width=0.8\textwidth]{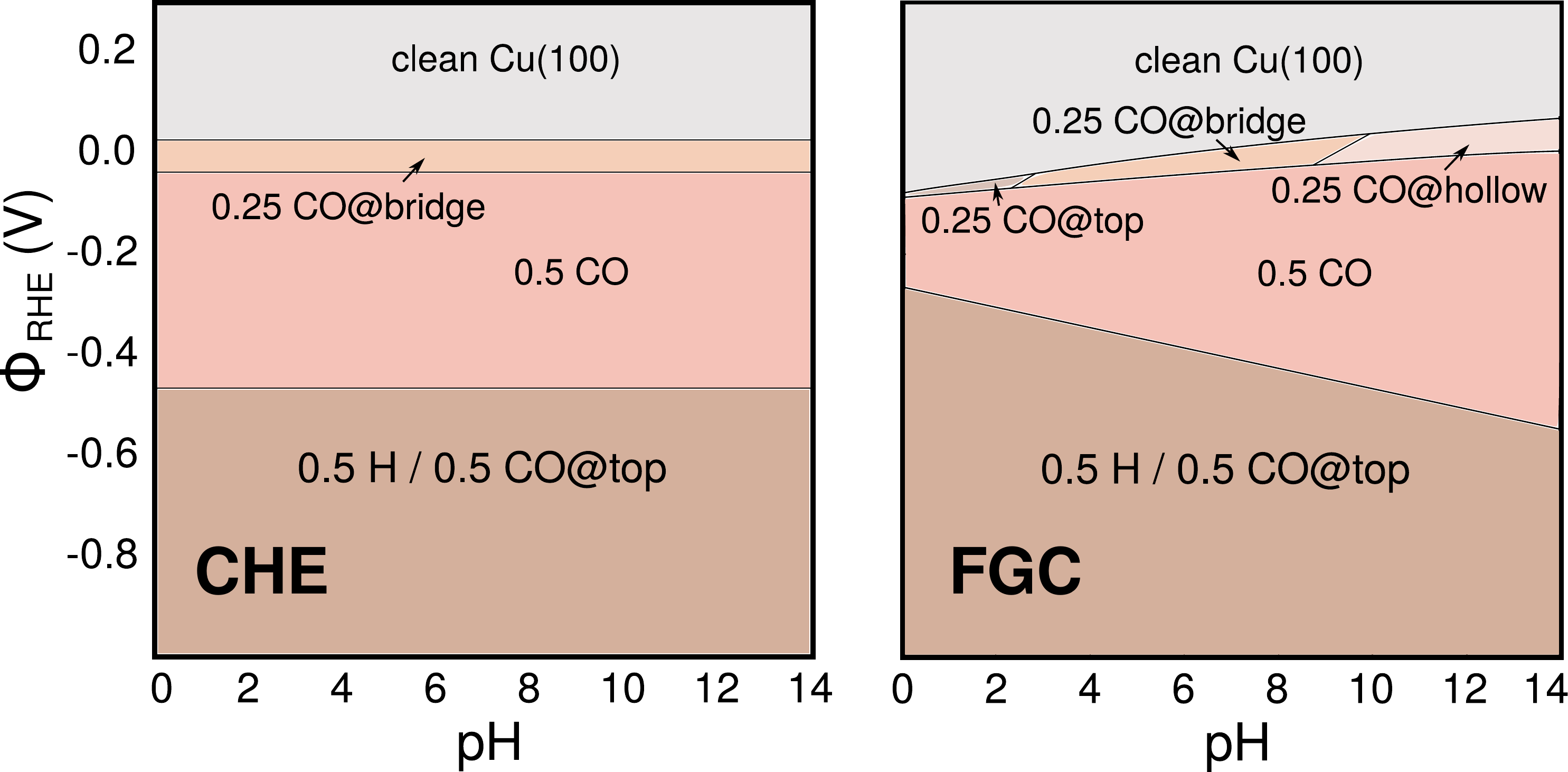}
\caption{{\bf Theoretical surface Pourbaix diagram of Cu(100) in implicit water considering H and CO adsorbates.
} The diagram obtained within the CHE approximation (left panel) shows only a trivial Nernstian pH-dependence, which vanishes on the here employed RHE scale. In contrast, non-trivial pH-dependencies are obtained with constant-potential aka FGC calculations (right panel). Figure created from data published in ref.~\citenum{huang2018potential}.}
\label{fig:PourbaixFGC}
\end{figure}

Compared to the CHE, changes in the potential-dependence of adsorption free energies as obtained in emerging constant-potential or constant-charge calculations seem indeed generally largely be understandable in terms of dipole-field interaction, even for larger molecules, such as e.g.~CO$_2$ reduction intermediates\cite{zhang2018importance}. Changes in this dipole-field interaction with applied potential can then lead to a wide range of conceptual physics that was outside of the realm of CHE theory. This includes a potential-induced switching of the most stable adsorption site\cite{hormann2019grand,weitzner2020toward,clark2019influence} or altered adsorbate geometries or adsorption motives\cite{steinmann2016assessing,goodpaster2016identification}, including e.g.~interfacial water\cite{filhol2013an}. Recent corresponding results have for instance also helped to clarify the impact of different cationic species\cite{waegele2019how,ringe2019understanding} on the interfacial capacitance\cite{garlyyev2018influence,ringe2019understanding,resasco2017promoter}, and how this can influence in an indirect way via the variation in the dipole-field interaction a variety of electrochemical observables such as the stability of adsorbed CO$_2$\cite{zhang2018importance,ringe2019understanding,vijay2020dipole,weitzner2020toward} and the Stark shifts of CO\cite{dabo2010towards,bonnet2014chemisorbed,ringe2019understanding}. In addition, the now available possibility to appropriately account for effects of the applied potential beyond the CHE has enabled significant progress in the simulation of electrochemical reaction barriers\cite{kastlunger2018controlled,gauthier2019unified} and bridged the gap to works that employ explicit charging strategies.\cite{skulason2007density,rossmeisl2008modeling,skulason2010modeling,bjorketun2013avoiding,ping2017reaction,gauthier2019unified,van2019assessment,chan2015electrochemical,chan2016potential}
Further works reported potential-induced surface reconstructions or lifting of those\cite{kolb1996reconstruction,hormann2019grand} as well as non-Nernstian dependencies for surface coverages\cite{bonnet2013first,hormann2019grand,sundararaman2017grand,weitzner2017voltage,weitzner2017quantum,mathew2019implicit}, nanoparticle shapes\cite{bonnet2013first,hormann2019grand}, Pourbaix diagrams\cite{huang2018potential,karmodak2020catalytic,kopac2021building}, and last but not least (thermodynamic) cyclic voltammograms \cite{hormann2019grand,hormann2021thermodynamic,hormann2021thermodynamiccyclic,jinnouchi2017simulated}, where peak positions and shapes can indeed be extremely sensitive to the electrochemical conditions\cite{rizo2015towards,mccrum2016first}. These developments are all quite recent and we expect significant further progress in our understanding of interfacial electrocatalysis to emerge from such constant-potential or constant-charge calculations\cite{abidi2021atomistic}. As one final example we only highlight in Fig.~\ref{fig:PourbaixFGC} how constant-potential calculations help to overcome the trivially Nernstian pH-dependence of the CHE approach, cf.~eq.(\ref{eq:RHE}). In the Pourbaix diagram for Cu(100) shown on the RHE potential scale, there is correspondingly no further pH dependence within the CHE, but significant structure when computed with the constant potential approach.

\section{Conclusions and outlook}

Predictive-quality first-principles calculations based on DFT have undoubtedly become a cornerstone in modern materials, catalysis and energy research. In the specific context of catalysis at electrified interfaces, this development is largely connected with the ingenious computational hydrogen electrode (CHE) approach of Rossmeisl, N{\o}rskov and coworkers.\cite{norskov2004origin,rossmeisl2005electrolysis} It is difficult to understate the impact that this single approach has made on the design of electrocatalysts or the unraveling of electrochemical reaction mechanisms.\cite{seh2017combining, norskov2009towards,zeng2015towards,bagger2020fundamental,callevallejo2012first,abidi2021atomistic} By the very nature of its approximation, the CHE puts the predominant emphasis on the electrode site. Over the last decade or so, first-principles electrocatalysis research at solid-liquid interfaces (SLIs) was correspondingly dominated by finding optimum catalyst materials that lie at the top of reaction volcanos or gaining mechanistic understanding in terms of surface chemical bonds, yet without much caring for the electrolyte side of the SLI. 

It is only within the last few years that an ever increasing understanding of electrified SLIs starts to trigger a return to this foundational pillar of electrochemistry, namely the influence of the electrolyte at the SLI.\cite{debye1923theory,chapman1913theory,helmholtz1853ueber,stern1924zur,gouy1910constitution,gouy1916sur} Unfortunately, it is also only when one starts to devise strategies of how to actually do so within the realm of present-day DFT and supercomputer capabilities, that one really starts to appreciate the ingenuity of the CHE approximation and the simplicity of the calculations it enables. Any real consideration of the extended double layer (DL) with its intricate long-range electrostatics and inherent dynamics quickly leads to excessive computational costs. In this respect, implicit solvation methodology forms a unique compromise. As we have surveyed in this review, consideration of corresponding methodology within the {\em ab initio} thermodynamics framework commonly employed in surface catalysis anyway, immediately gives rise to multiple avenues beyond the CHE. At the same time, the computational cost of corresponding constant potential or constant charge calculations stays not too different from the one of the CHE approach.

While thus highly promising, this approach is not without its own challenges. The implementation of implicit solvation functionality into a series of powerful DFT software packages that can describe extended SLIs typically within the frame of periodic boundary condition supercells has been the enabler for this new field and a great community effort. However, in these implementations the methodological framework, historically developed to assess solvation effects on molecular solutes, has largely been left unaltered. To one end, this concerns the usage of parametrizations derived from molecular experimental reference data. To the other end, functional expressions for the effectively treated explicit electrode - implicit electrolyte interactions have if at all only marginally been modified, for instance if they contained quantities like a cavity volume that is not accessible at an extended SLI. As we have seen in the course of this review, the primary advance brought about by implicit solvation for the SLI context is more the flexibility with which one can represent the counter charges in the DL, rather than the actual account of solvation effects. For this purpose, the present state-of-the-art may largely be sufficient---and in addition to the already obtained massive insight into catalysis at electrified interfaces, we expect truly grand-canonical results (like the discussed constant potential or constant charge calculations) on the basis of existing implicit electrolyte models to continue carving out important electrochemistry that was not accessible within the CHE framework. 

However, this can only be a first step. At present, the community is at a crossroad. One route is to focus efforts towards mixed explicit/implicit solvation approaches. The other is to refine the implicit solvation technology itself. For both case, what will centrally be required is a re-thinking of the functional expressions, in particular of the non-electrostatic terms, and reference data that is more pertinent for extended SLIs. Regarding the latter, we have seen throughout the review, that many experimentally accessible quantities commonly measured in electrochemistry are not ideal for this task, as their computation intricately mixes solvation effects with the specificities of the employed {\em ab initio} thermodynamics ansatz. In this respect, more, systematic and accurate measurements of PZCs and interfacial capacitances for well-defined model electrodes would certainly be helpful. In our view, also contact angles could be another highly useful class of quantities. Without any such data, it is largely impossible to develop highly parametrized and thus potentially more accurate implicit solvation methods without running into overfitting. From this perspective, the actual developments of first-principles machine-learned interatomic potentials are probably most exciting.\cite{ko2021fourth,csanyi2021gaussian} Explicit AIMD data\cite{cheng2009redox,cheng2010aligning,cheng2012alignment,le2017determining,le2018structure,le2020molecular,le2021modeling,sakong2016structure, sakong2018electric,ludwig2019solvent,heenen2020solvation,ambrosio2016structural,bouzid2018atomic,ambrosio2018absolute,goldsmith2021effects,surendralal2018first} has long been used to validate and improve implicit solvation methodology. If machine-learned interatomic potentials allow to generate comparably accurate, but orders of magnitude longer trajectories and in larger simulation cells, then this will be an invaluable asset that might even ultimately enable to validate and refine implicit solvation schemes for application outside the domain of {\em ab initio} thermodynamics, notably the modeling of kinetic reaction barriers.

\section{Acknowledgements}

S.R. acknowledges funding from the National Research Foundation (NRF-2019R1A2C2086770), (NRF-2021R1C1C1008776) funded by the Korea Ministry of Science, ICT \& Future Planning. N.G.H., H.O., and K.R.~gratefully acknowledge funding by the Deutsche Forschungsgemeinschaft (DFG, German Research Foundation) under Germany's Excellence Strategy – EXC 2089/1 – 390776260, as well as under project RE1509/35-1. N.G.H. acknowledges financial support through the EuroTech Postdoc Programme which is co-funded by the European Commission under its framework programme Horizon 2020 and Grant Agreement number 754462. 
H.O. further acknowledges funding by the Deutsche Forschungsgemeinschaft (DFG, German Research Foundation) through its Heisenberg program, grant No.~OB 425-9/1.
The authors acknowledge the core developers of the DFT program packages that helped in confirming the correctness of the content in Table~\ref{tab:implicit_solvation_codes}, as well as thank the developers of the implicit solvation functionality in these codes for their great service to the community.

\clearpage
\bibliography{bibliography}

\end{document}